\documentclass[reprint,
unsortedaddress,
nobibnotes,
amsmath,amssymb,
aps,
prb,
]{revtex4-2}

\usepackage{graphicx}
\usepackage{dcolumn}
\usepackage{bm}
\usepackage{braket}
\usepackage{graphicx}
\usepackage{latexsym}
\usepackage{amsfonts}
\usepackage{bm}
\usepackage{multirow}
\usepackage{color}
\usepackage{xcolor}
\usepackage{calrsfs}
\usepackage{dutchcal}
\newcommand{\be}{\begin{equation}}
\newcommand{\ee}{\end{equation}}
\newcommand{\bea}{\begin{eqnarray}}
\newcommand{\eea}{\end{eqnarray}}

\usepackage{graphicx}
\usepackage{amsthm}

\usepackage{amsfonts} 
\usepackage{fancyhdr} 
\usepackage{comment}
\usepackage[caption=false]{subfig}
\usepackage{physics}
\usepackage{float}
\usepackage{soul}
\usepackage{indentfirst}
\usepackage{oplotsymbl}
\usepackage{tcolorbox}
\usepackage{appendix}
\usepackage[hidelinks, colorlinks=true,
linkcolor=blue, citecolor=blue, pdfauthor={Leandro M. Chinellato},
pdftitle={Quantum spin dynamics in spin-1/2 soliton chain}]{hyperref}

\renewcommand{\eqref}[1]{Eq.~\hyperref[#1]{(\ref*{#1})}}

\newcommand{\Mod}[1]{\ (\mathrm{mod}\ #1)}
\begin{document}

\title{Dynamics of  Quantum Chiral Solitons}
\author{Leandro M. Chinellato}
\thanks{\url{lchinell@vols.utk.edu}}
\affiliation{Department of Physics and Astronomy, University of Tennessee, Knoxville, TN 37996, USA}
\author{Oleg A. Starykh}
\thanks{\url{oleg.starykh@utah.edu}}
\affiliation{Department of Physics and Astronomy, University of Utah, Salt Lake City, UT 84112, USA}
\author{Cristian D. Batista}
\thanks{\url{cbatist2@utk.edu}}
\affiliation{Department of Physics and Astronomy, University of Tennessee, Knoxville, TN 37996, USA}
\affiliation{Neutron Scattering Division, Oak Ridge National Laboratory, Oak Ridge, Tennessee 37831, USA}

\date{\today}

\begin{abstract}
We introduce a nonperturbative framework for quantizing chiral solitons in interacting quantum spin chains. This approach provides a direct lattice extension of the well-established $S$-duality between the sine-Gordon and Thirring models, thereby bridging the gap between continuum dualities and their lattice counterparts. By constructing the quantum chiral-soliton operators explicitly, we show how their unconventional dynamics appear in the excitation spectrum and correlation functions across the full Brillouin zone. A key result is that the dominant soliton tunneling amplitude alternates in sign, $\operatorname{sgn}(t_{1+}) = (-1)^{2S+1}$, sharply distinguishing half-odd-integer from integer spin chains. We further identify characteristic signatures of these chiral excitations in the dynamical spin structure factor, demonstrating their visibility in inelastic neutron scattering. Our results open a route to experimentally probing nonperturbative features of dual quantum field theories in condensed-matter settings.
\end{abstract}

\maketitle

\section{Introduction}

Topological excitations play a central role across many areas of physics,
from nonlinear wave equations to quantum field theory and condensed-matter systems.
Classically, solitons are localized, particlelike textures whose proliferation
generically produces a periodic soliton crystal~\cite{Kishine2015,chaikin2000}.
Quantum mechanics fundamentally alters this picture.
Zero-point motion can melt the classical crystal into a quantum liquid, as happens in the quantum spin-1/2 chain considered in this work.
This result raises the
following questions: How
do topological textures become
genuine quantum 
quasiparticles, and 
how do quantum fluctuations reshape the phases that emerge from their condensation?

Continuum approaches based on the sine-Gordon model and its dual massive Thirring theory
provide an elegant long-wavelength description of quantum solitons,
where solitons map to fermionic excitations under $S$-duality \cite{Rajaraman1982}.
However, such treatments rely on gradient expansions
and are not designed to yield quantitative, momentum-resolved predictions
for microscopic lattice magnets or their experimentally measured response functions.
In particular, they do not directly establish how topological excitations
manifest in the dynamical spin structure factor
$\tilde{\mathcal{S}}(q,\omega)$ probed by inelastic neutron scattering (INS).

Magnetic chains provide an especially fertile ground for exploring soliton physics in quantum materials~\cite{Dzyaloshinskii1964,Izyumov1984,Kosevich1990,Kishine2015,Tereshchenko2024,Bostrem2025}.

In this work, we develop a fully quantum, nonperturbative, lattice-resolved
framework for chiral solitons in a ferromagnetic spin-$1/2$ chain with
Dzyaloshinskii-Moriya interactions and a magnetic field applied parallel to the easy plane.
Our approach constructs momentum-resolved many-body wave functions
for quantum chiral solitons directly from the microscopic Hamiltonian,
without invoking any small expansion parameter,
such as $1/S$ or the continuum ratio $a/\ell_s$
(where $a$ denotes the lattice spacing and $\ell_s$ the soliton size).
Consequently, the theory remains quantitatively controlled
even in the strongly quantum regime,
including the physically relevant case $S=1/2$,
where soliton cores extend over only a few lattice spacings.
Within this framework, solitons emerge as bona fide quantum quasiparticles
with a well-defined dispersion across the full Brillouin zone,
a calculable spectral weight in $\tilde{\mathcal{S}}(q,\omega)$,
and a systematically derived description of their interactions.

As the magnetic field approaches the saturation value $H=H_c$ from above,
quantum chiral solitons become the lowest-energy excitations over a finite
field window $H_c < H < H^\ast$.
At $H=H_c$, the soliton chemical potential vanishes,
while the conventional magnon mode remains gapped.
The persistence of a finite magnon gap implies that interactions between
solitons, mediated by virtual magnon excitations, are short ranged
and decay exponentially with distance.
In this dilute regime, topology and quantum fluctuations reorganize
the excitation hierarchy of the system:
Topological solitons, rather than spin-wave modes,
constitute the relevant low-energy degrees of freedom.

Through a Jordan-Wigner transformation,
these excitations admit a fermionic description.
Microscopically, we extend the sine-Gordon $S$-duality to the lattice
by constructing localized soliton creation and annihilation operators
directly from the spin Hamiltonian.
A Wannierization procedure yields orthonormal quantum soliton states,
systematically incorporating quantum fluctuations
and capturing the tunneling processes that endow solitons with mobility,
thereby generating a well-defined and nontrivial dispersion relation.
Remarkably, the dominant tunneling amplitude exhibits a spin-parity effect,
alternating in sign between half-odd-integer and integer spin chains.
The resulting dynamics is described by an effective, weakly coupled,
tight-binding fermionic Hamiltonian with a nontrivial dispersion relation.
Near $H=H_c$, the short-range interactions are irrelevant in the
renormalization-group sense,
rendering the critical point a free-fermion fixed point.
At finite soliton density, however,
the residual interactions become marginally relevant,
driving the system into a gapless Tomonaga-Luttinger liquid -- the quantum-melted counterpart of the classical soliton lattice phase.

A central question is whether these quantum solitons can be directly observed
above $H_c$ and how their dynamics intertwine with the gapped magnons.
Because INS measures the dynamical spin structure factor
$\tilde{\mathcal{S}}(q,\omega)$, a purely topological excitation would carry
negligible spectral weight unless it hybridizes with a single magnon mode.
Our lattice-resolved framework demonstrates that such visibility
is enabled by hybridization between chiral solitons and single-magnon modes.
This mixing transfers spectral weight to the soliton branch,
thereby rendering it experimentally accessible.

The hybridization becomes particularly pronounced for $H \gtrsim H^\ast$,
where the soliton and magnon dispersions approach each other and cross.
In this regime, level repulsion enhances the overlap between the two sectors,
producing a characteristic, field-dependent redistribution of spectral weight.
The resulting intensity of the soliton branch in $\tilde{\mathcal{S}}(q,\omega)$
provides a direct and quantitative measure of soliton-magnon hybridization.
The predicted dispersion and spectral-weight redistribution are
quantitatively benchmarked against density matrix renormalization group (DMRG)
and matrix product state (MPS) simulations, showing excellent agreement.
Together, these results establish a concrete experimental detection mechanism
for quantum chiral solitons and a lattice-level connection between
topological quasiparticles and measurable dynamical response functions.

More broadly, the procedure introduced here for promoting classical topological textures to quantum quasiparticles directly at the lattice level
provides a systematic route toward deriving effective low-energy theories
of interacting topological defects.
While demonstrated in one dimension,
the framework does not rely on integrability or continuum approximations
and can, in principle, be extended to higher-dimensional systems,
including skyrmion-hosting  magnets \cite{Bogdanov1989,Muehlbauer2009,Yu2010,Zhang2023,Williams2025}.
In this perspective, quantum liquids may emerge
as the quantum-melted phases of interacting topological textures,
offering an alternative microscopic pathway
toward strongly correlated topological matter.

\section{\label{Sec:Model}Model and Classical limit}

The simplest model that supports quantum chiral solitons is provided by the
one-dimensional spin-$\tfrac{1}{2}$ ferromagnetic Heisenberg chain of length $L$, oriented along the $x$ axis and subject to a Dzyaloshinskii–Moriya (DM) interaction characterized by a vector ${\bf D} = D \hat{z}$.  
An external magnetic field ${\bf H} = H \hat{x}$ is applied perpendicular to the DM vector. 
The Hamiltonian of the system is
\begin{equation}
\hat{H}_S = -\sum_{j=1}^{L} \Big[ 
    J\, \hat{{\bm S}}_{j} \cdot \hat{{\bm S}}_{j+1} 
    + D \big(\hat{S}_j^x \hat{S}_{j+1}^y - \hat{S}_j^y \hat{S}_{j+1}^x\big) 
    + H \hat{S}_j^x 
\Big],
\label{Eq:Ham}
\end{equation}
where $J > 0$ and $H = g \mu_B B$, with $\mu_B$ the Bohr magneton and $g$ the gyromagnetic factor. 
It is important to note that the magnetic field is applied transverse to the \(U(1)\)-invariant axis set by the DM interaction. Consequently, the full Hamiltonian has only discrete symmetries. 

We begin by summarizing key facts about the {\em classical} limit of \eqref{Eq:Ham}, $S \to \infty$.
It is obtained by replacing the spin operators with their expectation values over SU(2) spin coherent states $\ket{\Omega(\theta, \varphi)}$, which represent points on the unit sphere $S^2$ parametrized by the polar and azimuthal angles $\theta$ and $\varphi$. In other words we substitute $\hat{{\bf S}}_j \to S {\bf n}_j =  S (\sin\theta_j \cos\varphi_j,  \sin\theta_j \sin\varphi_j, \cos\theta_j)$. Since the Dzyaloshinskii–Moriya interaction favors spin alignment in the plane perpendicular to $\hat{z}$, the polar angle can be fixed to $\theta=\pi/2$, leading to a configuration of the form $S{\bf n}_j = S(\cos \varphi_j, \sin \varphi_j, 0)$.   For small pitch angles, $|\varphi_{j+1} - \varphi_j| \ll 1$, it is possible to take the long-wavelength (continuum) limit of the lattice model. In this regime, the system is effectively described by a static chiral sine-Gordon model for the angular field variable $\varphi = \varphi(x)$, with the Hamiltonian density given by
\begin{eqnarray}
    {\cal H}_{\text sG} =   JS^2a \left[ \frac{1}{2}(\partial_x \varphi)^2 - q_0\partial_x\varphi + m^2 (1-\cos{\varphi}) \right].
    \label{Eq:Hamcon}
\end{eqnarray}
Here  $q_0 \equiv D/(J a)$ is the wave number of the zero-field spiral, $a$ the lattice parameter, $m^2 \equiv H/(JSa^2)$, and the full Hamiltonian is given by $H_{sG} = \int dx \, {\cal H}$. A constant $m^2$ is added to make the energy of the uniform vacuum state equal to zero. 
The Hamiltonian functional  in \eqref{Eq:Hamcon} is minimized  by solving for the extreme condition $\delta {\cal H} = 0$,
\begin{equation}
    \frac{d}{dx}\frac{\partial {\cal H}}{\partial (\partial_x\varphi)} - \frac{\partial {\cal H}}{\partial\varphi} = 0.
\end{equation}
The variational equation is satisfied by the solutions of the static sine-Gordon equation,
\begin{equation}
    \frac{d^2\varphi}{dx^2} = m^2\sin{\varphi}
    \label{Eq:sineg}
\end{equation}
Clearly, in the absence of a magnetic field ($m=0$), the ground state is a chiral helix~\cite{Dzyaloshinskii1964} given by $\varphi(x) = q_0x + \varphi(0)$, with the sign of $q_0$ determining the helix chirality.

\begin{figure}[!ht]
    \centering
    \includegraphics[width=\columnwidth]{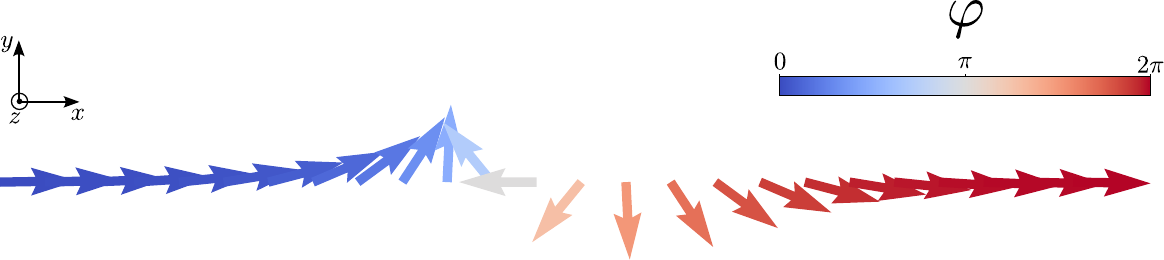}
    \caption{Profile of the classical chiral soliton. Arrow directions indicate spin orientation, while their color encodes the value of the angle $\varphi$. }
    \label{Fig:soliton_profile}
\end{figure}

\begin{figure}[!ht]
  \centering
    \includegraphics[width=\columnwidth]{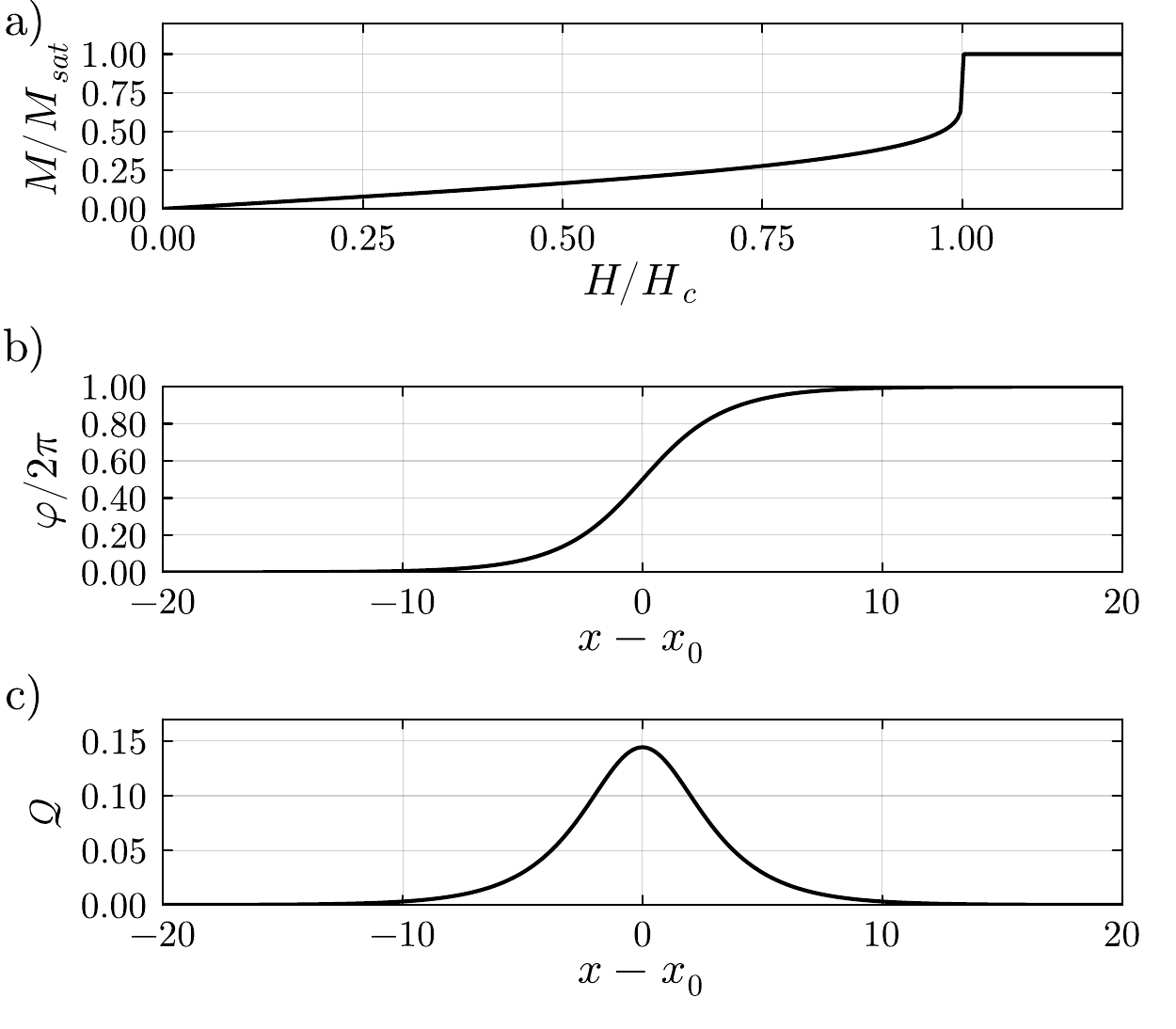}
    \caption{a) Field dependence of the magnetization. b) Spatial profile of the soliton solution $\varphi_{+}(x)$ in units of $2\pi$. c) Spatial profile of the topological charge density $\mathcal{Q}(x)$ associated with a single soliton at the critical field $H = H_c$. }
    \label{Fig:cont_mz}
\end{figure}

When an external magnetic field is applied ($m \neq 0$), soliton solutions appear. They represent a compromise between the exchange and Dzyaloshinskii–Moriya interactions, which favor a uniform spiral, and the transverse magnetic field, which tends to align the spins along a fixed direction within the spiral’s polarization plane. These competing tendencies are balanced by the soliton configuration. The single soliton $(\tau = +1)$ and antisoliton $(\tau = -1)$ solutions are described by
\begin{equation}
\varphi_{\tau}(x) = \tau \, 4 \arctan\left[e^{m(x - x_0)}\right] = - 2 \tau i\, \ln\frac{1+ i e^{m(x-x_0)}}{1- i e^{m(x-x_0)}}.
\label{eq:classol}
\end{equation}
The corresponding soliton spin configuration is shown in Fig.~\ref{Fig:soliton_profile}, while its field profile, $\varphi_{\tau = +1}(x)$, is plotted in Fig.~\ref{Fig:cont_mz}(b). Solitons have lower energy than antisolitons due to the presence of the DM term in \eqref{Eq:Ham}. To verify that our definition is consistent with this condition, consider a spin at site $i$ pointing along the $x$ axis and another spin at site $j$ rotated around the $z$ axis. The corresponding DM energy contribution is $\sim -\sin\varphi$. For the excitation to lower the energy, we require $\varphi > 0$, which corresponds to a counterclockwise rotation (with $\varphi$ measured from the $x$ axis under the standard convention). Since our operator produces precisely this type of rotation, we conclude that it generates excitations with the energetically preferred chirality. 

The size of the soliton is determined by the magnetization $\Delta M^x$ it carries with respect to the fully polarized (FP) state,
\begin{equation}
\Delta M^x = S \int_{-\infty}^{\infty} \left[1 - \cos{\varphi_{\tau}(x)}\right] dx = \frac{4S}{m}.
\label{Eq:cont_mag}
\end{equation}
The characteristic length is therefore $\ell_s(H) = \Delta M^x/S = 4/m(H)$, making explicit its dependence on the external magnetic field $H$.

The energy of a single soliton is also easy to calculate, ${\cal E}_1 = J S^2 a (8 m - 2\pi q_0)$. Therefore, the energetic cost of the soliton vanishes at the quantum critical point $H_c$~\cite{Kishine2015}:
\begin{equation}
    H_c = \left(\frac{\pi D}{4aJ}\right)^2 JSa^2.
    \label{Eq:Hc}
\end{equation}
For $H < H_c$, the solitons proliferate and form an ordered chiral soliton lattice (CSL). This phase emerges from the competition between 
the energetic gain of soliton condensation and the repulsive intersoliton interaction, which decays exponentially with the distance between solitons \cite{chaikin2000}. The chiral soliton lattice \cite{Kishine2015} is described by
\begin{equation}
\varphi(x) = 2\mathrm{am}(\overline{x}) + \pi,
\label{Eq:csl}
\end{equation}
where $\overline{x} = (m/\kappa)x$ and $\mathrm{am}$ denotes Jacobi’s amplitude function with elliptic modulus $0 < \kappa < 1$. By substituting \eqref{Eq:csl} into the spin components and applying trigonometric identities, we obtain $S{\bf n} = S(2\text{sn}^2(\overline{x})-1, -2\text{sn}^2(\overline{x})\text{cn}^2(\overline{x}),0)$. For sufficiently strong external fields, the FP state ($\varphi = 0$) is recovered as the lowest-energy configuration. The magnetization along the field direction is given by:
\begin{equation}
    M^x =  S \left( \frac{2}{\kappa^2} - \frac{2E}{\kappa^2K}-1\right),
\end{equation}
where $K \equiv K(\kappa)$ and $E \equiv E(\kappa)$ are the complete elliptic integrals of the first and second kinds, respectively~\cite{Kishine2015}.
The elliptic modulus $\kappa$ is related to the external magnetic field through $\kappa = \tfrac{4E}{\pi q_0} \sqrt{\tfrac{H}{JSa^2}}$ ~\cite{Kishine2015}. As shown in Fig.~\ref{Fig:cont_mz} (a), the magnetization curve exhibits a distinctive profile; most notably, it evolves continuously with the applied field approaching saturation with an infinite derivative. The magnetization is directly related to the number of solitons because the ``addition'' of a single soliton to the FP state reduces the total magnetization by a finite amount $\Delta M^x$. The continuity of the magnetization curve thus indicates that the density of chiral solitons vanishes as the quantum critical point is approached from below, $H \to H_c^{-}$, in agreement with the soliton-condensation mechanism: the single-soliton energy vanishes continuously at $H=H_{c}$ [\eqref{Eq:Hc}].

This continuous evolution stands in sharp contrast to other topological spin systems---such as conventional skyrmion crystals---where the field-driven transition from a topological to a polarized phase is first order, and the density of topological solitons cannot be reduced continuously \cite{Bogdanov1989,Muehlbauer2009,Yu2010,Zhang2023,Williams2025}.

In the present case, the field-induced commensurate-incommensurate transition in the classical model is continuous and reflects the gradual suppression of the soliton density as the magnetic field increases. In the quantum spin-$1/2$ chain, this transition is replaced by a continuous quantum phase transition between a Tomonaga-Luttinger liquid and a gapped, fully polarized phase (see \cite{Giamarchi2004} and Appendix \ref{Appe:sine-Gordon}). 

The defining feature of the soliton is its topological structure. The angle $\varphi_\tau(x)$ winds by $2\pi$ as $x$ goes over the chain from $-\infty$ to $+\infty$, as Fig.\ref{Fig:cont_mz}(b) illustrates. Mathematically, in the continuum limit, the projection of the spin configuration onto the $xy$ plane defines a mapping $f:S^1\to S^1$. The base manifold $S^1$ corresponds to the spatial coordinate $x$ of the chain with periodic boundary conditions (PBC), while the target manifold, also $S^1$, represents the possible orientations of the spin projection in the $xy$ plane. The net soliton charge $\mathcal{C}$ can therefore be identified with the topological degree of this mapping, i.e., the winding number, which is classified by the fundamental homotopy group $\pi_1(S^1)\cong\mathbb{Z}$ \cite{Manton2004,Nitta2022}. 

The total topological charge is
\begin{equation}
\mathcal{C}_\tau \equiv \int_{-\infty}^\infty \mathcal{Q}_\tau(x) dx = \int_{-\infty}^\infty \frac{\partial_x \varphi_\tau(x)}{2\pi} dx ,
\label{eq:10}
\end{equation}
where the integrand $\mathcal{Q}_\tau(x)$ defines 
topological charge density. It is immediately apparent that the term $q_0 \partial_x \varphi$ in the Hamiltonian density \eqref{Eq:Hamcon} arising from the DM interaction, can be interpreted as a chemical potential coupled to the topological charge ${\cal C}$. Individual soliton or antisoliton solutions \eqref{eq:classol}, are characterized by the integer charge ${\cal C}_\tau = \tau$. 

It is important to emphasize here that solitons are \emph{not} the ferromagnetic domain walls, or $\varphi^4$ kinks \cite{Egami1973,Egami1973b,Shibata2000,Chudnovsky1992}, which are classified by a different homotopy $\pi_0(\mathbb{Z}_2)\cong\mathbb{Z}_2$ \cite{Manton2004,Nitta2022}.

Returning to the lattice problem, the underlying physics remains essentially the same as in the continuum limit~\cite{Kishine2015}. On a discrete chain, the winding number associated with a given spin configuration can be computed through a geodesic interpolation,
\begin{equation}
{\cal C} \equiv \frac{1}{2\pi} \sum_{i=1}^{L} \arcsin\left[\hat{x}\cdot({\bm S}_i\times {\bm S}_{i+1})\right],
\end{equation}
where the identification $L + 1 \equiv 1$ enforces periodic boundary conditions, ${\bm S}_{L+1} = {\bm S}_1$. These boundary conditions are essential for $\cal C$ to be well defined, since the base manifold must be topologically equivalent to $S^1$. From this point onward, we assume periodic boundary conditions throughout. The sign of the topological charge carried by each soliton is fixed by the chirality imposed by the Dzyaloshinskii–Moriya interaction, and the total number of solitons in the system is simply given by $N_s = |{\cal C}|$.

\begin{figure}[!ht]
    \centering
    \includegraphics[width=\columnwidth]{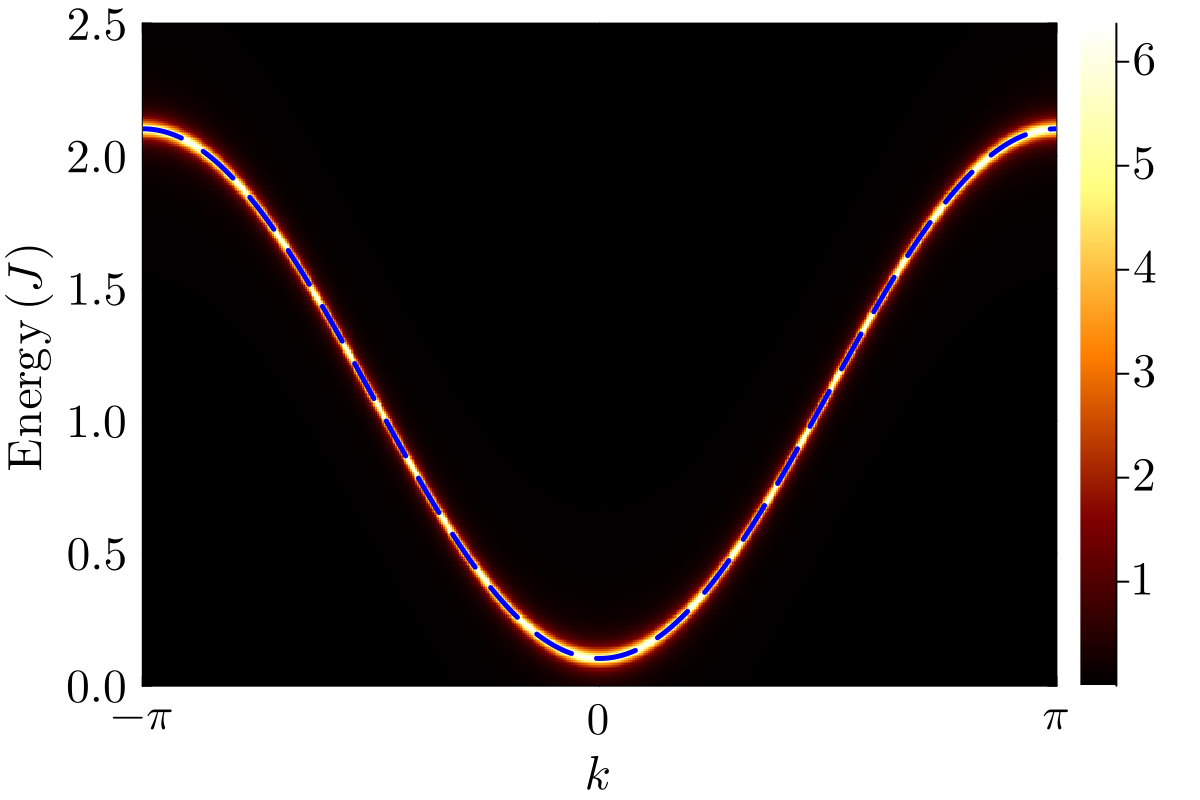}
    \caption{Dynamical spin structure factor and magnon dispersion (blue dashed line) obtained from linear spin wave theory (LSWT) calculations. The spectral intensities were computed using \texttt{Sunny.jl} library (v0.7)~\cite{Sunny2025}.}
    \label{Fig:lswt_disp}
\end{figure}

As stated in the Introduction, the central goal of this work is to explore the physics of the commensurate-incommensurate transition near the quantum critical point, with particular emphasis on the character of the low-energy excitations. For $H > H_c$, the system is fully polarized and the classical elementary excitations correspond to spin waves with a gapped dispersion centered at $k = 0$, as presented in Fig.~\ref{Fig:lswt_disp}. The magnon dispersion is given by the familiar expression 
\begin{equation}
\omega(k) = -J \left[\cos(ka) - 1\right] + H.
\label{Eq:magnon_disp}
\end{equation}
At the critical field $H = H_c$, the spin-wave spectrum remains gapped, with a gap $\Delta E = H_c$. This exact statement, proven in Appendix \ref{Appe:analitical_states}, once again reflects the fact that the commensurate–incommensurate transition is driven by the condensation of chiral solitons, and not of magnons, as in conventional magnetically ordered systems~\cite{zapf2014}. 

In the classical limit, the resulting CSL and the spiral magnetic state represent distinct types of order. However, once quantum fluctuations are included, both phases evolve into a Tomonaga–Luttinger liquid (TLL), and the distinction between them becomes less pronounced. As we show below, these two scenarios can nevertheless be clearly discriminated in the quantum regime by analyzing the nature of the low-energy excitations for $H \gtrsim H_c$. In this regime, the QCP marks the softening of a chiral soliton mode that drives the transition between the TLL and the fully polarized phase.

We are now in position to specify the parameter regime of our study. Since our objective is to investigate emergent solitonic physics, we focus on configurations in which solitons extend over several lattice sites, ensuring a clear distinction from magnons, which remain localized on the scale of a single site. In other words, we target the mesoscopic regime, defined by $a < \ell_s \ll L$, where $a$ is the lattice spacing and $\ell_s$ is the soliton size. For comparison, the semiclassical regime—considered a limiting case of the mesoscopic one—corresponds to $a \ll \ell_s \ll L$, where quantum corrections are perturbative and the soliton mass remains much larger than the magnon mass. To access the mesoscopic regime in our numerical simulations, we set $J = 1$ and $D = 1/\sqrt{3}$. This choice allows us to study system sizes that are both computationally tractable and large enough to capture the essential solitonic behavior. The qualitative features discussed in the following are robust in a wide range of $D$ values.

\section{\label{Sec:conti} Quantization and S-duality}
The profound connection between the bosonic quantum sine-Gordon model and the fermionic Thirring model is based on the identification of the soliton and antisoliton excitations of the quantum sine-Gordon model with the single-particle fermions of the Thirring model. It maps the C-IC transition of the sine-Gordon model to the metal-insulator transition of the Thirring model driven by the chemical potential. We summarize key arguments of this correspondence here; more details are provided in Appendix \ref{Appe:sine-Gordon}.

The quantum sine-Gordon Hamiltonian density is given by
\begin{eqnarray}
\hat{{\mathcal{H}}}_{\rm sG}=\frac{c \kappa}{2}\hat{\Pi}^2  + \frac{c}{2 \kappa} (\partial_x\hat{\varphi})^2 - \frac{c q_0}{\sqrt{4\pi} \kappa} \partial_x \hat{\varphi} 
- \frac{H S}{a} \cos[\sqrt{4\pi}\hat{\varphi}]. \nonumber \\
\label{eq:H4-main}
\end{eqnarray}
Here, $\hat{\Pi}$ denotes the canonical momentum conjugate to the field $\hat{\varphi}$, satisfying $[\hat{\varphi}(x,t),\hat{\Pi}(x',t)] = i {\delta}(x-x')$, and the parameter $\kappa$ is defined in ~\eqref{eq:H4}. Equation~\eqref{eq:H4-main} corresponds to the quantum sine-Gordon model describing the dynamics of kink excitations (solitons and antisolitons).

A remarkable discovery by Coleman \cite{Coleman1975} and Mandelstam \cite{Mandelstam1975}, and, independently, by Luther and Emery \cite{Luther1974}, is that \eqref{eq:H4-main} can be exactly mapped to the Thirring model of massive Dirac fermions. 
The mapping is based on the observation that the exponential operator,
\begin{equation}
    \hat{O}(x) = e^{i \zeta \int_{-\infty}^x dx' \hat{\Pi}(x')}
    \label{eq:o1}
\end{equation}
creates a soliton of height $\zeta$ centered at $x$. This result is proven with the help of the operator identity $[\hat{A}, e^{\hat{B}}] = \hat{C} e^{\hat{B}}$ which holds if $\hat{C}=[\hat{A},\hat{B}]$ commutes with both $\hat{A}$ and $\hat{B}$. Applied to our situation, the identity,
\begin{equation}
[\hat{\varphi}(y),\hat{O}(x)] = - \zeta \Theta(x-y) \hat{O}(x)
\label{eq:com1-main}
\end{equation}
means that the action of $\hat{O}(x)$ on an eigenstate $|\varphi\rangle$ of the operator $\hat{\varphi}$ changes it to the state $|\varphi'\rangle$ with the eigenvalue $\varphi(y) - \zeta \Theta(x-y)$ which is exactly the kink of height $\zeta$ at position $x$.

This point underlies the bosonization ``miracle'' \cite{Shankar2017}. As reviewed in
Appendix~\ref{app:field-theory}, the operator
\begin{equation}
    \hat{\psi}_{\pm}^\dagger(x)
    =
    \frac{1}{\sqrt{2\pi\alpha}}\,
    \exp\!\Big[\mp i \sqrt{\pi}\big(\hat{\varphi}(x) \mp \!\int_{-\infty}^x\! dx'\,\hat{\Pi}(x')\big)\Big]\,
    e^{-i p_0 x}
    \label{eq:o2}
\end{equation}
creates right- and left-moving Dirac fermions with the correct Fermi
statistics~\cite{Mandelstam1975}.  Notice that $\hat{\psi}_\pm$ contains the
exponential soliton operator of ~\eqref{eq:o1} with
$\zeta=\sqrt{\pi}$, implying that these fermions are precisely the
quantum solitons of the sine-Gordon field $\hat{\varphi}$ appearing in
~\eqref{eq:H4-main}.

The remarkable feature of ~\eqref{eq:o2} is the emergence of a
finite momentum shift
\[
p_0 = \frac{2\pi S}{a},
\]
which distinguishes integer- from half-integer-spin chains.  For
integer $S$, the soliton dispersion minimum is located at the Brillouin-zone
origin because $p_0 = (2\pi/a)\times \mathrm{integer} \equiv 0$.
In contrast, for half-integer $S$, the soliton band minimum is shifted to the
zone boundary, $p_0 \sim \pi/a$.  This topological spin-parity effect
originates from the Berry-phase term in \eqref{eq:L} and was
previously obtained in the instanton analysis of the quantum
sine-Gordon model~\cite{Loss1996,Kato2023}.  Our lattice-based
calculation of the soliton dispersion, together with the DMRG results
presented in the next sections, provides a direct and independent
confirmation of this striking prediction.

Subsequent steps of the bosonization technology are well documented [see, e.g., Ref. \cite{Shankar2017} and  Appendix \ref{Appe:sine-Gordon}] and provide a rigorous way to map the sine-Gordon Hamiltonian \eqref{eq:H4-main} to a massive  Thirring model (mT), i.e., a  Thirring model  with massive Dirac fermions \eqref{eq:thirring}.

The fermionic formulation of the problem becomes particularly transparent at the special value $\kappa = 1$, known as the Luther–Emery point~\cite{Luther1974}, where the mapping yields a noninteracting Hamiltonian that can be readily diagonalized. In this limit, the Hamiltonian describes two independent fermionic bands,
\begin{equation}
    \hat{H}_{\rm mT}^{\kappa=1} = \sum_k \epsilon_k (\hat{u}_k^\dagger \hat{u}_k + \hat{h}_k^\dagger \hat{h}_k) 
    - \frac{c q_0}{2} (\hat{u}_k^\dagger \hat{u}_k - \hat{h}_k^\dagger \hat{h}_k),
    \label{eq:HmT1}
\end{equation}
where $k$ is measured from $p_0$, $\epsilon_k= \sqrt{c^2 k^2 + \lambda^2}$ is the dispersion relation of solitons and antisolitons, $\lambda = H S \alpha/a$ is the mass gap in the fermion spectrum and positive (negative) $q_0$ plays the role of chemical potential that adds solitons  $\hat{u}_k^\dagger$ (antisolitons $\hat{h}_k^\dagger$) to the system provided $c |q_0| > 2 \lambda$ \cite{Giamarchi2004}.

It is worth noting the emergence of a \(U(1)\) symmetry in the effective low-energy Hamiltonian,~\eqref{eq:HmT1}, which captures the commensurate-incommensurate transition of the lattice spin model~\eqref{Eq:Ham}.

\begin{figure}[!ht]
    \centering
    \includegraphics[width=0.8\columnwidth]{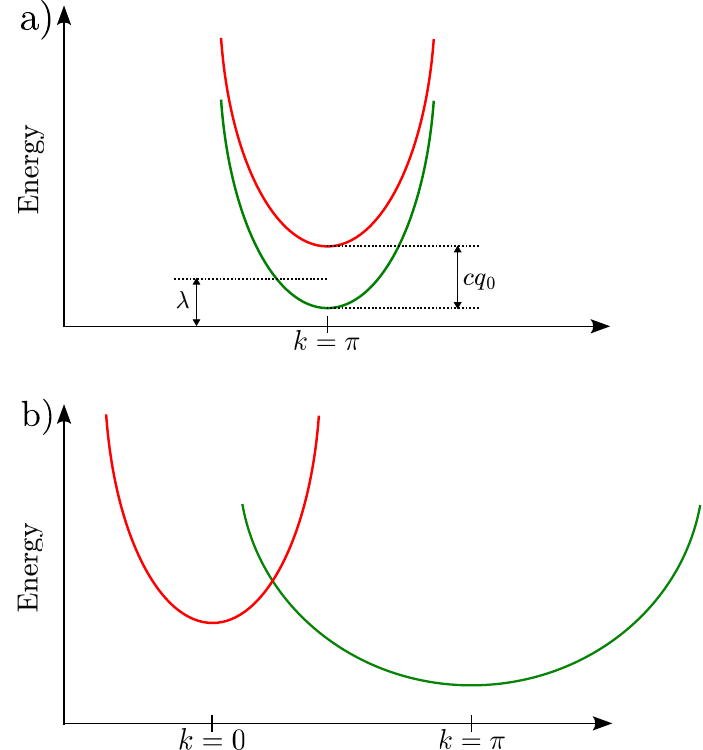}
    \caption{a) Schematic representation of the fermionic soliton and antisoliton bands in the
massive Thirring model, both featuring minima at $k=\pi$ and possessing
identical effective masses. The sine–Gordon vacuum corresponds to the
half-filled case, where the chemical potential lies within the band gap
[$c|q_0| < 2\lambda$ in \eqref{eq:HmT1}] and energies of solitons and antisolitons are positive. (b) Same as in panel (a) but for the effective lattice model, \eqref{eq:Heff},
where the dispersion minima occur at different momenta ($k=\pi$ for the
soliton band and $k=0$ for the antisoliton band) and the effective masses are
not equal (the soliton mass is significantly larger than that of the
antisoliton). }
    \label{Fig:field_bands}
\end{figure}

\section{\label{Sec:Quantum_solitons}Quantum solitons}

We now turn to the description of quantum solitons in the lattice model \(\hat{H}_S\).
Our starting point is an operator that creates a classical soliton---represented as a product state---when acting on the fully polarized reference state.
Quantum fluctuations are incorporated through a momentum-space (Bloch-space) orthogonalization procedure, analogous to the construction of Wannier orbitals: The soliton states are first normalized in momentum space and then transformed back to real space.
This method produces a set of mutually orthogonal soliton states, enabling us to compute the quantum tunneling of a soliton between different lattice sites.
We emphasize that these ``Wannierized'' soliton states are no longer product states; the orthogonalization procedure necessarily generates entanglement in the spin degrees of freedom.

Within this framework, we can then formulate an effective tight-binding Hamiltonian, which makes explicit how quantum fluctuations 
generate soliton delocalization and thereby enrich the system’s low-energy dynamics.

Finally, we compare the predictions of this semiclassical framework with the quantum dynamics obtained using time-evolving block decimation (TEBD) techniques and discuss the relevance of these results for potential inelastic neutron scattering experiments.

\subsection{Soliton operator}

As shown in Appendix~\ref{Appe:analitical_states}, the exact ground state for
$H \ge H_c$ is the fully polarized configuration along the $\hat{x}$ direction.
Since this state contains no solitons, we refer to it as the vacuum
state and denote it by
\begin{equation}
\ket{0} \equiv \bigotimes_l \ket{\rightarrow}_l,
\end{equation}
where $\ket{\rightarrow}_l = (\ket{\uparrow}_l+\ket{\downarrow}_l)/\sqrt{2}$ is the maximal-weight eigenstate of the operator $\hat S^x_l$.

The operators that create a classical soliton ($\tau=+1$) or antisoliton ($\tau=-1$) centered at  site $j$ on top of the fully polarized ground state are
\begin{equation}
    \hat{T}^{\dagger}_{j\tau} \equiv (\hat{S}^z_j + i \hat{S}_j^y) \prod_{l\neq j}  \exp{-i \varphi_{\tau}(l-j) \hat{S}^z_l}.
    \label{Eq:SolOp}
\end{equation}
The corresponding annihilation operators are
\begin{equation}
    \hat{T}^{}_{j\tau} =  \prod_{l \neq j}  \exp{i\varphi_{\tau}(l-j) \hat{S}^z_l} (\hat{S}^z_j - i \hat{S}_j^y).
\end{equation}    
As expected, these operators annihilate the vacuum: $\hat{T}^{}_{j\tau}\ket{0}=0$. In terms of the local spin degrees of freedom, the creation operators $\hat{T}^{\dagger}_{j\tau}$ rotate the spins around the $\hat{z}$ axis by the angle $\varphi_{\tau}(x)$ given by \eqref{eq:classol},
which corresponds to the profile of the single-soliton (or antisoliton) solution of the classical sine-Gordon Hamiltonian~\eqref{Eq:Hamcon} for $m^2 = H / (J S a^2)$.  To properly account for periodic boundary conditions, we define $x \equiv (l - j + L/2) \Mod L - L/2$.

As shown in Appendix~\ref{Appe:su2algebra}, the operators 
$\{\hat{T}_{\tau}^\dagger, \hat{T}_{\tau}, \hat{S}^x\}$ satisfy the following commutation relations: 
\begin{equation}
    \begin{split}
        [ \hat{T}^{}_{j'\tau}, \hat{T}^\dagger_{j\tau} ] &= 2\,\Tilde{\delta}_{j'j\tau}\, \hat{S}^x_{j}, \\ 
        [ ~\hat{T}^{\dagger}_{j\tau}, \hat{S}^x_{j~} ] &= 2\,\hat{T}^\dagger_{j\tau}, \\ 
        [ ~\hat{T}^{}_{j\tau}, \hat{S}^x_{j~} ] &= -2\,\hat{T}_{j\tau},
    \end{split}
\end{equation}
where the function $\Tilde{\delta}_{j'j\tau}$ plays the role of an approximate Kronecker delta in a coarse-grained description. 
Equivalently, upon choosing the soliton size as the unit of length, the set 
$\{\hat{T}_{\tau}^\dagger, \hat{T}_{\tau}, \hat{S}^x\}$ provides an approximate realization of the 
$\mathfrak{su}(2)$ Lie algebra.

As follows from the algebraic structure, the $\hat{T}$ operators obey bosoniclike commutation relations. 
In a coarse-grained description, solitons can therefore be interpreted as hard-core bosons. 
Moreover, in the dilute limit, that is, when the soliton density is sufficiently small, this identification becomes exact.

\subsection{Quantum soliton operator: Wannierization of the soliton operator}

A drawback of the operators $\hat T_{j\tau}$ introduced above is that the single-particle states  $\hat{T}^{\dagger}_{j\tau}\ket{0}=\ket{\varphi_{\tau}^{(j)}}$ that they generate 
\begin{equation}
\begin{split}
    \ket{\varphi_{\tau}^{(j)}} 
    = \bigotimes_{l} \frac{1}{\sqrt{2}} \Big( e^{-i \varphi_\tau(l-j)/2} |\uparrow\rangle_l + e^{i \varphi_\tau(l-j)/2} |\downarrow\rangle_l \Big)
\end{split}
    \label{eq:sol0}
\end{equation}
are not mutually orthogonal, 
\begin{equation}
\begin{split}
    \langle {\varphi_{\tau}^{(j)}} |{\varphi_{\tau}^{(j')}}\rangle &= \prod_{l=1}^L \cos{\left[\frac{\varphi_{\tau}(l-j)-\varphi_{\tau}(l-j')}{2}\right]} \neq \delta_{jj'},
\end{split}
\label{Eq:soliton_ov}
\end{equation}
for $j,j' = 1,\dots,L$. By using a standard Wannierization procedure, we can construct a new set of single-particle orthonormal states  $\{\hat{\mathfrak{T}}_{j\tau}^{\dagger} \ket{0} \}$ given by
\begin{equation}
    \hat{\mathfrak{T}}_{j\tau}^{\dagger} \ket{0} \equiv \frac{1}{\sqrt{L}} \sum_{k \in \text{1BZ}} e^{-ikj}~ \frac{\hat{\tilde{T}}_{k\tau}^{\dagger}}{\norm{\hat{\tilde{T}}_{k\tau}^{\dagger}\ket{0}}} ~\ket{0},
    \label{Eq:reop}
\end{equation}
where
\begin{equation}
    \hat{\tilde{T}}_{k\tau}^{\dagger} = \frac{1}{\sqrt{L}} \sum_{l=1}^L e^{ikl}~ \hat{{T}}_{l\tau}^{\dagger}
\end{equation}
is the Fourier transform of the soliton operator with $k\in[-\pi ,\pi)$. Equation \eqref{Eq:reop} can be rewritten in the form

\begin{equation}
    \begin{split}
        \hat{\mathfrak{T}}^{\dagger}_{j\tau} \ket{0} &= \frac{1}{\sqrt{L}} \sum_k e^{-ikj}~ \frac{\hat{\tilde{T}}_{k\tau}^{\dagger}\ket{0}}{\norm{\hat{\tilde{T}}_{k\tau}^{\dagger}\ket{0}}} \\ 
        &= \sum_{l=1}^L \left(\frac{1}{L}\sum_k \frac{ e^{-i(j-l)k}}{\norm{\hat{\tilde{T}}_{k\tau}^{\dagger}\ket{0}}} \right) \hat{T}_{l\tau}^{\dagger} \ket{0} \\ 
        &\equiv \sum_{l=1}^L w_{j-l\tau} \hat{T}_{l\tau}^{\dagger} \ket{0}.
    \end{split}
    \label{Eq:reop_2}
\end{equation}
From the last term, it is evident that the coefficients
\begin{equation}
w_{j-l\tau} = \frac{1}{L}\sum_k \frac{ e^{-i(j-l)k}}{\norm{\hat{\tilde{T}}_{k\tau}^{\dagger}\ket{0}}}
\label{eq:wc}
\end{equation}
represent the components of the Wannierized operator in the original basis spanned by the $\hat{T}^{\dagger}_{j \tau}$ operators. 

The norm of the Fourier–transformed soliton operator,
\(N_{k\tau} \equiv \bigl\lVert \hat{\tilde T}_{k\tau}^{\dagger}\ket{0} \bigr\rVert\),
can be evaluated analytically:
\begin{align}
N_{k\tau}^2
&= \frac{1}{L}\sum_{l,l'=1}^{L} e^{-ik(l-l')}
   \prod_{p}^{}
   \cos\!\left[\frac{\varphi_{\tau}(p-l)-\varphi_{\tau}(p-l')}{2}\right] 
   \nonumber\\
&= \sum_{l=1}^{L} e^{-ik l}
   \prod_{p}^{}
   \cos\!\left[\frac{\varphi_{\tau}(p)-\varphi_{\tau}(p-l)}{2}\right],
\end{align}
where, in the second line, we have used translation invariance of the chain (i.e., periodic boundary conditions).

Let us reflect on the previous construction. We began with a well-localized sine–Gordon soliton (or antisoliton) creation operator \(\hat T^\dagger_{j\tau}\).
Through a Wannierization procedure, we constructed a new set of operators as linear combinations of solitons centered at all sites,
\begin{equation}
\hat{\mathfrak T}_{j\tau}^{\dagger}=\sum_{l=1}^L w_{j-l\tau}\,\hat T^{\dagger}_{l\tau},
\end{equation}
with coefficients \(w_{j-l\tau}\) that, formally, are the overlaps \(w_{j-l\tau}\propto \mel{0}{\hat{\mathfrak T}_{j\tau} \hat T^\dagger_{l\tau}}{0}\). As shown in Appendix~\ref{Appe:soliton_ov}, these overlaps decay exponentially,
\(w_{j-j'}\sim e^{-2m|j-j'|}\) for large separation  $|j-j'|$.

\begin{figure}[!ht]
    \centering
    \includegraphics[width=\columnwidth]{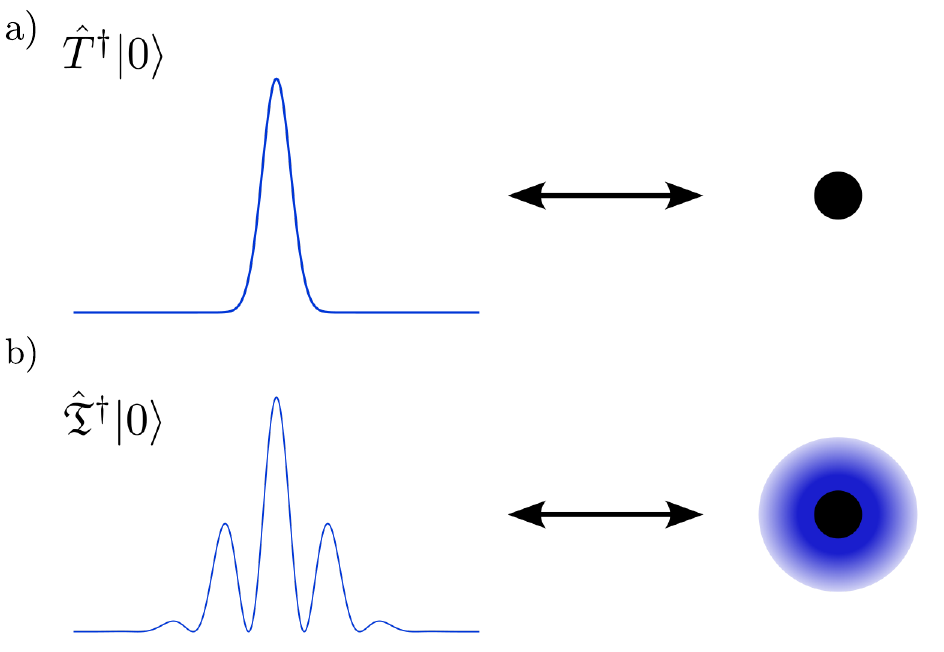}
    \caption{Schematic representation of the action of the soliton operator (a) and the Wannierized soliton operator (b).}
    \label{Fig:Reno_sol_op}
\end{figure}

The Wannier-like operators create a soliton wave packet centered at site \(j\) that incorporates quantum delocalization via the generation of entangled single-soliton states that are orthonormal for solitons centered around different sites. By contrast, the bare states generated by \(\hat T_{j\tau}^\dagger\ket{0}\) are nonorthogonal product states. The structure of the Wannierized soliton state  is sketched in Fig.~\ref{Fig:Reno_sol_op}. Henceforth---with a slight abuse of terminology---we refer to these objects as \emph{quantum solitons}, since their strongly localized wave packets still support a quasiparticle interpretation. Equivalently, the Wannierization procedure furnishes a variational construction of the single-soliton states of the spin-\(\tfrac{1}{2}\) chain. Related ``quantum skyrmion'' operators have been introduced in Refs.~\cite{Haller2024,Bhowmick2025,Takashima2016}; similarities and differences with our construction are discussed in Sec.~\ref{Sec:Discussion}.

\subsection{\label{Sec:sol_kin} Soliton kinematics}

\begin{figure}[!ht]
    \centering
    \includegraphics[width=\columnwidth]{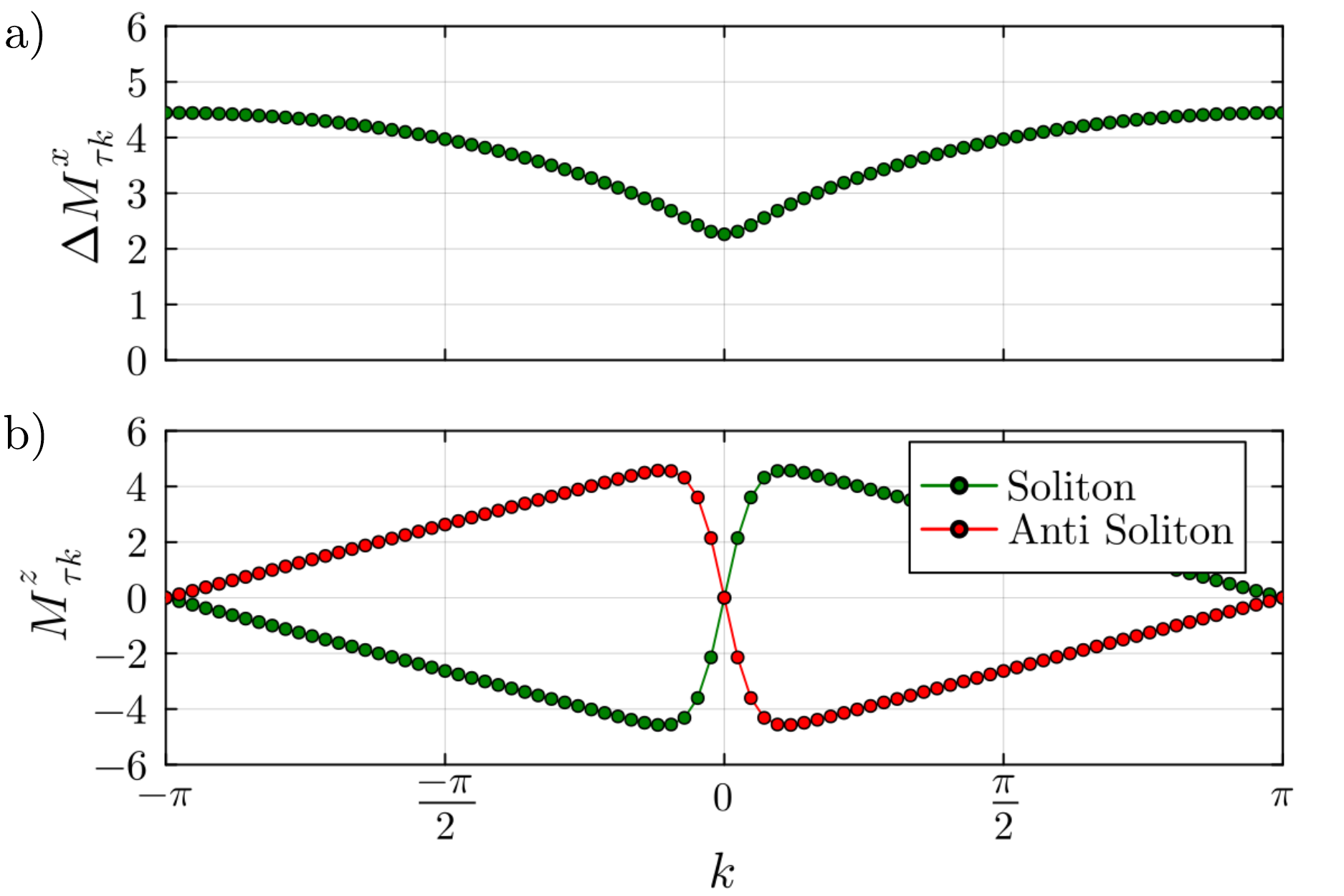}
    \caption{Field-induced magnetization of the soliton (green) and antisoliton (red). a) Longitudinal magnetization difference $\Delta M^x_{\tau k}$ as a function of momentum $k$. Only the soliton branch is shown, as the antisoliton results are identical. b) Transverse magnetization $M^z_{\tau k}$ along the DM direction. The quantity $M^z_{\tau k}$ changes sign with the momentum and vanishes at $k = 0$ and $k = \pi$. These calculations are at $H=0.11$.}
    \label{Fig:kinematics}
\end{figure}

Before addressing the dynamics and deriving an effective model for the solitons, let us first analyze their kinematic properties.
The Fourier transform of the Wannierized soliton operator is simply given by:
\begin{equation} \hat{\Tilde{\mathfrak{T}}}^{\dagger}_{k\tau} \ket{0} = \frac{1}{\norm{\hat{\tilde{T}}_{k\tau}^{\dagger}\ket{0}}} \hat{\tilde{T}}_{k\tau}^{\dagger}\ket{0},
\end{equation}
that is, the normalized Fourier transform of the classical soliton operator.
As expected, the solitons (antisolitons) carry a finite magnetization along the $x$ direction, $\Delta M^x_{\tau k}  = M^x_{\text{FP}}-M^x_{\tau k} $, which varies continuously with the momentum $k$ as presented in Fig.~\ref{Fig:kinematics} (a). More interestingly, the solitons carry a finite magnetization component \(M^z_{\tau,k}\) along the \(z\) axis (the DM direction), with a sign set by the sign of momentum, see Fig.~\ref{Fig:kinematics} (b). At the special (inversion-symmetric) momenta \(k=0\) and \(k=\pi\), however, a mirror symmetry about the midpoint of a bond, which maps $x \to -x$,  $k \to -k$ and $S^x \to S^x, S^{y,z} \to - S^{y,z}$, forces this magnetization to vanish.
Namely, this symmetry relates \((k,M^{z})\leftrightarrow(-k,-M^{z})\) and guarantees a symmetric spectrum \(E(k)=E(-k)\).

An analogous behavior holds for the antisoliton branch [see Fig.~\ref{Fig:kinematics} (b)], with the opposite sign of the \(z\) magnetization at a given \(k\). This momentum-dependent polarization sharply contrasts with single-magnon excitations, whose magnetization along the field axis is fixed and essentially \(k\) independent.

\subsection{\label{Sec:tb_ham} Effective tight-binding Hamiltonian}

Now, assume that the chain hosts a single soliton centered at site $j$. Since it is a quantum-mechanical soliton, it can tunnel to the neighboring sites. The tunneling or hopping amplitude between sites $j$ and  $j'$ is given by the matrix element:
\begin{equation}
    t_{jj'\tau} = \langle \Phi_{\tau}^{(j)} | {\hat{H}_S} | {\Phi_{\tau}^{(j')}} \rangle,
\end{equation}
with $\ket{\Phi_{\tau}^{(j)}} = \hat{\mathfrak{T}}^{\dagger}_{j\tau}\ket{0}$. By substituting~\eqref{Eq:reop_2} into the previous expression, we obtain

\begin{equation}
  t_{jj'\tau}   = \sum_{l,l'=1}^L (w_{j-l'\tau})^* w_{j'-l\tau} \mel{\varphi_{\tau}^{(l')}} {\hat{H}_S}{\varphi_{\tau}^{(l)}}.
  \label{Eq:hop_cal}
\end{equation}

The explicit form of these hopping amplitudes is obtained by inserting the expressions  of the Wannier coefficients {$w_{j-l\tau}$} given in \eqref{eq:wc}  and reordering the factors:
\begin{equation}
    t_{jj'\tau} = \frac{1}{L^2} \sum_{k, k'} \frac{e^{-ik'j} e^{-ik j'}}{N_{k\tau} N_{k'\tau}} 
         \sum_{l, l'=1}^L e^{ik' l'} e^{ik l} \mel{\varphi_{\tau}^{({l'})}} {\hat{H}_S} {\varphi_{\tau}^{({l})}}.
\end{equation}

Now, we can exploit  the translational invariance of ${\hat{H}_S}$ by defining $\mel{\varphi_{\tau}^{({l'})}}{\hat{H}_S}{\varphi_{\tau}^{(l)}} \equiv h_{\tau}(n - n')$, then 
\begin{eqnarray}
\sum_{l, l'=1}^L e^{ik l} e^{ik' l'} h(l - l') 
&=& \sum_{s=1}^L h_{\tau}(s) e^{ik s} \sum_{s'=1}^L e^{i(k + k')s'}
\nonumber \\
&=& L \sum_s h_{\tau}(s) e^{ik s} \delta_{k, - k'},
\end{eqnarray}
to obtain:
\begin{equation}
t_{jj'\tau} = \frac{1}{L} \sum_k \frac{e^{i k (j - j')}}{N_{k\tau}^2} \sum_s h_{\tau}(s) e^{ik s}
= \frac{1}{L} \sum_k \frac{e^{ik(j - j')}}{N_{k\tau}^2} \tilde{h}_{\tau}(k)
\end{equation}
where $\tilde{h}_{\tau}(k) = \sum_s h_{\tau}(s) e^{iks}
$ is the Fourier transform of $h(s)$.
In conclusion, 
\begin{equation}
t_{\delta\tau} = \frac{1}{L} \sum_k \frac{\tilde{h}_{\tau}(k)}{N_{k\tau}^2} \, e^{ik\delta}
\label{Eq:calc_hop}
\end{equation}
with  $\delta = j-j'$, are the hopping amplitudes of the quantum chiral soliton. The next step is to evaluate the matrix elements $h_{\tau}(s)$. Since the resulting expression is lengthy and offers no direct physical insight, we defer this calculation to Appendix~\ref{Appe:Matrix_elems}. As shown there, the matrix elements $h_{\tau}(s)$ are real, implying that $\tilde{h}_{\tau}(k) =  \tilde{h}^*_{\tau}(-k)$. 
This finding in turn implies that the hopping amplitudes $t_{\delta\tau}$ are real.

\begin{figure}[!ht]
    \centering
    \includegraphics[width=\columnwidth]{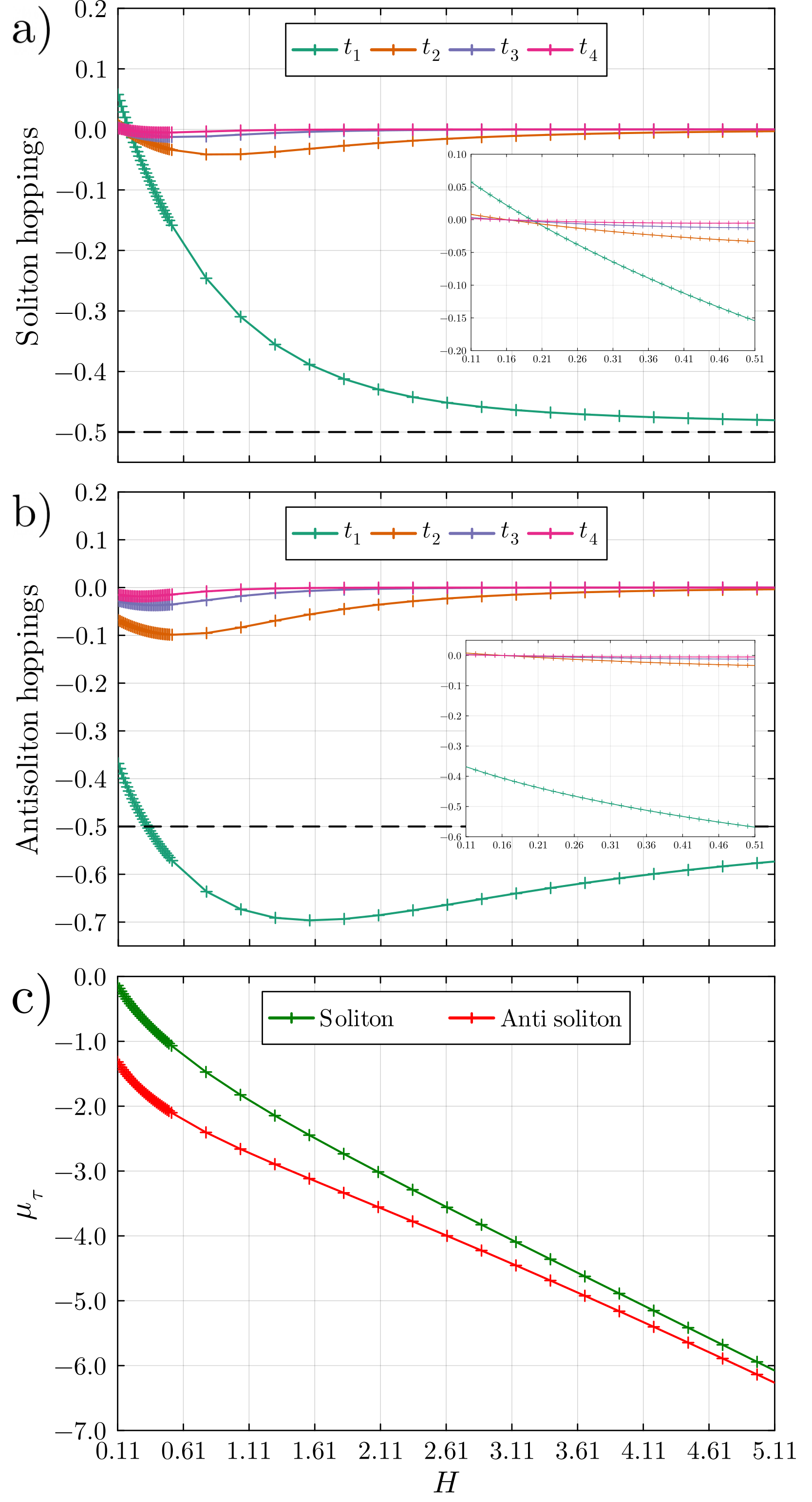}
    \caption{Soliton (a), antisoliton (b) hopping amplitudes and (c) chemical potentials as a function of the Zeeman field. The inset in each of the first two panels is a zoom of a small field region. The dashed line in a) and b) panels represents the single magnon hopping amplitude.}
    \label{Fig:hopping}
\end{figure}

With these hopping amplitudes, we can now construct an effective tight-binding Hamiltonian for the solitons on the chain. The last necessary ingredient is the associated soliton creation chemical potential, which can be computed as 
\begin{equation}
    \mu_{\tau} = -\mel{\Phi_{\tau}^{(j)}} {\hat{H}_S} {\Phi_{\tau}^{(j)}} + E_0,
\end{equation}
where $E_0/L = -J/4 - H/2$ is the ground-state energy density, and $j$ can be any site on the chain due to the translational invariance of the Hamiltonian. The chemical potential as a function of field is presented in Fig.~\ref{Fig:hopping}~(c).

The resulting tight-binding Hamiltonian is given by
\begin{equation}
    \hat{H}_{\rm eff} = \sum_{j, \delta, \tau} (t_{\delta\tau} \, \hat{b}^{\dagger}_{j\tau} \hat{b}_{j+\delta\tau} + {\rm H. c.}),
\label{eq:rspaceH}
\end{equation}
where the hopping amplitudes $t_{\delta\tau}$ and the chemical potential for each flavor, $\mu_{\tau}= - t_{0 \tau}$, depend  on the external field $H$. Note that the bosonic operators $\hat{b}_{\tau}$ and $\hat{b}_{\tau}^{\dagger}$ also implicitly depend on the field because the soliton size decreases with increasing $H$.

For periodic boundary conditions ($L+1 \equiv 1$), the Hamiltonian is diagonalized in reciprocal space, 
\begin{equation}
\hat{b}_{k\tau} = \frac{1}{\sqrt{L}} \sum_{l=1}^L e^{i l k} \hat{b}_{l\tau},
\end{equation}
to obtain
\begin{equation}
    \hat{H}_{\rm eff} = \sum_{k, \tau} [\omega^{\tau}_k  -\mu_{\tau}] \hat{b}^{\dagger}_{k\tau}\hat{b}_{k\tau},
    \label{eq:Heff_b}
\end{equation}
where
\begin{equation}
    \omega^{\tau}_k = \sum_{\delta=1}^n t_{\delta\tau}\cos{ k \delta}. 
\end{equation}
The number \(n\) of neighbors retained in the calculation is determined by the decay rate of the hopping amplitudes \(|t_{\delta\tau}|\) with the separation \(\delta\). 
We also note that \(\omega^{\tau}_{k} = \omega^{\tau}_{-k}\), a direct consequence of the reality of the hopping amplitudes \(t_{\delta\tau}= t^*_{\delta\tau} \).

We defer the discussion of the soliton dispersion \(\omega^{\tau}_{k}\) to the next section and conclude here by analyzing the dependence of the hopping amplitudes on the external magnetic field. Figures~\ref{Fig:hopping} a) and b) show the hopping amplitudes up to fourth neighbors as a function of the field.

For the soliton, the nearest-neighbor hopping amplitude dominates over the entire field range, except within a narrow window around \(H = H^\ast \simeq 0.2\), where it changes sign: \( t_{1+}>0\) for \(H<0.2\) and \(t_{1+}<0\) for \(H>0.2\). This sign reversal shifts the minimum of the soliton dispersion from \(k=\pi\) (for \(H<0.2\)) to \(k=0\) (for \(H>0.2\)). In addition, all hopping amplitudes are strongly suppressed near \(H^\ast\), leading to an almost flat band. The change of sign of \(t_{1+}\) is expected, since the soliton evolves continuously into a magnon as the field increases, and the magnon dispersion attains its minimum at \(k = 0\), as shown in Fig.~\ref{Fig:lswt_disp}.

In contrast, for the antisoliton the nearest-neighbor hopping also dominates but remains negative throughout the entire field range. At higher fields, the nearest-neighbor term essentially becomes the only appreciable contribution for both the soliton and anti-soliton, with its magnitude approaching the expected magnon-hopping value in the fully polarized limit.

While our main discussion focused on the $S=\tfrac{1}{2}$ version of $\hat{H}_S$, Appendix~\ref{App:sign_hop} shows that the lattice formulation for arbitrary $S$ correctly reproduces the integer and half-integer alternation of the soliton-band minimum discussed in Sec.~\ref{Sec:conti} and reported in  Ref.~\cite{Kato2023} for a particular limit of the model. This alternation is caused by a spin Berry phase \cite{Loss1996}.

\subsection{Jordan-Wigner transformation}

Since the effective model derived in the previous section describes a dilute gas of hard-core bosons in one dimension (1D), it is convenient to employ a generalized Jordan–Wigner transformation~\cite{Batista2001}, which maps  hard-core bosons with two flavors $(\tau=\pm)$ to constrained ($\hat{c}^{\dagger}_{j,+} \hat{c}^{\dagger}_{j,-}=0$) spin-$1/2$ fermions:
\begin{eqnarray}
\hat{c}_{j,+} \; &=& \; e^{i\pi \sum_{\ell<j} \hat{n}_\ell^{(b)}}\, \hat{b}_{j+}, \quad 
\hat{c}^{\;}_{j,-} \; = \; e^{i\pi \sum_{\ell<j} \hat{n}_\ell^{(b)}}\, \hat{b}_{j-},
\nonumber \\
\hat{c}_{j, +}^\dagger \;&=& \; \hat{b}_{j+}^\dagger\, e^{-i\pi \sum_{\ell<j} \hat{n}_\ell^{(b)}} \quad
\hat{c}_{j,-}^\dagger \;= \; \hat{b}_{j-}^\dagger\, e^{-i\pi \sum_{\ell<j} \hat{n}_\ell^{(b)}},
\nonumber 
\end{eqnarray}
where,
\begin{equation}
\hat{n}_\ell^{(b)} = \sum_{\tau} \hat{b}_{j\tau}^\dagger \hat{b}_{j\tau}=   \hat{n}^{(c)}_{\ell, +} + \hat{n}^{(c)}_{\ell, -}\equiv \hat{n}_\ell^{(c)},
\end{equation} 
and $\hat{n}^{(c)}_{\ell, \tau}\equiv \hat{c}^{\dagger}_{\ell,\tau} \hat{c}_{\ell,\tau}$. Note that the constraint satisfied by the fermionic operators, $\hat{c}^{\dagger}_{j,+} \hat{c}^{\dagger}_{j,-}=0$, implies that double-occupied states are excluded from the Hilbert space.

The real-space effective Hamiltonian, \eqref{eq:rspaceH}, can now be expressed in terms of the fermionic operators:
\begin{equation}
\hat{H}_{\mathrm{eff}}=\sum_{j, \delta, \tau} t_{\delta,\tau} \left[ \hat{c}_{j,\tau}^{\dagger} \hat{\mathcal{S}}_{j, \delta} \hat{c}^{}_{j+\delta,\tau} + {\rm H. c.} \right],
\end{equation}
with, 
\begin{equation}
\hat{\mathcal{S}}_{j, \delta}=\prod_{j \leq \ell<j+\delta}  [1-2 \hat{n}^{(c)}_{\ell} ]
\label{eq:JWstr}
\end{equation}
being the Jordan-Wigner string. 

By separating the noninteracting part of $\hat{H}_{\mathrm{eff}}$ from the interaction terms, we obtain 
\begin{equation}
\hat{H}_{\mathrm{eff}}= \sum_{j, \delta, \tau} \left[ t_{\delta,\tau} \hat{c}_{j,\tau}^{\dagger}  \hat{c}_{j +\delta,\tau} + {\rm H. c.} \right]+ \hat{H}^{\rm int}_{\mathrm{eff}},
\end{equation}
where  $\hat{H}^{\rm int}_{\mathrm{eff}}$ corresponds to correlated hopping terms arising from the density operators in \eqref{eq:JWstr}. In reciprocal space, 
\begin{equation}
\hat{c}_{k,\tau} = \frac{1}{\sqrt{L}} \sum_{l=1}^{L} e^{i l k} \hat{c}_{l,\tau},
\end{equation}
the noninteracting part of $\hat{H}_{\mathrm{eff}}$ becomes diagonal,
\begin{equation}
\hat{H}_{\mathrm{eff}}= \sum_{k, \tau}   [{\omega}^{\tau}_k  -\mu_{\tau}] \hat{c}^{\dagger}_{k,\tau}\hat{c}_{k,\tau} + \hat{H}^{\rm int}_{\mathrm{eff}},
\end{equation}
and the connection with Thirring's model with $\kappa=1$ becomes evident after relabeling operators:
\begin{equation}
\hat{u}_k \equiv \hat{c}_{k,+}, \quad \hat{h}_k \equiv \hat{c}^{\;}_{k, -},
\end{equation}
which leads to
\begin{eqnarray}
\hat{H}_{\mathrm{eff}} &=& \sum_{k, \tau}   [\epsilon^+_k  -\mu ] \hat{u}^{\dagger}_{k}\hat{u}_{k} +
[\epsilon^{-}_k  - \mu] \hat{h}^{\dagger}_{k}\hat{h}_{k} + \hat{H}^{\rm int}_{\mathrm{eff}},
\label{eq:Heff}
\end{eqnarray}
with
\begin{equation}
\mu = \frac{\mu_{+} + \mu_{-} }{2}, \nonumber \ \
\delta \mu = \frac{\mu_{+} - \mu_{-} }{2}, \nonumber \ \
\epsilon^{\pm}_k = \omega^{\pm}_k \mp  \delta \mu. 
\end{equation}
Similar to the Thirring model, the noninteracting sector of  
\(\hat{H}_{\mathrm{eff}}\) describes two gapped fermions 
with the chemical potential determined by the magnetic field. A quantum phase transition occurs when the  
chemical potential reaches the bottom of the soliton band. Owing to the  
\(\mathrm{U}(1)\) invariance of \(\hat{H}_{\mathrm{eff}}\), associated with the  
conservation of soliton charge, the corresponding critical point is a  
free-fermion fixed point (\(\kappa = 1\)) characterized by a correlation-length  
exponent \(\nu = 1/2\) and a dynamical exponent \(z = 2\), reflecting the  
quadratic dispersion of the free fermions.

The interaction terms in  
\(\hat{H}_{\mathrm{eff}}^{\mathrm{int}}\) are irrelevant under the  
renormalization group because the no–double-occupancy constraint forces them to involve spatial derivatives in the continuum limit. Beyond the critical point, however, once the bottom of the conduction band is populated and a finite fermion density develops, these interactions become marginally relevant. The system then enters a Tomonaga-Luttinger liquid phase with  
Luttinger parameter \(\kappa < 1\) in the presence of repulsive interactions.

While the massive Thirring model in ~\eqref{eq:HmT1} and in 
$\hat{H}_{\mathrm{eff}}$~\eqref{eq:Heff} share the same universal low-energy
physics, it is nevertheless instructive to examine their qualitative and
quantitative differences. Such a comparison clarifies the role of quantum
fluctuations and of the higher-order terms in the gradient expansion leading
to the sine–Gordon field theory, which is not fully captured by traditional
semiclassical treatments. As we will  
show in the following sections, these nonuniversal aspects play a crucial  
role in achieving a quantitative description of the soliton dynamics and in  
guiding the experimental characterization of the phenomena under study.

The first qualitative difference is that, in the massive Thirring model, the
soliton and antisoliton have identical effective masses—their parabolic
dispersions differ only by a constant vertical shift
[Fig.~\ref{Fig:field_bands} (a)]. This property, which is a consequence of neglecting cubic and higher-order terms in the gradient expansion of the DM interaction, is naturally absent in the lattice
effective Hamiltonian $\hat{H}_{\mathrm{eff}}$, where the two bands generally
acquire distinct curvatures, as shown in
Fig.~\ref{Fig:field_bands} (b). Moreover, the minimum of the soliton band can
shift to $k=\pi$ because the nearest-neighbor hopping amplitude becomes
positive near the saturation field, as illustrated in
Fig.~\ref{Fig:hopping}. A further qualitative distinction is that the
effective masses themselves exhibit a pronounced dependence on the applied
field, as shown in the same figure.

Beyond these qualitative differences, \(\hat{H}_{\mathrm{eff}}\) provides the  
full soliton and antisoliton dispersions across the entire Brillouin zone. As  
we discuss in Sec.~\ref{sec:exp}, the field dependence of these dispersions is  
crucial for identifying experimental signatures of solitons and antisolitons,  
both in inelastic neutron scattering spectra and in specific-heat measurements.

\section{\label{Sec:Numerical_results} Numerical Results}

Throughout this section, we present numerical results for the full quantum model to test and validate the arguments and effective theory developed in the preceding section. Although the classical picture of well-defined solitons in real space is blurred by quantum fluctuations, we show that the essential classical features survive, most notably the continuous field-driven phase transition evident in the magnetization curve. We further demonstrate that, over a finite range of magnetic fields, the lowest-lying excited states above the ground state are indeed solitons. Building on this, we compute their dynamics by evaluating a soliton-soliton dynamical correlator, which, as  anticipated, is accurately captured by the soliton effective Hamiltonian  
constructed earlier. Finally, we discuss the implications of these soliton modes for the full spin-spin dynamical correlator.

All numerical results are obtained using 
DMRG and TEBD calculations implemented in  
the \texttt{ITensor.jl} library (v0.9)~\cite{itensor,itensor-r0.3} under PBC. Although DMRG with PBC is computationally demanding  
due to the rapid growth of the bond dimension, open boundary conditions (OBC)  
would explicitly break translational symmetry and introduce an artificial  
attractive potential that localizes solitons near the edges. To prevent such  
artifacts, we perform all simulations on finite rings of manageable size. Our  
main results correspond to systems with \(L = 48\) and \(L = 84\) sites, for  
which no appreciable finite-size effects are observed.

\subsection{Static properties}

\begin{figure}[!ht]
    \centering
\includegraphics[width=1\columnwidth]{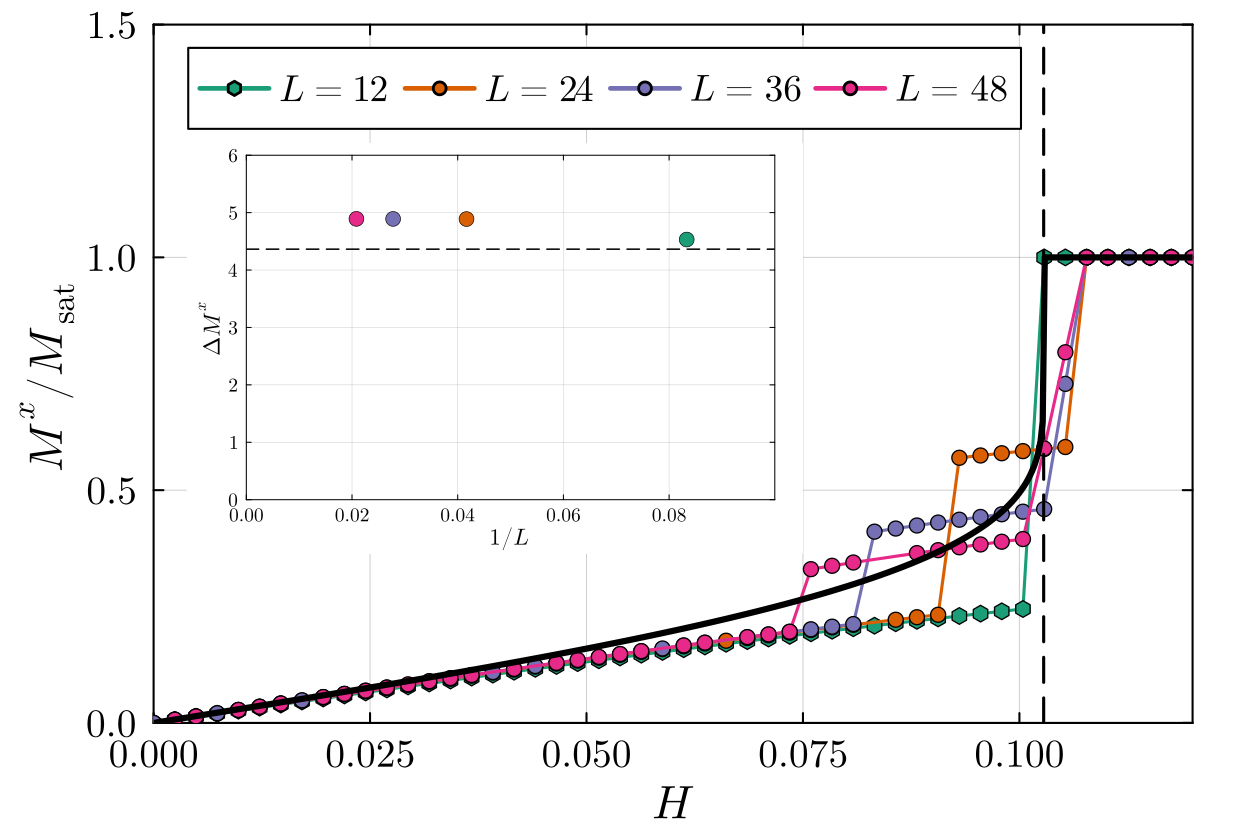}
    \caption{Magnetization curves obtained from DMRG calculations for different system sizes. The dashed vertical line marks the critical field in the classical continuum limit, while the solid black curve represents the classical and continuum limit. Inset: absolute magnetization jump, $\Delta M$, as a function of the inverse system size. The dashed horizontal line shows the continuum-limit prediction for classical solitons. }
    \label{Fig:dmrg_mag_curve}
\end{figure}

To assess the correspondence between the quantum soliton solution and the  
classical result presented in the previous section, we begin by examining the  
magnetization curve shown in Fig.~\ref{Fig:dmrg_mag_curve}. The agreement  
between the classical prediction and the quantum numerical data is remarkably  
good, indicating that the classical picture provides a reliable foundation for  
understanding the properties of the quantum soliton.

The numerically obtained saturation field, \(H_{c}=0.105\), is in very good  
agreement with the classical estimate \(H_{c}=0.1028\) given in \eqref{Eq:Hc}. As expected, the magnetization curve becomes progressively  
smoother for larger system sizes, since the discrete jumps associated with the  
successive addition of individual solitons diminish in relative magnitude. In  
the thermodynamic limit, the curve is expected to approach the smooth behavior  
predicted by the classical continuum solution. An important distinction,  
however, is that the logarithmic divergence of the slope of the classical  
magnetization curve as the saturation field is approached from below is  
replaced, in the quantum case, by a square-root divergence, reflecting the  
free-fermion fixed point that governs the critical behavior.

The size of the soliton at the critical field can be inferred from the height  
of the first magnetization jump. From the inset of  
Fig.~\ref{Fig:dmrg_mag_curve}, we estimate a soliton size  
\(\ell_{s} \approx 10\,a\), corresponding to \(\Delta M^{x} \approx 5\), in  
excellent consistency with the continuum prediction  
\(\Delta M^{x} \approx 4.36\) obtained from \eqref{Eq:cont_mag}.

While the magnetization curve shows good agreement with classical  
predictions, it is important to emphasize that, in the chiral-soliton phase,  
the solitons become fully delocalized. As a consequence, the expectation  
values of the spin components develop an essentially flat spatial profile  
along the chain, illustrating how quantum fluctuations melt the classical  
soliton crystal and restore the translational invariance that is broken in the classical solution.

\begin{figure}[!ht]
 \centering
    \includegraphics[width=\columnwidth]{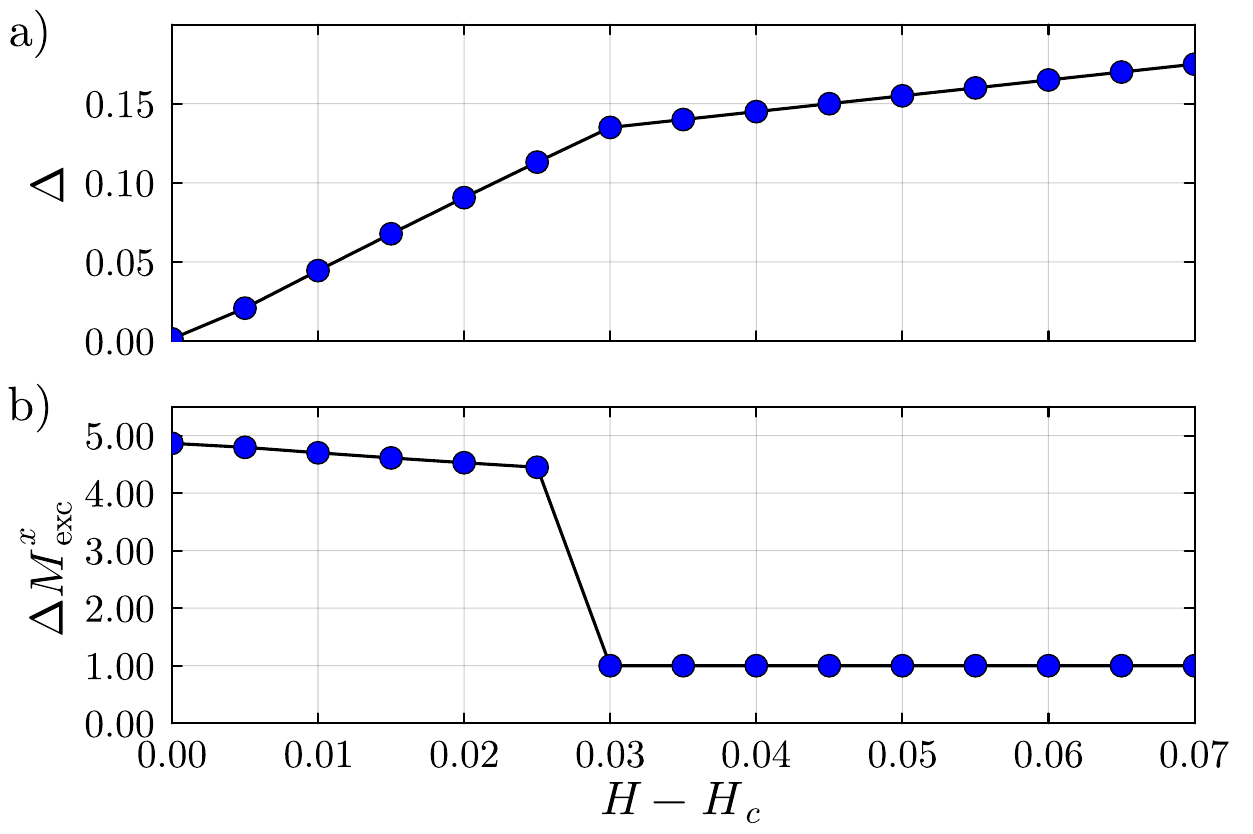}
     \caption{
     a) Energy gap, $\Delta = E_1 - E_0$, and (b) difference between the ground-state magnetization and that of the first excited state, $\Delta M^x_{\mathrm{exc}}$, for a chain of $L=48$ sites, plotted as a function of the external field $H - H_c$. Here, $H_c$ denotes the critical field obtained from DMRG calculations.}
     \label{Fig:gap_vs_E}
\end{figure}

To further characterize the excitations, we now examine the lowest excited  
state above the fully saturated phase (\(H > H_{c}\)).  
Figure~\ref{Fig:gap_vs_E} (a) shows the evolution of the energy gap  
\(\Delta = E_{1} - E_{0}\) as a function of the magnetic field for a chain of  
length \(L = 48\), where \(E_{0}\) and \(E_{1}\) denote the ground-state and  
first excited-state energies, respectively.

As shown in Fig.~\ref{Fig:gap_vs_E} (a), the slope of the energy gap \(\Delta\)  
changes discontinuously around \(h \equiv H - H_{c} \approx 0.03\), signaling  
the expected qualitative change in the nature of the lowest-energy  
excitations. This transition is further corroborated by the magnetization  
difference \(\Delta M^{x}_{\mathrm{exc}} = M^{x}_{\mathrm{FP}} - M^{x}_{1}\),  
plotted in Fig.~\ref{Fig:gap_vs_E} (b). For fields  
\(h > 0.03\), the excitations carry  
\(\Delta M^{x}_{\mathrm{exc}} = \pm 1\), consistent with single-magnon modes.  
In contrast, for \(0 < h < 0.03\), the first excited state exhibits a  
magnetization jump of  
\(\Delta M^{x}_{\mathrm{exc}} \approx 5\), consistent with the soliton size identified earlier.

We also note that the energy gap scales linearly with \(H - H_{c}\), in  
agreement with the emergent \(\mathrm{U}(1)\) symmetry of the free-fermion  
fixed point:  
\(\Delta \propto (H - H_{c})^{\nu z}\) with \(\nu = 1/2\) and \(z = 2\), as discussed in the previous section.

\subsection{Dynamical spin structure factor}

As shown previously, for sufficiently large values of $H$, the lowest excitations of the system are $S=1$ magnons. In this regime, we expect to recover the LSWT prediction with a magnon dispersion given by \eqref{Eq:magnon_disp}. To test this expectation, we evaluate the real-space, real-time spin–spin correlation function,
\begin{align}
{\cal S}^{\alpha\alpha}(j-j',t) &= \mel{0}{\hat{S}^{\alpha}_j(t)\hat{S}^{\alpha}_{j'}(0)}{0},
\label{Eq:dssf}
\end{align}
where $\ket{0}$ denotes the fully polarized ground state and $\alpha=x,y,z$. These correlations are computed using the TEBD method, as described in Appendix~\ref{Appe:dmrg}. By applying spatial and temporal Fourier transforms, we obtain the dynamical spin structure factor (DSSF), $\tilde{\mathcal{S}}(q,\omega) = \sum_{\alpha} \tilde{\mathcal{S}}(q,\omega)$, with
\begin{align}
\tilde{\cal S}^{\alpha\alpha}(k,\omega) &= 
\frac{1}{2\pi L}\sum_{l,l'}\int_{-\infty}^{\infty} dt \, e^{i[\omega t -  k(l-l')]} 
{\cal S}^{\alpha\alpha}(l-l',t).
\label{Eq:DSSF_FT}
\end{align}
The most relevant contributions are the transverse components with $\alpha = y, z$, which correspond to excitations perpendicular to the fully polarized order along the $\hat{x}$ direction. In contrast, the longitudinal component ($\alpha = x$) probes fluctuations parallel to the polarization, which are not expected in the fully polarized state. This expectation is confirmed by our numerical results, which show a negligible response in the longitudinal channel.

\begin{figure}[!ht]
    \centering
    \includegraphics[width=\columnwidth]{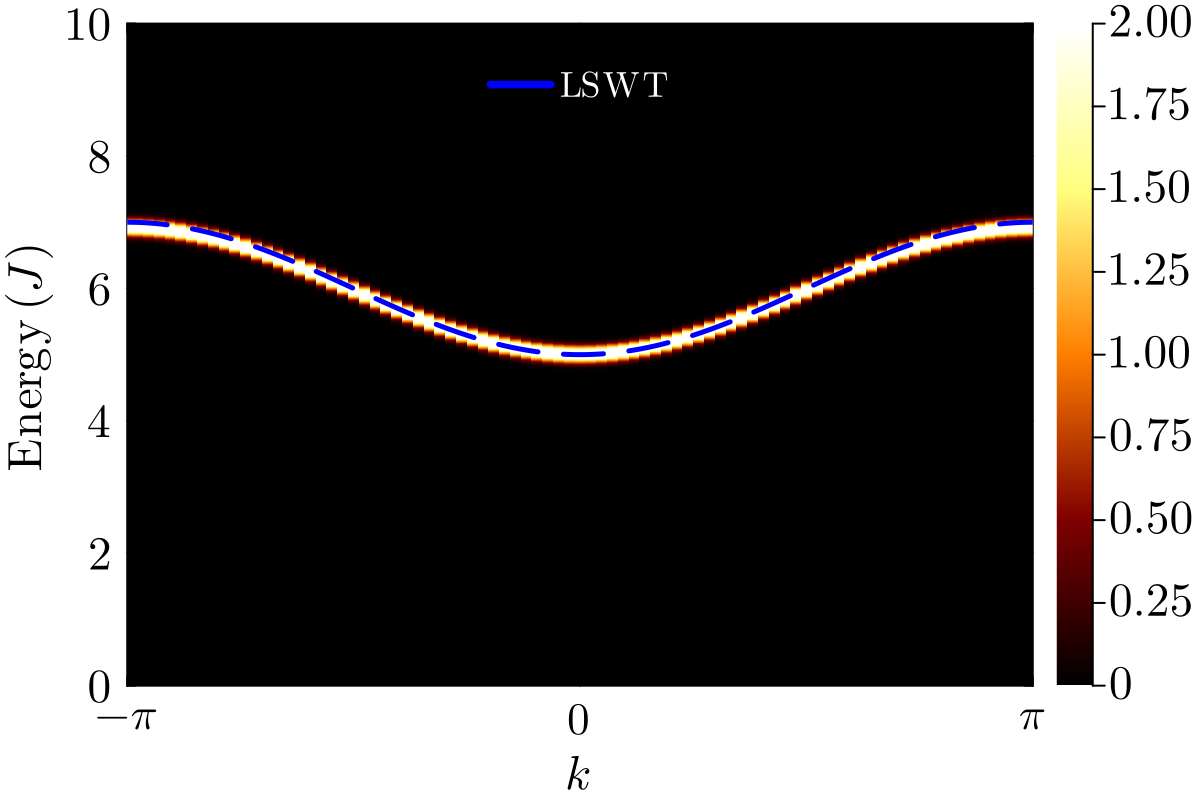}
    \caption{DSSF for a chain of $L=48$ and $H=5 \approx 50H_c$ computed by TEBD. The blue dashed lines show the LSWT magnon dispersion. The time step used for the TEBD calculation is $\delta t = 0.01$ with a truncation error of $10^{-12}$ for each application of the time evolution operator; the total evolution time is $T_f=40$ (in units of $J^{-1}$). }
    \label{Fig:dssf_vs_lwst}
\end{figure}

In Fig.~\ref{Fig:dssf_vs_lwst}, we present a direct comparison between the DSSF obtained from TEBD and the magnon dispersion predicted by LSWT at $H = 5.0$. The results show excellent agreement, demonstrating that LSWT provides a complete description of the excitation spectrum in this high-field regime.

\subsection{Dynamical soliton structure factor}

In Sec.~\ref{Sec:Quantum_solitons}, we introduced an effective tight-binding Hamiltonian describing the propagation of a soliton (or antisoliton) excitation over the fully polarized background. Although this construction is based on simple physical arguments, it is crucial to test its validity by confronting its predictions with numerical results. To this end, we define the real-space and real-time soliton--soliton ($\tau=+$) or antisoliton-antisoliton ($\tau=-$) correlation function,
\begin{equation}
{\cal T}_{\tau}(j-j',t) \equiv \mel{0}{\hat{T}_{j\tau} (t)\,\hat{T}^{\dagger}_{j'\tau}(0)}{0},
\label{Eq:dynamic_sol}
\end{equation}
which measures the probability amplitude of finding a soliton at site $j$ and time $t$, given that a soliton was created at site $j'$ at $t=0$ on top of the ground state. 

The corresponding dynamical soliton or antisoliton spectral function  is obtained from the space–time Fourier transform of ${\cal T}_{\tau}(j-j',t)$,  
\begin{equation}
\tilde{\cal T}_{\tau}(k,\omega) = \frac{1}{2\pi L}\sum_{l,l'}\int_{-\infty}^{\infty} dt\;
e^{i[\omega t - k (l-l')]}\,
{\cal T}_{\tau}(l-l',t),
\label{Eq:dyn_spec}
\end{equation}
which directly encodes the soliton dispersion and lifetime. This representation provides the natural quantity for comparison with the predictions of the effective tight-binding Hamiltonian.  

Here we employ the original soliton creation operator [see \eqref{Eq:SolOp}] rather than its Wannierized form. It is worth noting that using the Wannierized form of the soliton operators in \eqref{Eq:dynamic_sol} would only affect the spectral weight of the $\delta$-function peaks in $\tilde{\cal T}_{\tau}(k,\omega)$. Since our primary interest lies in extracting the soliton dispersion relation---that is, the frequency of the peak for each value of $k$---from numerical calculations, and because $\tilde{\cal T}_{\tau}(k,\omega)$ is not directly observable, it suffices to define ${\cal T}_{\tau}(j-j',t)$ in terms of the non-Wannierized soliton operators.

\begin{figure*}[!ht]
    \centering
    \includegraphics[width=1\linewidth]{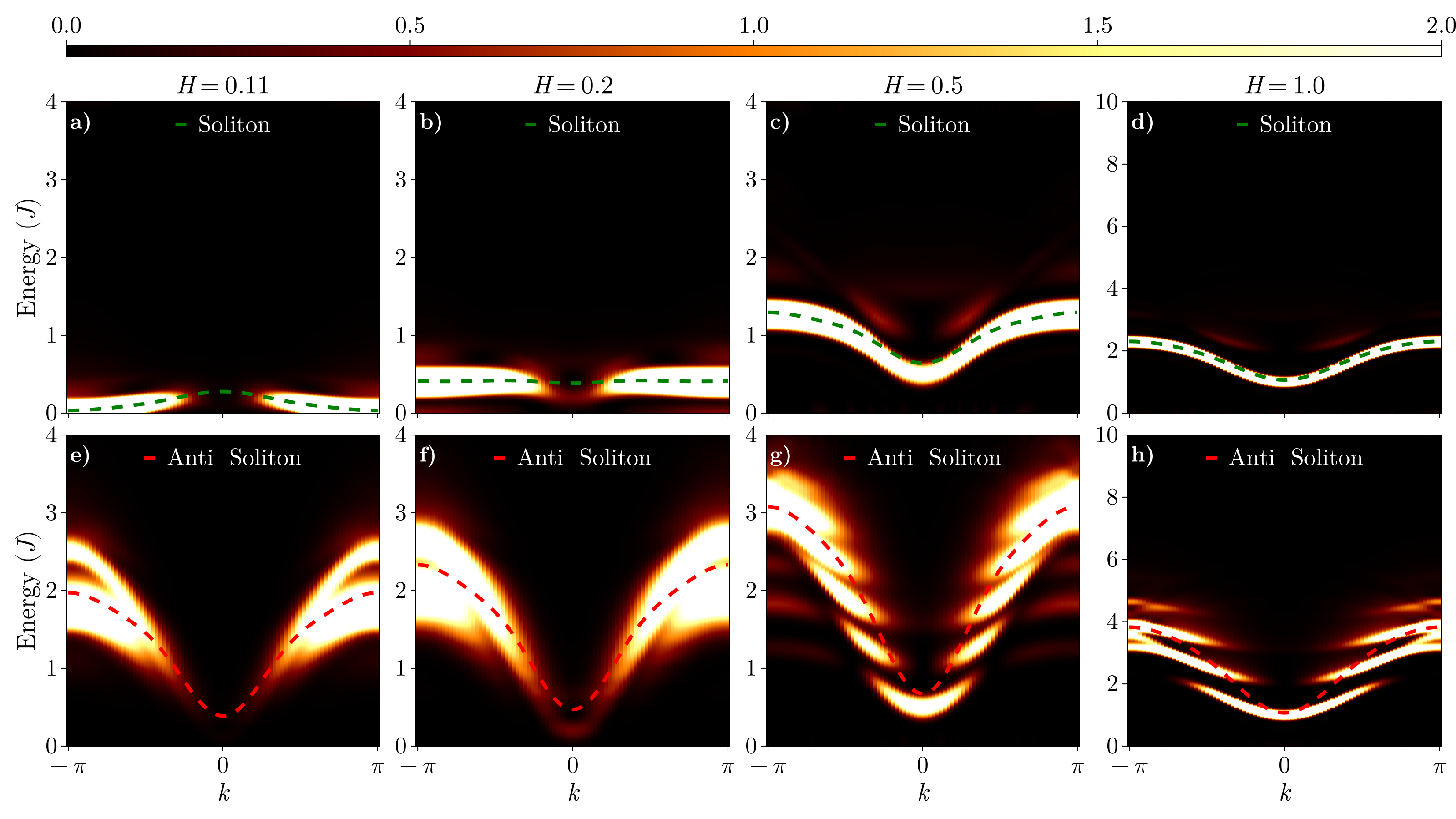}
    \caption{Dynamical soliton-soliton  (top) and antisoliton-antisoliton structure factor (bottom) for a chain of $L=84$ sites at (a,e) $H=0.11$, (b,f) $H=0.2$, (c,g) $H=0.5$, and (d,h) $H=1.0$. Note that the vertical axis scale changes for panels (d,h). The TEBD parameters are the same as in Fig.~\ref{Fig:dssf_vs_lwst}. }
     \label{Fig:Tkw}
\end{figure*}

The top panel of Fig.~\ref{Fig:Tkw} presents the numerical results for the dynamical soliton structure factor  at several values of the magnetic field, together with the soliton bands predicted by the analytical theory. The agreement is nearly perfect: The predicted soliton bands reproduce the numerically obtained dispersions with high accuracy. The apparent broadening of the numerical spectra originates primarily from the finite frequency resolution imposed by the truncation of the time evolution.

The bottom panel of Fig.~\ref{Fig:Tkw} shows the results for the dynamical antisoliton structure factor. In contrast to the soliton case, the spectrum exhibits a more intricate structure. Although the numerically computed intensities broadly follow the analytical antisoliton dispersion, additional modes appear, reconstructing the spectrum and giving rise to richer features. These extra modes are faintly visible in the dynamical soliton structure factor for fields $H=0.5$ and $1.0$ [see Figs.~\ref{Fig:Tkw} (e) and \ref{Fig:Tkw} (d)]. This behavior provides a clear signature of hybridization between antisoliton (or soliton) modes and other excitations. In particular, because the antisoliton is always the most energetically costly excitation, it tends to strongly overlap with, for instance, the magnon mode, thereby generating the enhanced spectral complexity observed. Far from being a drawback, such mode hybridization offers a pathway to experimentally ``illuminate’’ solitonic excitations, as discussed in the following section.

\section{\label{sec:exp}Experimental signatures }

In the previous section, we validated our effective low-energy theory against  
numerical calculations, showing that the sine-Gordon model provides an  
accurate description of the low-energy physics.  
To summarize, (i) the magnetization curve is well captured by the classical  
sine-Gordon prediction; (ii) the classical soliton solution offers a  
meaningful starting point for constructing a ``quantum soliton'' state;  
(iii) at the critical field, solitons soften and condense, giving rise to a  
Tomonaga-Luttinger liquid; and (iv) over a finite field range above  
saturation, solitons constitute the lowest-energy excitations.

Building on these results, we now identify experimental signatures of chiral  
solitons in real materials. The central finding is that magnon excitations, 
directly accessible through INS, acquire a  
substantial hybridization with the soliton bands over a finite window of  
magnetic fields, making the solitons observable in INS spectra. In addition,  
we show that the presence of soliton bands strongly modifies the field  
dependence of thermodynamic properties, most notably the specific heat.

\subsection{Magnon-soliton hybridization and inelastic neutron scattering}
\label{Sec:hybridization}

\begin{figure*}[!ht]
\centering
\includegraphics[width=0.8\linewidth]{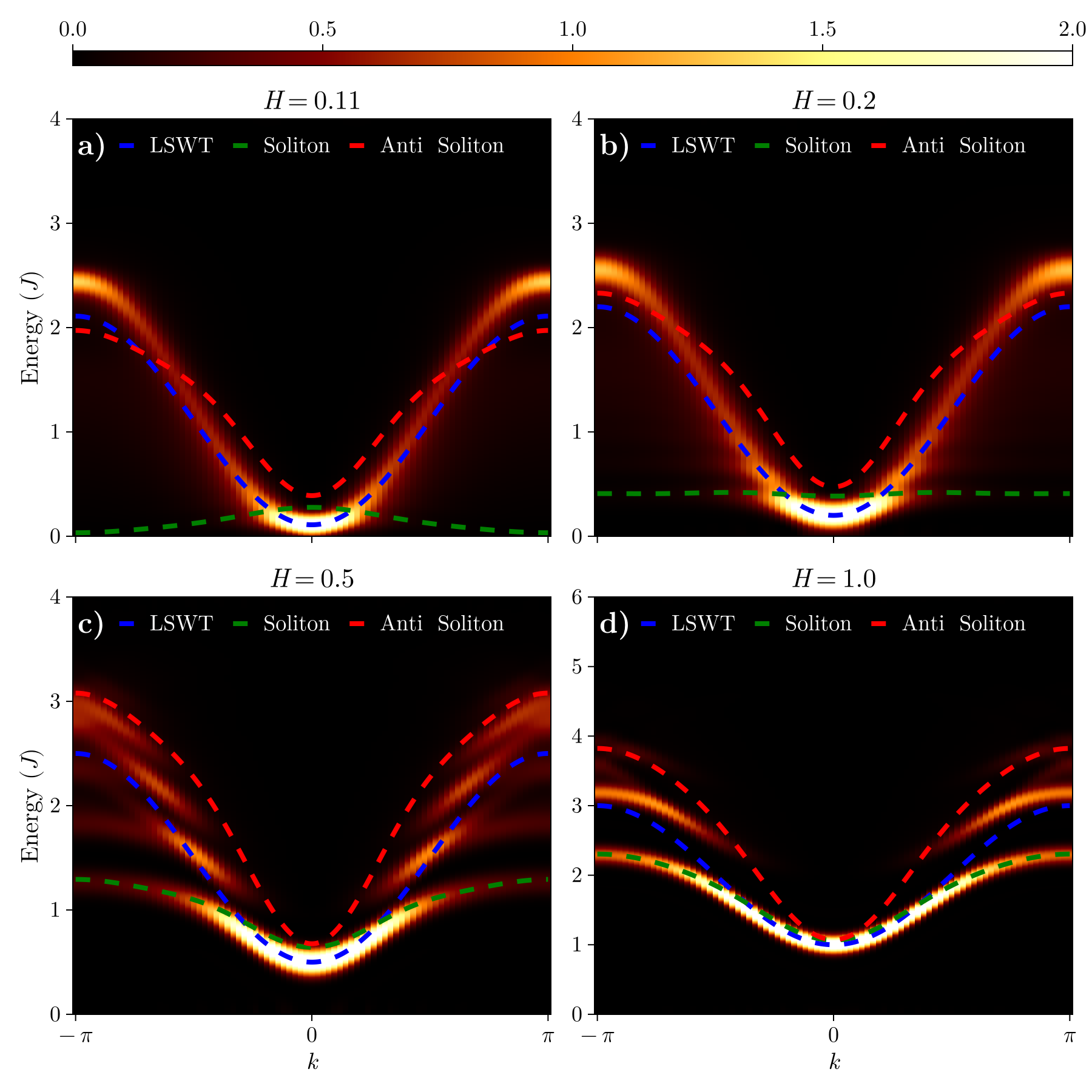}
\caption{Dynamical spin structure factor $\tilde{\mathcal{S}}(q,\omega)$ for a chain of $L=84$ sites at (a) $H=0.11$, (b) $H=0.2$, (c) $H=0.5$, and (d) $H=1.0$. The TEBD parameters are the same as in Fig.~\ref{Fig:dssf_vs_lwst}. Close to the critical field ($H\approx H_c$), hybridization between magnon and soliton modes is negligible because magnons with $k\approx0$ remain nearly exact eigenstates of the Hamiltonian. At $H\approx0.2$, incipient magnon-soliton hybridization becomes visible as the two dispersions intersect away from $k=0$. Hybridization between the magnon band and the multisoliton continuum is also observed at larger momenta. At $H=0.5$, the full soliton band becomes visible through its hybridization with the magnon mode. At $H=1$, the soliton and magnon excitations become indistinguishable as the soliton size approaches the lattice spacing.}
\label{Fig:Skw}
\end{figure*}

We begin by summarizing the main results of this section. Figure~\ref{Fig:Skw} displays the DSSF, defined in ~\eqref{Eq:DSSF_FT}, obtained from TEBD calculations for a chain of $L=84$ sites at magnetic fields $H = 0.11$, $0.2$, $0.5$, and $1.0$. As anticipated in the previous sections, the numerical spectra exhibit pronounced deviations from the predictions of linear spin-wave theory (LSWT). The resulting excitation spectrum reveals several distinctive features that we now analyze and relate to the effective theory of quantum solitons developed earlier.

INS probes excitations with $\Delta S^x = \pm 1,0$. In the present system, the Dzyaloshinskii-Moriya (DM) interaction breaks the U(1) symmetry of $\hat{H}_S$; thus, $[\hat{H}_S,\hat{M}^x]\neq0$, and the magnetization along the field direction is no longer conserved. Consequently, eigenstates of $\hat{H}_S$ generally involve superpositions of sectors with different magnon numbers. Two notable exceptions are the fully polarized ground state $\ket{0}$ above the saturation field and the zero-momentum single-magnon state $\hat{S}^z_{k=0}\ket{0}$, which remain exact eigenstates with magnetizations $L/2$ and $L/2-1$, respectively (see Appendix~\ref{Appe:analitical_states}).

At sufficiently large magnetic fields the U(1) symmetry is approximately restored, since the energy separation between different magnon-number sectors greatly exceeds the matrix elements of the DM interaction. In this regime the excitations probed by INS essentially correspond to single magnons throughout the Brillouin zone (see Fig.~\ref{Fig:dssf_vs_lwst}). As the magnetic field is reduced, however, the situation changes qualitatively.

The quantum soliton wave function constructed in Sec.~\ref{Sec:Quantum_solitons} can be viewed as a variational state consisting of a coherent superposition of single-magnon and multimagnon bound states, with the finite chirality encoded in the coefficients of the superposition. Although multi-magnon bound states incur a larger Zeeman energy cost due to the presence of multiple spin flips, this cost is partially compensated by their mutual binding energy. As a consequence, the bound-state branches are shifted downward in the spectrum. In an intermediate-field regime, the energies of single-magnon and multimagnon bound states approach one another, enabling strong hybridization mediated by the DM interaction. This hybridization transfers spectral weight from the dominant single-magnon modes to the bound states, making the latter directly visible in the DSSF.

We now analyze the excitation spectrum for representative magnetic fields, proceeding from high fields toward the critical field $H_c$.

\paragraph{High field: $H = 1.0 \approx 10 H_c$.}

 Two principal bands are clearly visible in this regime [Fig.~\ref{Fig:Skw} (d)]. The lower-energy branch shows excellent agreement with the soliton dispersion derived in Sec.~\ref{Sec:Quantum_solitons}. At this field strength, the classical soliton size—estimated from the magnetization jump relative to the ground state—is $\Delta M^{x} \approx 1.41$, placing it between the one- and two-magnon bound-state sectors. Although the spatial extent of this excitation becomes comparable to the lattice spacing at such large fields, we retain the term ``soliton'' to emphasize its continuous evolution into the semiclassical soliton mode as $H$ approaches $H_c$.

A natural next step is to quantify the hybridization between the two branches. As shown in Appendix~\ref{Appe:bound}, this analysis yields, within controlled approximations, the following band structure:
\begin{align}
E_1(k) &\approx \varepsilon_1(k) - \frac{D^2}{H}\sin^4\!\left(k/2\right) \\
E_2(k) &\approx \varepsilon_2(k) + \frac{D^2}{H}\sin^4\!\left(k/2\right) 
\label{Eq:two_bounds_band}
\end{align}
with
\[
\epsilon_{1}(k) = H + 2J \sin^{2}(k/2), \qquad
\epsilon_{2}(k) = 2H + J \sin^{2}(k/2).
\]

\begin{figure}[!ht]
\centering
\includegraphics[width=\columnwidth]{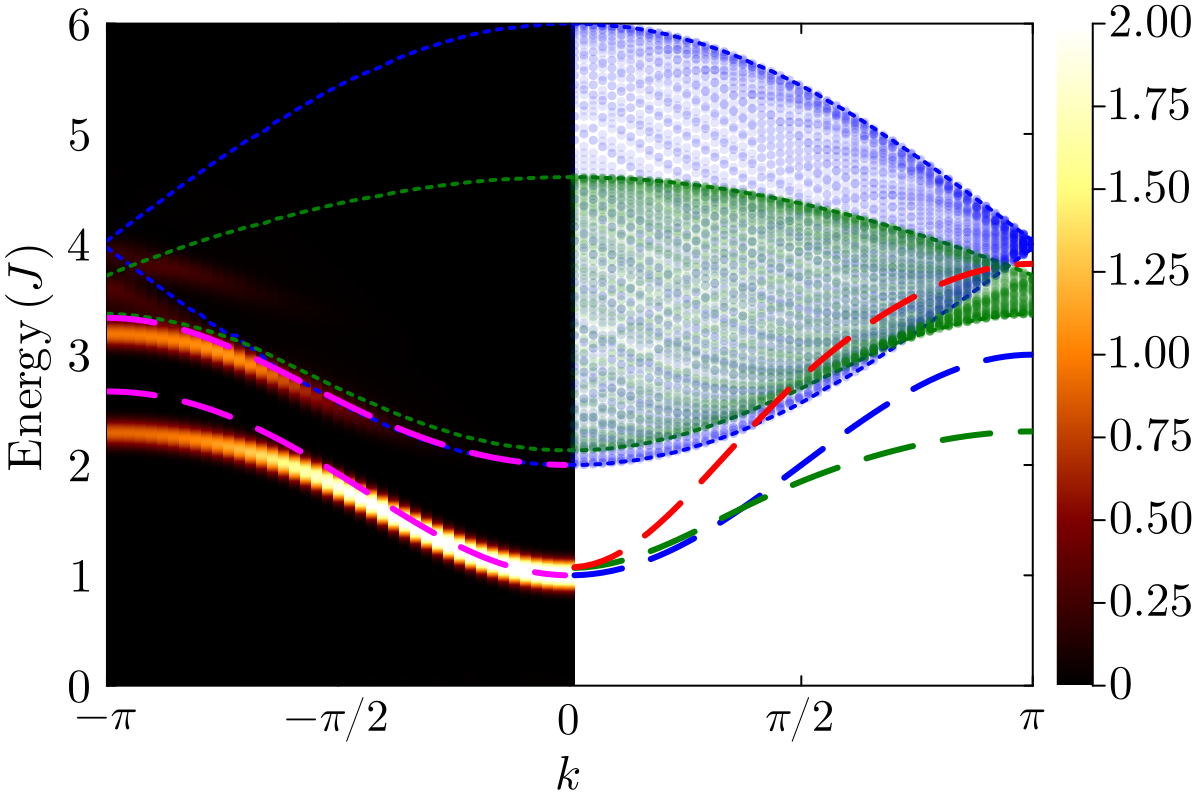}
\caption{Left: DSSF showing the bands obtained from~\eqref{Eq:two_bounds_band} (magenta dashed line). Right: two-magnon continuum (blue dots) and two-soliton continuum (green dots), together with the single-magnon dispersion (blue dashed), soliton dispersion (green dashed), and antisoliton band (red dashed). Results correspond to $H=1$.}
\label{Fig:two_cont_H1}
\end{figure}

As shown in Fig.~\ref{Fig:two_cont_H1}, this simple description reproduces the high-energy band remarkably well, while slightly overestimating the energy of the low-energy branch for $k>\pi/2$. The high-energy magenta band corresponds to the lower edge of the two-soliton continuum (green dots), with its spectral weight vanishing as it approaches the two-magnon continuum (blue dots). As discussed previously, only the magnon excitation is visible in INS at $k=0$, since $\hat{\tilde{S}}^z_{k=0}\ket{0}$ is an exact eigenstate.

\paragraph{Intermediate field: $H = 0.5 \approx 5 H_c$.}

Upon lowering the field [Fig.~\ref{Fig:Skw} (c)], the excitation spectrum becomes more intricate and displays a richer structure. Nevertheless, the lowest branch remains in excellent agreement with the soliton dispersion predicted by the effective theory. To account for the additional bands, we extend the previous analysis to include hybridization among a larger set of states: the soliton, the antisoliton, the single-magnon excitation, and a composite magnon-soliton state formed by a $k=0$ magnon and a soliton of momentum $k$,
\begin{equation}
\ket{\alpha_k}=
\frac{\hat{\tilde{S}}^z_{k=0}\hat{\tilde{T}}^\dagger_k\ket{0}}
{\sqrt{\bra{0}\hat{\tilde{T}}_k(\hat{\tilde{S}}^z_{k=0})^2\hat{\tilde{T}}^\dagger_k\ket{0}}}.
\end{equation}

These four states span the low-energy subspace $\mathcal{S}_0$. Because the corresponding variational states are not mutually orthogonal, we construct an orthonormal basis by means of L\"owdin symmetric orthogonalization~\cite{Lowdin1950,Mayer2002}, which preserves, in a least-squares sense, the geometric structure of the original subspace (see Appendix~\ref{Appe:Lowdin}). Physically, this procedure allows each orthogonalized state to retain a well-defined dominant character—single soliton, single antisoliton, single magnon, or composite magnon-soliton. The Hamiltonian projected onto this basis then reads
\begin{equation}
h^{\mathrm{multi}}_{jj'}(k)
=
\mel{\psi_j}{\hat{H}_S}{\psi_{j'}} - E_0\delta_{jj'} .
\end{equation}

\begin{figure}[!ht]
\centering
\includegraphics[width=\columnwidth]{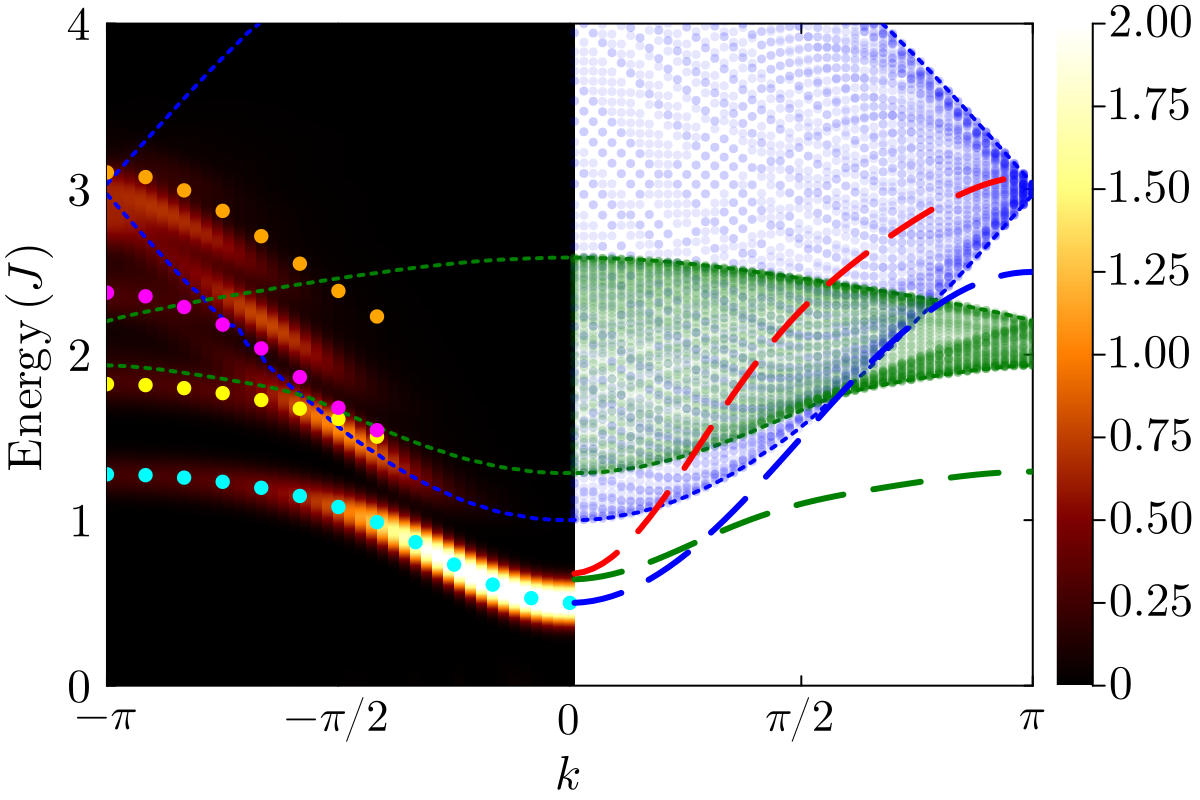}
\caption{Left: DSSF with hybridized bands obtained from the projection procedure described in the text. Right: two-magnon continuum (blue dots) and two-soliton continuum (green dots), together with the single-magnon, soliton, and antisoliton dispersions.}
\label{Fig:two_contH05}
\end{figure}

Diagonalizing this matrix yields the hybridized modes shown in Fig.~\ref{Fig:two_contH05}. The dispersive branches are accurately reproduced until they merge into the two-magnon and two-soliton continua. In this field range all four excitation branches carry observable spectral weight in the DSSF, making both the soliton and antisoliton dispersions experimentally accessible.

\paragraph{Near the critical field: $H \gtrsim H_c$.}

As the field approaches $H_c$ [Figs.~\ref{Fig:Skw} (a) and~\ref{Fig:Skw} (b)], the lowest excited state evolves into a coherent superposition of several multimagnon bound states that can be identified with the semiclassical soliton. The soliton gap decreases continuously and vanishes at $H=H_c$, where a finite density of solitons condenses into the ground state, driving the commensurate-to-incommensurate transition into the Tomonaga-Luttinger liquid phase discussed earlier.

\begin{figure}[!ht]
\centering
\includegraphics[width=\columnwidth]{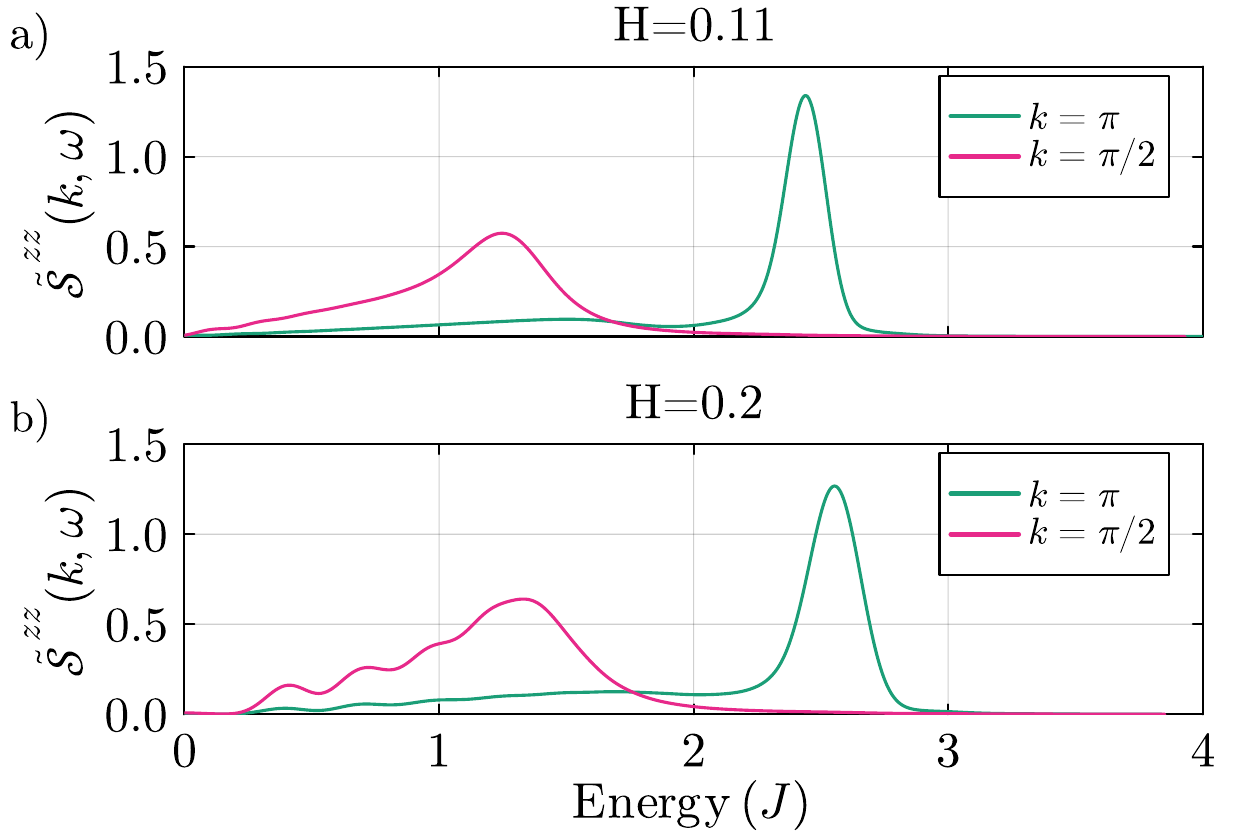}
\caption{Constant-momentum cuts of the DSSF $\tilde{\mathcal{S}}(k,\omega)$ at $k=\pi$ and $k=\pi/2$ for magnetic fields $H=0.11$ (a) and $H=0.2$ (b).}
\label{Fig:cuts}
\end{figure}

From a spectroscopic standpoint, the overlap between the soliton and single-magnon excitations becomes very small near $H_c$, rendering the soliton mode nearly invisible in INS measurements. Nevertheless, careful inspection of Figs.~\ref{Fig:Skw} (a) and~\ref{Fig:Skw} (b) reveals clear solitonic signatures in the excitation spectrum. These features become particularly evident in the constant-momentum cuts shown in Fig.~\ref{Fig:cuts}, which exhibit well-defined peaks at energies consistent with the soliton dispersion.

\paragraph{Character fidelity across the field range.}

\begin{figure*}[!ht]
    \centering
    \includegraphics[width=\linewidth]{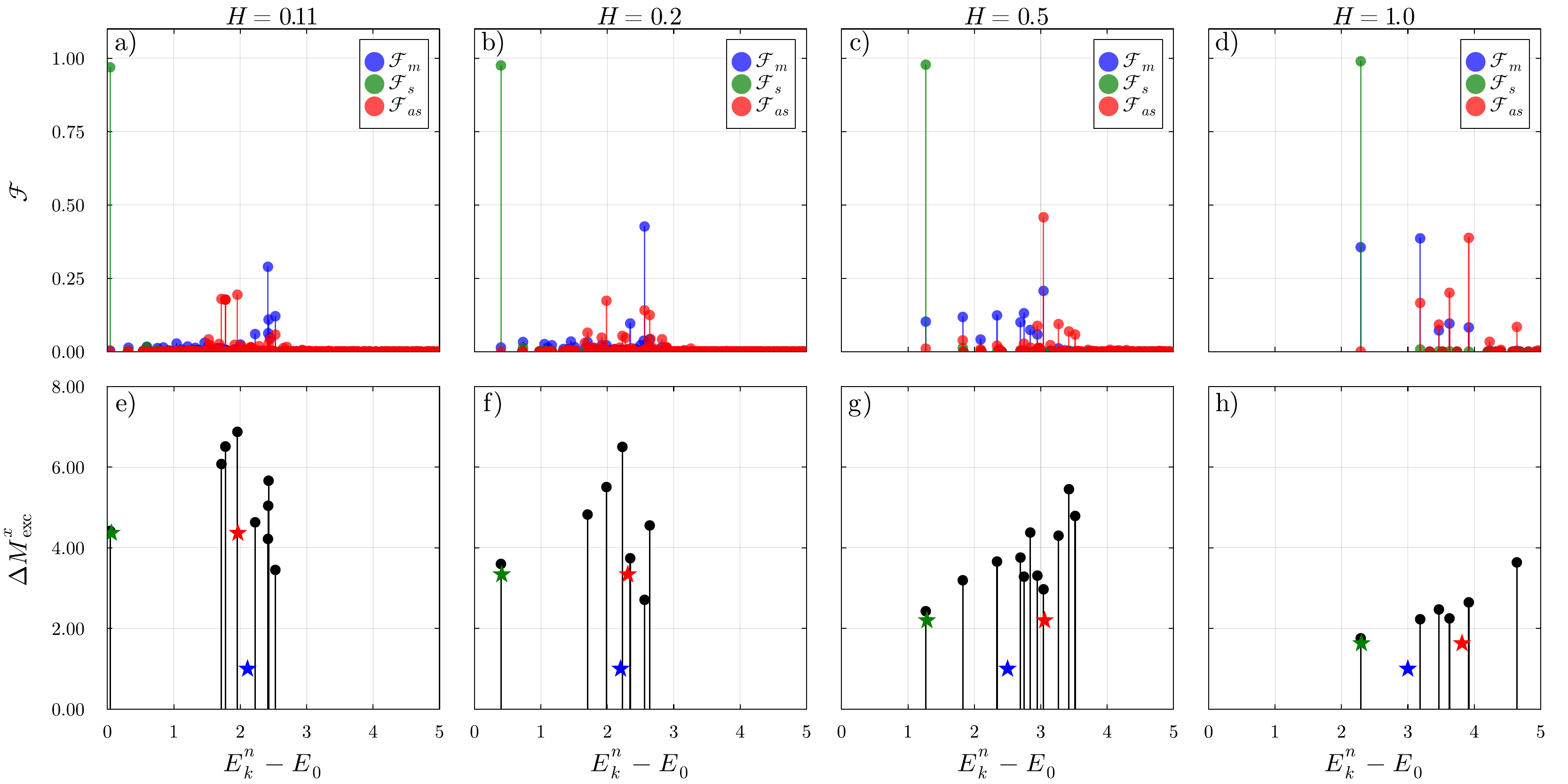}
    \caption{a-d)~Character fidelity, as defined in \eqref{Eq:character}. e-h)~Longitudinal magnetization difference $\Delta M^x_{\mathrm{exc}}$ of the lowest-lying excitations at momentum $k = \pi$ for different magnetic fields. The stars in the bottom panels indicate the expected energy and magnetization values corresponding to the variational magnon (blue), soliton (green), and antisoliton (red) states.}
    \label{Fig:Characteres}
\end{figure*}

As a final verification of the excitation character across the full field range, we compute the squared overlap between the exact eigenstates and the variational magnon, soliton, and antisoliton states,
\begin{equation}
\mathcal{F}_{\mathrm{m}}
=
|\mel{n}{\hat{\tilde{S}}^z_k}{0}|^2,
\qquad
\mathcal{F}_\tau
=
|\mel{n}{\hat{\tilde{T}}^\dagger_{k\tau}}{0}|^2 ,
\label{Eq:character}
\end{equation}
where $\ket{n}$ denotes exact eigenstates obtained from exact diagonalization for $L=12$. The operators $\hat{\tilde{S}}^z_k$ and $\hat{\tilde{T}}^\dagger_{k\tau}$ generate normalized states when acting on $\ket{0}$, so $\mathcal{F}\le1$. We refer to these quantities as the  fidelity. Figure~\ref{Fig:Characteres} shows the results for the same field values as above, evaluated at momentum $k=\pi$, together with the longitudinal magnetization $\Delta M^x_{\mathrm{exc}}$ of the corresponding eigenstates.

At $H=1$ [Fig.~\ref{Fig:Characteres} (d)], the lowest excitation displays a pronounced solitonic character with fidelity $\mathcal{F}_s\approx0.95$. Its longitudinal magnetization agrees closely with the value predicted for the variational soliton state. The same mode also contains a substantial magnon component, reflecting the smooth crossover between soliton and magnon excitations at lattice scales. This crossover is precisely what allows INS to detect the soliton mode, as the magnon admixture transfers spectral weight to the $\Delta S^x=\pm1$ channel.

At $H=0.5$ [Fig.~\ref{Fig:Characteres} (c)], all four excitation branches are clearly resolved. The intermediate bands are predominantly magnonic, whereas the lowest and highest branches exhibit mainly solitonic and antisolitonic character, respectively. At higher energies the situation becomes more complex because the magnon and antisoliton modes hybridize strongly with the two-soliton continuum, resulting in a less clearly defined excitation character.

As the field approaches $H_c$ [Figs.~\ref{Fig:Characteres} a) and b)], the overlap with the single-magnon state decreases progressively, consistent with the soliton becoming increasingly invisible in INS. Nevertheless, the lowest excitation retains its dominant solitonic character throughout the entire field range, and its longitudinal magnetization remains in excellent agreement with the variational prediction.

\subsection{Thermodynamic properties}

The thermodynamic consequences of the emergence of solitons as low-energy excitations of the system  are no less remarkable. As discussed throughout this work, for fields slightly above the saturation value, solitons constitute the lowest-energy excitations and exhibit a dispersive spectrum. Owing to the relatively flat soliton dispersion in this field range, a Schottky-like anomaly is expected to appear in the specific heat,
\begin{equation}
C_v = \frac{1}{L}\frac{dU}{dT},
\end{equation}
where \( U \) denotes the internal energy. This low-temperature feature should be followed by a broader magnon-induced contribution at higher energies. At larger fields, the Schottky anomaly is expected to merge with the magnon contribution, resulting in a single broad peak in the specific heat.

\begin{figure}[!ht]
    \centering
    \includegraphics[width=\columnwidth]{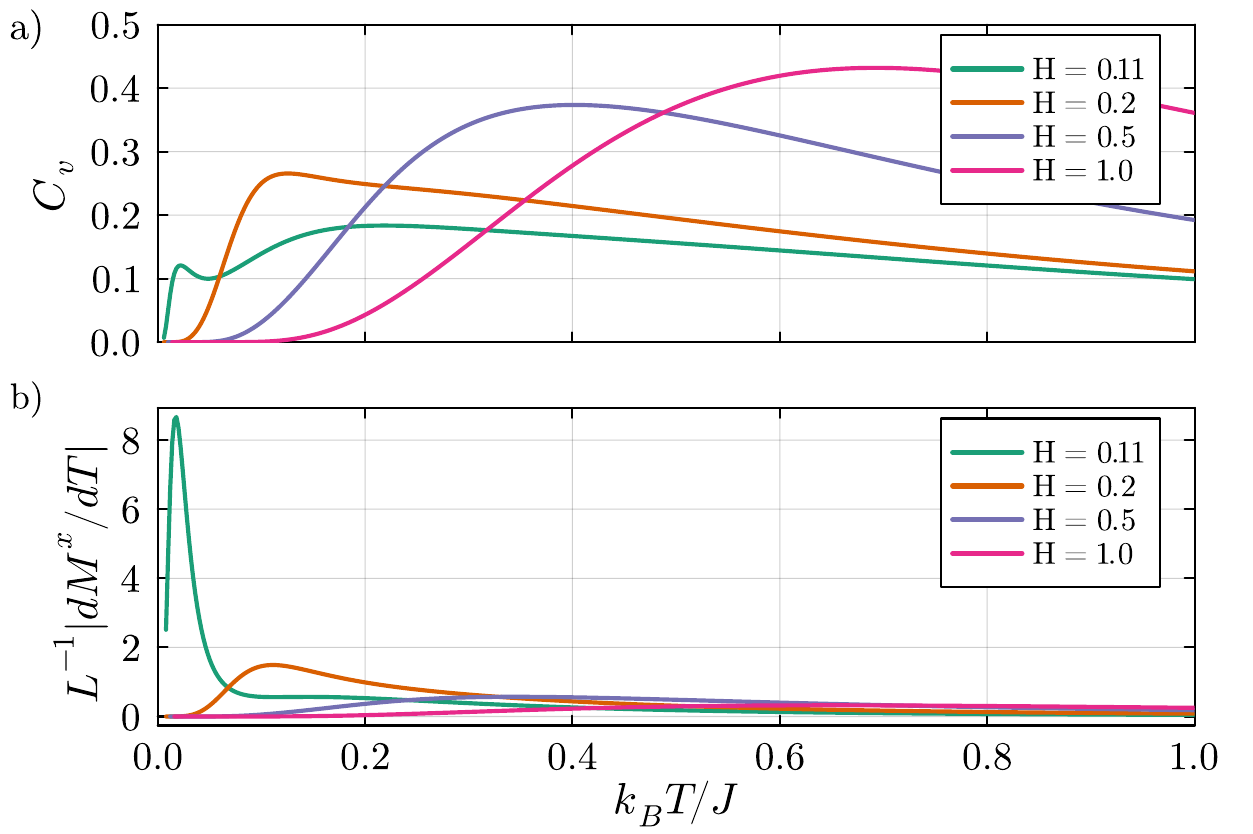}
    \caption{Temperature dependence of the specific heat (a) and of the absolute value of the temperature derivative of the total magnetization along the field direction (b) for a system with $L=12$ sites under different external magnetic fields.}
    \label{Fig:thermo}
\end{figure}

Furthermore, because creating a soliton involves a substantial change in the  
total magnetization, a pronounced response is also expected in the temperature  
derivative of the magnetization. At the critical field $H = H_{c}$, the free–fermion  
fixed point implies a characteristic divergence  
$ L^{-1} dM^{x}/dT \propto 1/\sqrt{T} $. To test this expectation, we perform exact diagonalization  
calculations for a system with $L = 12$ sites, and the resulting behavior is  
shown in Fig.~\ref{Fig:thermo}.

These observations, together with the magnetization curve shown in Fig.~\ref{Fig:dmrg_mag_curve}, provide clear and accessible signatures of solitonic excitations through relatively simple experimental measurements. 
Thus, they offer a practical strategy for identifying promising candidate materials for INS studies, where the presence of these excitations can be unambiguously established using the analysis presented in the previous section.

\section{\label{Sec:Discussion}Discussion}

This work establishes a fully quantum, lattice-resolved framework for the
dynamics and condensation of chiral solitons across the saturation field of a spin-$\tfrac12$ monoaxial chiral
helimagnet. The central message is that chiral solitons can be promoted from classical topological textures to microscopic quantum quasiparticles of
the lattice Hamiltonian, with a well-defined dispersion over the full
Brillouin zone, a systematically derived effective low-energy Hamiltonian in
the dilute regime, and, crucially, a quantitative connection to experimentally
measured response functions.

A key advantage of the lattice formulation is that it yields direct predictions
for the dynamical spin structure factor $\tilde{\mathcal{S}}(q,\omega)$ probed by
inelastic neutron scattering. Continuum quantum treatments based on the
sine-Gordon theory and its dual massive Thirring model provide an elegant
long-wavelength description of soliton excitations, but they do not
straightforwardly determine how those excitations appear in
$\tilde{\mathcal{S}}(q,\omega)$ for a microscopic magnet. By constructing
momentum-resolved many-body soliton wave functions and their associated
creation operators directly from the spin Hamiltonian, our approach makes the
lattice-operator content explicit and therefore allows one to compute not
only excitation energies but also their spectral weights. In this way, the
theory identifies a concrete detection mechanism for quantum chiral solitons:
Soliton-magnon hybridization transfers spectral weight into the soliton
sector, rendering solitons visible as sharp features in
$\tilde{\mathcal{S}}(q,\omega)$ above the saturation field.

From the viewpoint of the low-energy spectrum, the theory reveals how topology
and quantum fluctuations reorganize the hierarchy of excitations near the
transition. We show that quantum chiral solitons become the lowest-energy
excitations over a finite field window $H_c < H < H^\ast$, while the
single-magnon mode remains gapped at $H=H_c$. The persistence of a magnon gap
implies that soliton-soliton interactions mediated by virtual magnons are
short-ranged and exponentially suppressed in the dilute regime, enabling a
controlled effective description. The resulting dual Hamiltonian, obtained
directly from the microscopic model, recasts the system as a dilute gas of
interacting topological quasiparticles. In the classical limit, the
softening of the soliton mode at $H_c$ drives the crystallization into a
soliton lattice; in the quantum spin-$\tfrac12$ chain, zero-point motion melts
this classical crystal into a gapless Tomonaga-Luttinger liquid, providing a
fully quantum realization of soliton condensation.

Our construction also clarifies which aspects of the physics lie beyond the
standard sine-Gordon-Thirring paradigm. While the continuum theory predicts
symmetric soliton and antisoliton dispersions at leading order, the lattice
treatment naturally captures strong asymmetries between the two branches and
their distinct momentum dependence across the Brillouin zone. These effects
translate directly into measurable consequences through the momentum- and
field-dependent redistribution of spectral weight in
$\tilde{\mathcal{S}}(q,\omega)$. More broadly, the Wannierization-based construction
provides a systematic procedure to incorporate quantum fluctuations into
topological textures without invoking a gradient expansion or large-$S$
limit, and it exposes lattice-scale processes—such as the tunneling events
responsible for soliton mobility and the associated spin-parity effects.

On the experimental side, our results provide concrete guidance for
identifying quantum solitons in spectroscopy. The characteristic signatures
are (i) a well-defined soliton branch above $H_c$ with finite spectral weight
acquired through hybridization with magnons, (ii) a field-tunable level
repulsion and intensity transfer near the crossing of magnon and soliton
dispersions for $H\gtrsim H^\ast$, and (iii) a reorganization of the
low-energy spectral weight as the system approaches the soliton-condensation
point. These predictions are quantitatively benchmarked against DMRG
calculations, demonstrating that the lattice quasiparticle description remains
accurate in the strongly quantum regime where soliton cores span only a few
lattice spacings.

While the present work focuses on a ferromagnetic spin chain, it is natural
to ask about material realizations. Ferromagnetic chiral-chain systems are
less common than antiferromagnetic ones, but several candidate platforms are
known, including CrNb$_3$S$_6$~\cite{Shimamoto2022}, CsNiF$_3$~\cite{Mouritsen1984,Balucani1982,Karadamoglou2000},
and effective $S=\tfrac12$ chain compounds such as CoNb$_2$O$_6$~\cite{Coldea2010}
and CsCuCl$_3$~\cite{Adachi1980,Nikuni1993}. In CsCuCl$_3$, a
uniform DM interaction is allowed by the lack of inversion symmetry along the
chain direction, although interchain couplings are not negligible. Materials
with weaker interchain coupling would provide particularly clean platforms
for realizing the soliton-dominated regime described here. Another promising
direction is the $S=\tfrac12$ chiral compound YbNi$_3$Al$_9$, which exhibits
strong signatures of solitonic physics, although a definitive effective model
is still under active development~\cite{Aoki2018,WANG2025}.

Perhaps the most far-reaching implication of the present work is that it
provides a general nonperturbative strategy for deriving effective theories
of interacting \emph{topological} quasiparticles directly from microscopic
lattice Hamiltonians. Although demonstrated here in one dimension, the
framework does not rely on integrability or continuum approximations and is
therefore not inherently restricted to $1$D, suggesting an alternative
route toward topological quantum matter in higher dimensions: Rather than
starting from parton or gauge-theory constructions, one may begin with classical spin models whose ground states have
well-defined soliton crystals (e.g., skyrmion or meron crystals) and develop a controlled quantum description of their dynamics,
interactions, and condensation for the $S=1/2$ limit. In this perspective, quantum liquids can
emerge as the quantum-melted phases of interacting topological textures,
providing a microscopic pathway toward chiral quantum spin liquids and
related strongly correlated topological phases.

Recent works in $D=2+1$ have explored quantum effects on skyrmion textures using complementary strategies, including the explicit
construction of skyrmion creation operators and variational treatments of
quantum fluctuations~\cite{Lohani2019,Maeland2022,Haller2022,Takashima2016,Bhowmick2025,Haller2024,Petrovi2025}. 
A central challenge in many classical skyrmion-hosting models is that the
field-driven transition between the skyrmion crystal and the polarized state
is often strongly first order, which avoids access to  a dilute regime where
individual topological quasiparticles are well defined and their interactions are  relatively weak.
Identifying microscopic Hamiltonians or parameter regimes where a continuous,
density-tuned transition occurs—analogous to the commensurate-incommensurate
transition studied here—would therefore be especially valuable. In addition,
numerical simulations in two dimensions remain substantially more demanding,
and finite-size effects can pin textures and obscure intrinsic dynamics.
We expect that the present lattice-level quantum formalism, together with the
clear one-dimensional benchmark established here, will motivate further
efforts to overcome these obstacles.

In summary, our results provide a unified microscopic description of quantum
chiral solitons as observable quasiparticles and of their condensation as a many-body quantum critical phenomenon. We hope that this work will stimulate
experimental searches for quantum soliton signatures in spectroscopy and encourage the development of analogous lattice-based, non-perturbative
descriptions of topological excitations in higher-dimensional quantum
materials.

\section*{Acknowledgments}
The authors are grateful to Yusuke Kato for insightful discussions. We thank Kirill Povarov and Andrew Christianson for the discussion of the experimental situation. L.M.C. was supported by the U.S. Department of Energy, Office of Science, Basic Energy Sciences, Materials Science and Engineering Division.
C.D.B. acknowledges support from the U.S. Department of Energy, Office of Science, Office of Basic Energy Sciences, under Award No. DE-SC0022311.

\appendix

\section{\label{Appe:sine-Gordon} Semiclassical derivation of the sine-Gordon Hamiltonian}
\label{app:field-theory}

The classical continuum limit of $\hat{H}_S$ [\eqref{Eq:Ham}] is obtained by parametrizing the spin at site $j$ as ${\bf S}_j = S (\sin\theta_j \cos\varphi_j,  \sin\theta_j \sin\varphi_j, \cos\theta_j)$ and replacing lattice differences with spatial derivatives to obtain 
\begin{eqnarray}
\label{eq:H}
H = &&\int dx J S^2 a \Big(\frac{1}{2} (\partial_x \theta)^2 + \frac{1}{2}\sin^2\theta (\partial_x \varphi)^2 + q_0 \sin^2\theta \partial_x \varphi \nonumber\\
&&- m^2 \sin\theta \cos\varphi \Big),
\end{eqnarray}
where $a$ is the lattice spacing, $x = ja$ is the continuous coordinate, $q_0 = D/(J a)$, and $m^2 = H/(J S a^2)$. It proves convenient to remove the linear derivative term by defining $\tilde\varphi = \varphi - q_0 x$ so that,
\begin{eqnarray}
\label{eq:H2}
H = &&\int dx J S^2 a \Big(\frac{1}{2} (\partial_x \theta)^2 + \frac{1}{2}\sin^2\theta (\partial_x \tilde\varphi)^2 - \frac{1}{2} q_0^2 \sin^2\theta  \nonumber\\
&&- m^2 \sin\theta \cos(\tilde{\varphi}+q_0 x)\Big).
\end{eqnarray}
Time dynamics of the continuous fields $\theta, \tilde\varphi$ follow from the Landau-Lifshitz equations. An elegant way to incorporate them is via Euler-Lagrange equations of the associated Lagrangian \cite{Kosevich1990,Kim2023},
\begin{equation}
L = \int dx \Big(\frac{1}{g} (\cos\theta -1) \dot{\tilde{\varphi}} - {\cal H}\Big),
\label{eq:L}
\end{equation}
where ${\cal H}$ is the Hamiltonian density (the integrand of \eqref{eq:H2}) and the dot in the first Berry-phase term denotes the time derivative $\partial_t$. Fields $\theta$ and $\tilde\varphi$ depend on $x$ and $t$, and we abbreviate $1/g = S/a$.

An efficient way to obtain the result is to set $\theta=\pi/2 + \tilde\theta$ in \eqref{eq:L} and expand the Lagrangian up to quadratic terms in $\tilde\theta$, 
\begin{eqnarray}
L = &&\int dx \Big(-\frac{1}{g}\dot{\tilde{\varphi}} -\frac{1}{g}\tilde{\theta}\dot{\tilde{\varphi}} - J S^2 a\Big[ \frac{1}{2} q_0^2 \tilde\theta^2 + \frac{1}{2} (\partial_x \tilde\theta)^2 + \nonumber\\
&& + \frac{1}{2} (\partial_x \tilde\varphi)^2 - m^2 \cos(\tilde\varphi + q_0 x)\Big] \Big).
\label{eq:L2}
\end{eqnarray}
For small $|\partial_x \tilde\theta| \ll q_0 \tilde\theta$, the second term in square brackets can be dropped as well \cite{Mikeska1978}. 

The Lagrangian form in configuration space,  $L = L(\tilde\varphi, \dot{\tilde\varphi})$, is obtained by integrating out $\tilde\theta$.
Since the Lagrangian is quadratic in $\tilde{\theta}$, this approach is equivalent to replacing $\tilde{\theta}$ with the solution of the equation of motion:
\begin{equation}
\frac{\delta L}{\delta\tilde\theta} = -\frac{1}{g} \dot{\tilde{\varphi}}-J S^{2} a q_{0}^{2} \tilde{\theta}=0,
\end{equation}
which gives 
\begin{equation}
\tilde{\theta}(x)=-\frac{1}{g J S^{2} a} \frac{1}{q_{0}^{2}} \dot{\tilde{\varphi}}(x).
\end{equation}
The resulting Lagrangian form in configuration space is
\begin{eqnarray}
L = &&\int dx \Big( J S^2 a \Big[\frac{1}{2c^2}(\partial_t \tilde\varphi)^2  - \frac{1}{2}(\partial_x \tilde\varphi)^2 + m^2 \cos[\tilde\varphi + q_0 x]\Big]\nonumber\\
&&- \frac{1}{g} \partial_t \tilde\varphi\Big)
\label{eq:L3}.
\end{eqnarray}
with $c = g q_0 J S^2 a = g D S^2$.

Going back to phase space, the sine-Gordon Hamiltonian can now be expressed in terms of the field $\varphi$ and the conjugate momentum $\tilde\Pi = (J S^2 a/c^2) \partial_t\tilde{\varphi} - 1/g$:
\begin{equation}
\tilde{H}_{\rm sG}=\int dx \;\;\mathcal{H}_{\rm sG},
\end{equation}
where
\begin{eqnarray}
\mathcal{H}_{\rm sG}&=&\frac{c K}{2}(\tilde\Pi + \frac{1}{g})^2  + \frac{c}{2 K} (\partial_x \tilde\varphi)^2 \nonumber\\
&&- \frac{H S}{a} \cos[\tilde\varphi + q_0 x].
\label{eq:H3}
\end{eqnarray}
Here, $K = g q_0$, and we simplified the coefficient of the cosine term using the definition of $m^2$. Observe the persistence of a slightly unusual constant term $1/g$ that originates from the linear time-derivative term in \eqref{eq:L}.

Next, we quantize and rescale $\hat{\tilde\Pi} = \hat{\Pi}/\sqrt{4\pi}$, $\hat{\tilde\varphi} = \sqrt{4\pi} \hat{\varphi}$, and $K = 4\pi \kappa$, so that the commutation relation $[\hat{\tilde\varphi}(x), \hat{\tilde\Pi}(x')] = [\hat{\varphi}(x), \hat{\Pi}(x')] = i \delta(x-x')$ is preserved. In addition, we undo the position-dependent shift of the field $\hat{\varphi}$, $\hat{\varphi}(x) \to \hat{\varphi}(x) - q_0 x/\sqrt{4\pi}$, to obtain
\begin{equation}
\begin{split}
\hat{\mathcal{H}}_{\rm sG}&=\frac{c \kappa}{2}(\hat{\Pi} + \frac{\sqrt{4\pi}}{g})^2  + \frac{c}{2 \kappa} (\partial_x\hat{\varphi})^2 \\
&- \frac{c q_0}{\sqrt{4\pi} \kappa} \partial_x \hat{\varphi} 
- \frac{H S}{a} \cos[\sqrt{4\pi}\hat{\varphi}], 
\label{eq:H4}
\end{split}
\end{equation}
We observe that momentum $\hat{\Pi}$ appears in a sum
\begin{equation}
    \hat{\Pi} + \frac{\sqrt{4\pi}}{g} = \hat{\Pi} + \Pi_0,
    \label{eq:PiP}
\end{equation}
where the ``offset" momentum is $\Pi_0 = \frac{\sqrt{4\pi}S}{a}$.

A remarkable discovery by Coleman \cite{Coleman1975} and Mandelstam \cite{Mandelstam1975}, and, independently, by Luther and Emery \cite{Luther1974}, is that \eqref{eq:H4} can be exactly mapped to the Thirring model of massive Dirac fermions. 
The mapping is based on the observation that the exponential operator
\begin{equation}
    \hat{O}(x) = e^{i \zeta \int_{-\infty}^x dx' \hat{\Pi}(x')}
\end{equation}
creates a soliton centered at $x$. This result is proven with the help of the identity $[\hat{A}, e^{\hat{B}}] = \hat{C} e^{\hat{B}}$ that holds for any two operators $\hat{A}, \hat{B}$ such that their commutator $\hat{C}=[\hat{A},\hat{B}]$ commutes with both of them, $[\hat{C},\hat{A}]=[\hat{C},\hat{B}]=0$. 

The identity
\begin{equation}
[\hat{\varphi}(y),\hat{O}(x)] = i \frac{\delta \hat{O}(x) }{\delta \hat{\Pi}(y)}= - \zeta \Theta(x-y) \hat{O}(x)
\label{eq:com1}
\end{equation}
means that the action of $\hat{O}(x)$ on an eigenstate $|\varphi\rangle$ of the operator $\hat{\varphi}$ changes it to the state $|\varphi'\rangle$ with the eigenvalue $\varphi(y) - \zeta \Theta(x-y)$, i.e., 
\begin{equation}
\varphi(y)\;\longmapsto\;
\begin{cases}
\varphi(y)-\zeta & y<x\\[4pt]
\varphi(y) & y>x.
\end{cases}
\end{equation}
In other words, $\hat{O}(x)$ creates a kink of height $\zeta$ at position $x$ in the field $\hat{\varphi}(y)$. To describe solitons in \eqref{eq:H4}, we need to take $\zeta = 2\pi/\sqrt{4\pi} = \sqrt{\pi}$.

From here on, we make use of the powerful bosonization technique, which is described in vast detail in numerous reviews and textbooks. We chose to follow \cite{Shankar2017} and introduce right- and left-moving fermions, describing propagating solitons,  via
\begin{equation}
\begin{split}
\hat{\psi}_{+}(x) &= \frac{1}{\sqrt{2\pi\alpha}} e^{i \sqrt{4\pi} \hat{\Phi}_{+}(x)}, \\
\hat{\psi}_{-}(x) &= \frac{i}{\sqrt{2\pi\alpha}} e^{-i \sqrt{4\pi} \hat{\Phi}_{-}(x)},
\end{split}
\label{eq:fermion}
\end{equation}
where the additional factor of $i$ in $\hat{\psi}_{-}(x)$ is for future convenience. The chiral bosons are defined by
\begin{equation}
\hat{\Phi}_\pm(x) = \frac{1}{2} \left[ \hat{\varphi}(x) \mp \int_{-\infty}^x dx' \hat{\Pi}(x') \right]
\label{eq:chiralboson}
\end{equation}
Equations \eqref{eq:fermion} and \eqref{eq:chiralboson} show that $e^{-i \sqrt{\pi} \int^x dx' \hat{\Pi}(x')}$ is exactly the soliton operator, while the factors  $e^{\pm i \sqrt{\pi}\hat{\varphi}(x)}$ are needed to enforce the Fermi statistics of the $\hat{\psi}_\pm$ fields \cite{Mandelstam1975}.

Given that, in our formulation, momentum $\hat{\Pi}$ always appears together with ${\Pi}_0$, \eqref{eq:PiP}, we can now make the shift $\hat{\Pi} \to \hat{\Pi} - \hat{\Pi}_0$ to remove the offset. This shift changes \eqref{eq:chiralboson} by $\hat{\Phi}_\pm \to \hat{\Phi}_\pm \mp \sqrt{\pi} S (x-L)/a$, where $L \to -\infty$ is the (constant) contribution from the lower integration limit. Correspondingly, fermion operators $\hat{\psi}_\pm$ in \eqref{eq:fermion} change as 
\begin{equation}
    \hat{\psi}_\pm(x) \to \hat{\psi}_\pm(x) e^{-i \tfrac{2\pi S}{a} \, x}
    \label{eq:shift}
\end{equation}
Since $L/a$ is an integer, the phase factor $e^{2\pi S L/a} = \pm 1$ can always be removed by a unitary transformation. Equation \eqref{eq:shift} signifies the shift of the soliton momentum by $p_0 = 2\pi S/a$ relative to the origin. For the {\em integer} spin $S$, the shift is immaterial since $p_0$ is then a multiple of the reciprocal lattice momentum. However, in the case of {\em half-integer} spin $S$, the shift $p_0 = \pi/a$ actually describes the soliton that carries the Berry momentum $p_0 = \pi/a$. 

We therefore conclude that chiral solitons are distinguished by a nontrivial quantum number, namely, the momentum $p_0$. In the integer-$S$ chain, the soliton momentum is zero, $p_0 = 0$, while in the half-integer-$S$ chain, it is given by $p_0 = \pi/a$, the Brillouin zone boundary value. This effect is topological; it originates from the Berry phase in \eqref{eq:L}. This unusual finding agrees with the earlier semiclassical studies, Refs. \cite{Loss1996,Kato2023}.

Subsequent manipulations involve only the {\em slow} $\hat{\psi}_\pm(x)$ fields in the right-hand side of \eqref{eq:shift}. In particular, in \eqref{eq:bosonization} below, the {\em fast} oscillating factors $e^{\pm i p_0 x}$ are not present at all, which is in complete similarity to the standard bosonization procedure when the fast-oscillating exponentials $e^{\pm i k_F x}$ with Fermi momenta $\pm k_F$ are separated from the slow field $\hat{\psi}_\pm(x)$ describing right- and left-moving fermions that are bosonized according to the standard rules \cite{Giamarchi2004,Shankar2017}. 

Lengthy, but straightforward calculations \cite{Shankar2017} lead to the following set of bosonization identities,
\begin{align}
\hat{\psi}^\dagger_{+}(x) \hat{\psi}_{-}(x) + {\rm h.c.} = \frac{1}{\pi \alpha} \cos[\sqrt{4\pi}\hat{\varphi}(x)], \nonumber\\
:\hat{\psi}^\dagger_{+}(x) \hat{\psi}_{+}(x): = \frac{1}{\sqrt{\pi}} \partial_x \hat{\Phi}_{+}(x), \nonumber\\
:\hat{\psi}^\dagger_{-}(x) \hat{\psi}_{-}(x): = \frac{1}{\sqrt{\pi}} \partial_x \hat{\Phi}_{-}(x), \nonumber\\
:\hat{\psi}^\dagger_{+}(x) (-i\partial_x) \hat{\psi}_{+}(x): = (\partial_x \hat{\Phi}_{+}(x))^2,\nonumber\\
:\hat{\psi}^\dagger_{-}(x) (i\partial_x) \hat{\psi}_{-}(x): = (\partial_x \hat{\Phi}_{-}(x))^2.
\label{eq:bosonization}
\end{align}
Colons signify normal ordering. These identities allow us to rewrite \eqref{eq:H4} as the massive Thirring model of interacting fermions $\psi_\pm$,
\begin{eqnarray}
    \hat{\tilde{\mathcal{H}}}_{\rm mT} &=&\frac{c}{2}(\kappa + \frac{1}{\kappa}) \Big[\hat{\psi}^\dagger_{+}(x) (-i\partial_x) \hat{\psi}_{+}(x) + \hat{\psi}^\dagger_{-}(x) (i \partial_x)\hat{\psi}_{-}(x) \Big] \nonumber\\
   &+&  \pi c (\kappa - \frac{1}{\kappa}) \hat{\psi}^\dagger_{+}(x) \hat{\psi}_{+}(x) \hat{\psi}^\dagger_{-}(x) \hat{\psi}_{-}(x) \nonumber\\
  &-&  \frac{c q_0}{2\kappa} [\hat{\psi}^\dagger_{+}(x) \hat{\psi}_{+}(x) + \hat{\psi}^\dagger_{-}(x) \hat{\psi}_{-}(x)] \nonumber\\
  &-&   \lambda [\hat{\psi}^\dagger_{+}(x) \hat{\psi}_{-}(x) + \hat{\psi}^\dagger_{-}(x) \hat{\psi}_{+}(x)]
  \label{eq:thirring}
\end{eqnarray}
and $\lambda = \pi H S \alpha/a$. The first line represents the kinetic energy of Dirac fermions, while the second describes their interaction; notice that it vanishes for the special value of $\kappa = 1$. The third line shows that incommensuration $q_0$ plays the role of the fermion chemical potential, while the nonlinear cosine potential now describes backscattering of $\hat{\psi}_\pm$ fermions (the last line). In the field theory language, $\lambda$ is the mass term, responsible for the gap in the fermion spectrum.

It is worth noting once again that \eqref{eq:thirring} is written in terms of the slow $\hat{\psi}_\pm(x)$ fields that do not contain $e^{\pm i p_0 x}$ factors.

The essential physics of \eqref{eq:thirring} is understood by analyzing the noninteracting, $\kappa=1$, (Luther-Emery) point \cite{Giamarchi2004}. Here, $\tilde{\mathcal{H}}_{\rm mT}(\kappa=1)$ is diagonalized by
\begin{equation}
    \begin{pmatrix} 
    \hat{\psi}_{+}(k) \\
    \hat{\psi}_{-}(k)
    \end{pmatrix} = 
    \begin{pmatrix} 
    \alpha_k & \beta_k \\
    -\beta_k & \alpha_k
    \end{pmatrix} 
    \begin{pmatrix} 
    \hat{u}_k \\
    \hat{d}_k
    \end{pmatrix}
    \label{eq:diag1}
\end{equation}
where 
\begin{align}
    \alpha_k = \frac{1}{\sqrt{2}} \Big(1 + \frac{c |k|}{\sqrt{c^2k^2 + \lambda^2}}\Big)^{1/2}, \nonumber\\
    \beta_k = \frac{1}{\sqrt{2}} \Big(1 - \frac{c |k|}{\sqrt{c^2k^2 + \lambda^2}}\Big)^{1/2}.
    \label{eq:diag2}
\end{align}
Note again that, in accordance with our discussion above, $k$ is measured from $p_0$.
The $\hat{u}_k (\hat{d}_k)$ fermions describe upper (lower) bands with dispersion $\pm \epsilon_k - c q_0/2$, $\epsilon_k = \sqrt{\lambda^2 + c^2 k^2}$, representing solitons and antisolitons. As long as the chemical potential $c q_0/2$ is inside the band gap, $q_0 < q_c = 2\lambda/c$, the lower band is completely full and the upper one is completely empty. 
The density of solitons and antisolitons is zero, which is the commensurate vacuum, or high-field, state with $\langle \partial_x \hat{\varphi} \rangle  = 0$.

To describe soliton and antisoliton excitations on equal footing, it is convenient to make a particle-hole transformation for the occupied states: $\hat{d}_k \to \hat{h}_{-k}^\dagger$. The transformed Hamiltonian reads,
\begin{align}
    \hat{H}_{\rm mT}(\kappa=1) = \sum_k \sqrt{c^2 k^2 + \lambda^2} (\hat{u}_k^\dagger \hat{u}_k + \hat{h}_k^\dagger \hat{h}_k) \nonumber\\
    - \frac{c q_0}{2} (\hat{u}_k^\dagger \hat{u}_k - \hat{h}_k^\dagger \hat{h}_k)
    \label{eq:Hph}
\end{align}
Now, $\hat{u}_k (\hat{h}_k)$ describes excitations, which are solitons (antisolitons).

Once the DM interaction becomes strong enough, or the magnetic field is reduced below the critical value,  $|q_0| > q_c$, solitons enter the system. Positive $q_0$ induces solitons, while $q_0 < 0$ brings in antisolitons. The band structure in \eqref{eq:Hph} is illustrated in Fig.~\ref{Fig:field_bands}(a).

Let us choose $q_0 > 0$ for concreteness. 
The fermion density $\rho = k_F/\pi$ determines the Fermi momentum $k_F = \sqrt{q_0^2 - q_c^2}/2$. According to \eqref{eq:bosonization}, the fermion density $\hat{\rho} = \langle \hat{\psi}^\dagger_{+}(x) \hat{\psi}_{+}(x) + \hat{\psi}^\dagger_{-}(x) \hat{\psi}_{-}(x) \rangle$ is directly proportional to $\partial_x \hat{\varphi}$, $\langle \partial_x \hat{\varphi} \rangle  = \sqrt{\pi} \hat{\rho}$. Thus, $\langle \partial_x \hat{\varphi} \rangle  = \sqrt{(q_0^2 - q_c^2)/(4\pi)} \sim \sqrt{q_0 - q_c}$. The state with the finite soliton density is the critical Luttinger liquid state of soliton liquid. Classically, this is the soliton lattice state described in Sec. II. Qualitative features of the described commensurate-incommensurate transition from the vacuum to the Luttinger liquid state persist for all values of $\kappa$ \cite{Giamarchi2004}.

\section{\label{Appe:analitical_states} Exact eigenstates}

Let us show that the fully polarized state along the $\hat{x}$ direction is an eigenstate of the Hamiltonian in~\eqref{Eq:Ham}. The proof for the Heisenberg term is straightforward, as this term is SU(2) invariant; therefore, any fully polarized state is an eigenstate. Similarly, the Zeeman term is trivially satisfied.

We now focus on the Dzyaloshinskii-Moriya (DM) part of the Hamiltonian, which, when rewritten in terms of ladder operators, takes the form:

\begin{equation}
\hat{H}_{\rm DM} = -\frac{D}{2i} \sum_{l=1}^L \hat{h}_{l,l+1},
\end{equation}
with,
\begin{equation}
\hat{h}_{j,j+1} = \hat{S}^-_{j}\hat{S}^+_{j+1} - \hat{S}^+_{j}\hat{S}^-_{j+1}.
\end{equation}

The fully polarized state is given by:
$\ket{\Rightarrow} = \frac{1}{2^{L/2}} \bigotimes_{l=1}^L \Big[ \ket{\uparrow}_l + \ket{\downarrow}_l \Big]$, and the local action of $\hat{h}_{l,l+1}$ on this state is:
\begin{equation}
\hat{h}_{l,l+1} \Big[\ket{\uparrow}_l+\ket{\downarrow}_l\Big] \otimes \Big[\ket{\uparrow}_{l+1}+\ket{\downarrow}_{l+1}\Big]   
= \Big[\ket{\downarrow_j\uparrow_{l+1}} - \ket{\uparrow_l\downarrow_{l+1}} \Big] 
\end{equation}
which implies that
\begin{equation}
    \begin{split}
        \hat{H}_{\rm DM} \ket{\Rightarrow} &\propto \sum_{l=1}^L \ket{\dots, \Big[{\downarrow_l\uparrow_{l+1}} - {\uparrow_l\downarrow_{l+1}} \Big], \dots } \\ 
        &\propto \sum_{l=1}^L \ket{\dots, {\downarrow_l\uparrow_{l+1}} , \dots } - \ket{\dots, {\uparrow_l\downarrow_{l+1}} , \dots } 
    \end{split}
\end{equation}

Now, observe that the second term of the pair for index $l$ is $
-\ket{\dots \uparrow_l \downarrow_{l+1} \dots},
$ which is also exactly the same as the first term of the pair with index $l-1$. In other words, every basis configuration of the form $\ket{
\dots \uparrow_l \downarrow_{l+1} \dots}
$ appears twice in the sum: Once with a $(+)$ sign coming from $\hat{h}_{l,l+1}$, and once with a $(-)$ sign coming from $\hat{h}_{l-1,l}$. Therefore, the contributions cancel pairwise, and the complete sum vanishes, indicating that $\ket{\Rightarrow}$ is an eigenstate of the DM part of the Hamiltonian with a null eigenvalue. Finally, we have 

\begin{eqnarray}
    \frac{1}{L}\hat{H}_S \ket{\Rightarrow} = -\Big(\frac{J}{4} +\frac{H}{2}\Big) \ket{\Rightarrow}.
\end{eqnarray}

Now we can also show that $\hat{S}^z(k=0)\ket{\Rightarrow}$ is an eigenstate of $\hat{H}_S$. Indeed,  
\begin{equation}
    \hat{S}^z(k=0) = \frac{1}{\sqrt{L}} \sum_l \hat{S}^z_l = \frac{\hat{S}^z_{\text{tot}}}{\sqrt{L}} ,   
\end{equation}
which is proportional to the total $z$ component of the magnetization. From the SU(2) invariance of the Heisenberg term, we have:
\[
[\hat{H}_{J}, \hat{S}^z_{\text{tot}}] = 0 ,
\]
and by the U(1) invariance around the $z$ axis of the DM term,
\[
[\hat{H}_{\rm DM}, \hat{S}^z_{\text{tot}}] = 0 .
\]
Thus, $\hat{S}^z(k=0)$ acting on any eigenstate of $\hat{H}_J + \hat{H}_{\rm DM}$ also yields an eigenstate.  

For the Zeeman contribution,
\[
-H\sum_j \hat{S}^x_j = -H\hat{S}^x_{\text{tot}} ,
\]
the commutator gives
\begin{equation}
    \bigl[ \sqrt{L}\,\hat{S}^z(k=0), -H\hat{S}^x_{\text{tot}} \bigr]
    = -iH\sqrt{L}\,\hat{S}^y_{\text{tot}}.
\end{equation}
Therefore,
\begin{equation}
    \hat{S}^x_{\text{tot}} \,\hat{S}^z(k=0)\ket{\Rightarrow}
    = \Bigl(\hat{S}^z(k=0)\hat{S}^x_{\text{tot}} - i\hat{S}^y_{\text{tot}}\Bigr)\ket{\Rightarrow}.
    \label{Eq:eigen_Sx}
\end{equation}
From the first term on the rhs, we find
\[
\hat{S}^z(k=0)\hat{S}^x_{\text{tot}} \ket{\Rightarrow}
= \tfrac{L}{2}\,\hat{S}^z(k=0)\ket{\Rightarrow}.
\]
To compute the action of $\hat{S}^y_{\text{tot}}$, note that
\begin{equation}
\begin{split}
    i\hat{S}^y_l \,\frac{1}{\sqrt{2}}\bigl(\ket{\uparrow}_l + \ket{\downarrow}_l\bigr)
    &= \frac{1}{2\sqrt{2}}\bigl(\ket{\uparrow}_l - \ket{\downarrow}_l\bigr) \\ 
    &= \hat{S}^z_l \,\frac{1}{\sqrt{2}}\bigl(\ket{\uparrow}_l + \ket{\downarrow}_l\bigr) .
\end{split}
\end{equation}
By substituting this into~\eqref{Eq:eigen_Sx}, we obtain
\begin{equation}
    \hat{S}^x_{\text{tot}} \,\hat{S}^z(k=0)\ket{\Rightarrow}
    = \Bigl(\tfrac{L}{2}-1\Bigr)\hat{S}^z(k=0)\ket{\Rightarrow},
\end{equation}
which finally proves that $\hat{S}^z(k=0)\ket{\Rightarrow}$ is an eigenvector of $\hat{H}_S$, 
\begin{equation}
    \hat{H}_S \hat{S}^z(k=0) \ket{\Rightarrow} = -\Big(\frac{JL}{4} +H(\frac{L}{2}-1)\Big) \hat{S}^z(k=0)\ket{\Rightarrow}.
\end{equation}
The energy difference between the fully polarized state and $\hat{S}^z(k=0)\ket{\Rightarrow}$ is simply given by $\Delta E = H$, that is, merely the magnon gap in the fullypolarized phase.

\vspace{0.5cm}

\section{\label{Appe:su2algebra}\texorpdfstring{$\mathfrak{su}(2)$}{su(2)} algebra of soliton operators}

\begin{figure}[!ht]
    \centering
    \includegraphics[width=\columnwidth]{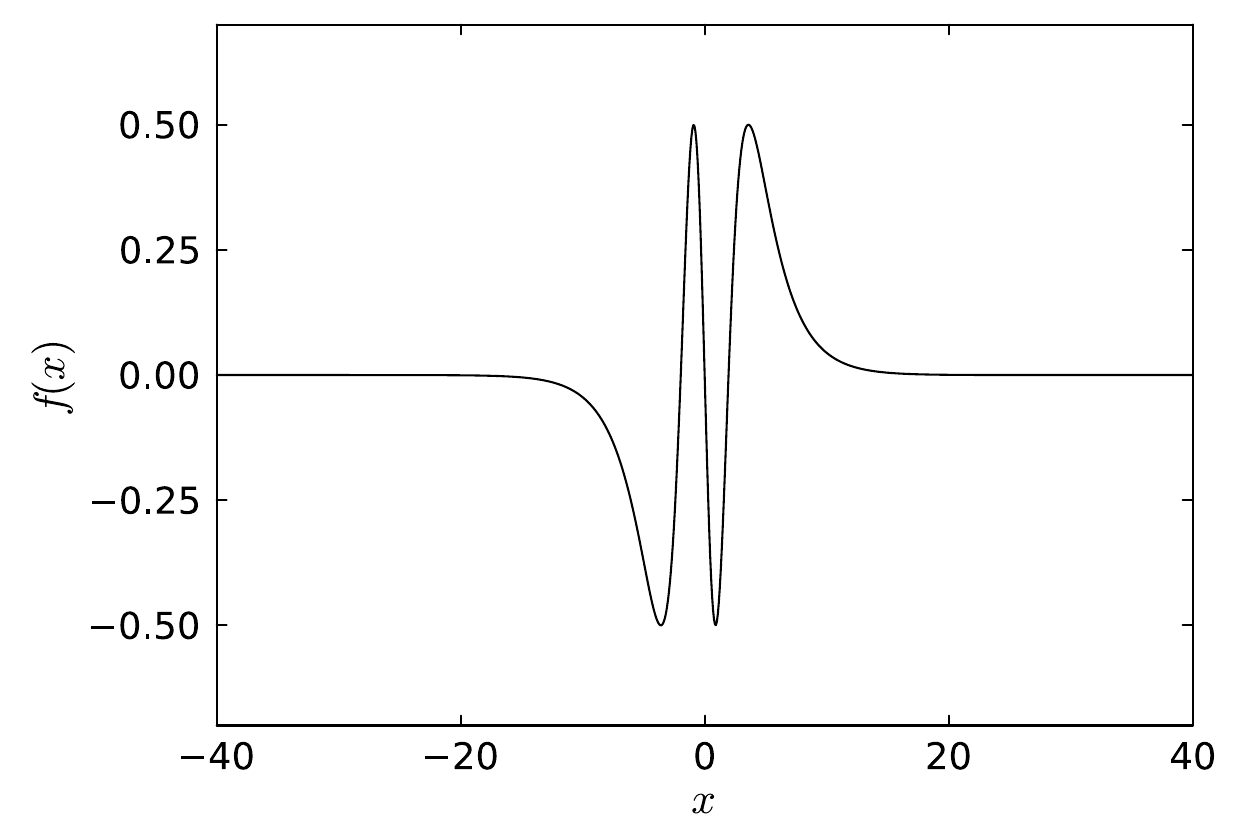}
    \caption{We show \( f(x) = \cos{\varphi(x)}\sin{\varphi(-x)} \) as a function of $x=|l-j|$ for $\tau=1$.}
    \label{Fig:fxcom}
\end{figure}

This appendix presents the explicit calculations of the commutators involving
the soliton and quantum–soliton operators. We begin with the former, evaluating
\begin{widetext}
\begin{eqnarray}
    \hat{T}^{\;}_{j\tau}\hat{T}^{\dagger}_{j\tau} &=& \prod_{l\neq j}  e^{i\varphi_{\tau}(l-j) \hat{S}^z_l} (\hat{S}^z_j - i \hat{S}_j^y)(\hat{S}^z_j + i \hat{S}_j^y) 
\prod_{l\neq j}  e^{-i\varphi_{\tau}(l-j) \hat{S}^z_l} = \frac{1}{2} +i[\hat{S}^z_j,\hat{S}^y_j] =\frac{1}{2} + \hat{S}^x_j, 
\nonumber \\ 
\hat{T}^{\dagger}_{j\tau}\hat{T}^{\;}_{j\tau} &=&  (\hat{S}^z_j + i \hat{S}_j^y) \prod_{l\neq j}  e^{-i\varphi_{\tau}(l-j) \hat{S}^z_l}\prod_{l\neq j}  e^{i\varphi_{\tau}(l-j) \hat{S}^z_l} (\hat{S}^z_j - i \hat{S}_j^y) = \frac{1}{2} +i[\hat{S}^y_j,\hat{S}^z_j] =\frac{1}{2} - \hat{S}^x_j, 
\end{eqnarray} 
which implies that the equal-site commutator is $\hat{T}^{\;}_{j\tau}\hat{T}^{\dagger}_{j\tau} - \hat{T}^{\dagger}_{j\tau}\hat{T}^{\;}_{j\tau} = 2\hat{S}^x_j$, while the  anticommutator is $\{\hat{T}^{\;}_{j\tau},\hat{T}^{\dagger}_{j\tau}\}=1$.

For different sites $j' \ne j$, we have
\begin{eqnarray}
[ \hat{T}^{\;}_{j'\tau}, \hat{T}^{\dagger}_{j\tau}] &=& (\hat{S}^z_j+i\hat{S}^y_j)(\hat{S}^z_{j'}-i\hat{S}^y_{j'}) \sum_{\sigma=\pm 1} \sigma e^{\sigma i\varphi_{\tau}(j'-j)\hat{S}^z_{j'}} e^{\sigma i\varphi_{\tau}(j-j')\hat{S}^z_{j}} 
\nonumber \\
&=& 2i(\hat{S}^z_{j}+i\hat{S}^y_j)(\hat{S}^z_{j'}-i\hat{S}^y_{j'}) \left[f_{\tau}(j'-j) \hat{S}^z_j + f_{\tau}(j-j') \hat{S}^{z}_{j'}\right],
\nonumber \\
\end{eqnarray}
\end{widetext}
where the function \( f_{\tau}(x) = \cos{\varphi_{\tau}(x)}\sin{\varphi_{\tau}(-x)} \) vanishes as \( x \) becomes much larger than the size of the chiral soliton [see Fig.~\ref{Fig:fxcom}]. Therefore, we can conclude that for all $j$ and $j'$,
\begin{equation}
    [ \hat{T}^{\;}_{j' \tau}, \hat{T}^{\dagger}_{j\tau}] = 2\Tilde{\delta}_{j'j\tau} \hat{S}^x_j
\end{equation}
with $\tilde{\delta}_{j'j\tau}$ denoting the coarse-grained version of the
Kronecker delta.  It is also straightforward to show that
\begin{equation}
    \begin{split}
        [\hat{T}^{\dagger}_{j\tau},\hat{S}^x_{j}] &= 2\hat{T}^{\dagger}_{j\tau} \\ 
        [\hat{T}^{}_{j'\tau},\hat{S}^x_j] &= -2\hat{T}^{}_{j'\tau}
    \end{split}
\end{equation}
Based on this result, one can construct an approximate $\mathfrak{su}(2)$ Lie algebra generated by $\{\hat{T}_{\tau}^{\dagger}, \hat{T}_{\tau}, \hat{S}^x\}$.

\section{\label{Appe:soliton_ov} Localization of the quantum soliton operator}

We analyze the large-distance behavior of the expectation value $\langle 0|\hat{\mathfrak T}^{}_{j\tau}\hat T_{j'\tau}^\dagger|0\rangle$. 
Using the definition:
\begin{equation}
\hat{\mathfrak T}_{j\tau}^\dagger=\sum_p w_{j-p,\tau}\,\hat T_{p\tau}^\dagger, \ \
\hat{\mathfrak T}_{j\tau}=\sum_p (w_{j-p,\tau})^*\,\hat T_{p\tau},
\end{equation}
the expectation value can be expressed as
\begin{equation}
\langle 0|\hat{\mathfrak T}^{}_{j\tau}\hat T_{j'\tau}^\dagger|0\rangle
= \sum_p (w_{j-p,\tau})^*O_{p,j'},
\end{equation}
where 
\begin{equation}
O_{p,j'}=\langle 0|\hat T^{}_{p\tau}\hat T^\dagger_{j'\tau}|0\rangle
\end{equation}
is given by \eqref{Eq:soliton_ov}. It is convenient to reexpress the overlaps as
\begin{equation}
O_{p,j'}= \sin^2\!\frac{\varphi_\tau(j'-p)}{2}
\prod_{l\neq p,j'}
\cos\Big[\frac{\varphi_\tau(l-p)-\varphi_\tau(l-j')}{2}\Big],
\label{eq:Fmn}
\end{equation}
which follows from the relations
$\varphi_\tau(0)=\pi$ and $\varphi_\tau(-x)=2\pi-\varphi_\tau(x)$.
By applying periodic boundary conditions, $x\mapsto {\rm mod}(x+L/2,L)-L/2$, both special factors
($l=p$ and $l=j'$) contribute positively because \(\cos[(\pi-\varphi_\tau(j'-p))/2]=\sin[\varphi_\tau(j'-p)/2]\).

For large separations $x = |j-j'|$ the soliton profile behaves as
\begin{equation}
\varphi(x) = 2\pi - 4 e^{-mx} + {\cal O}(e^{-3mx}),
\end{equation}
and therefore
\begin{equation}
\sin\!\frac{\varphi(x)}{2}
= \sin\!\bigl(\pi - 2 e^{-mx}\bigr)
\simeq 2 e^{-mx},
\label{eq:sin_decay}
\end{equation}
which exhibits the expected exponential decay controlled by the mass scale $m$.

The product of cosines in \eqref{eq:Fmn} remains of order unity,
for sites $l$ far from both centers $p$ and $j'$, 
$\varphi(l-p)\approx\varphi(l-j')$ and 
$\cos[(\varphi(l-p)-\varphi(l-j'))/2]\approx 1$.
The only deviations from unity arise from regions of width $\sim 1/m$ near the kink centers, producing a finite, ${\cal O}(1)$ prefactor $C(m)$ that does not depend on the distance $|l-p|$.

Combining these results, we obtain, for $x\gg 1$,
\begin{equation}
O_{p,j'} 
\simeq \pm 4C(m)e^{-2m|j'-p|}.
\label{eq:Fmn_asymptotic}
\end{equation}
Hence, $O_{p,j'}$ decays exponentially with rate $2m$. Substituting \eqref{eq:Fmn_asymptotic} into the definition of the full overlap yields
\begin{equation}
\langle 0|\hat{\mathfrak T}_{j\tau}\hat T_{j'\tau}^\dagger|0\rangle
\simeq
w_{j-j',\tau}^*
+ 4C(m)
\sum_{p\neq j'} 
w_{j-p,\tau}^*\,
e^{-2m|j'-p|}.
\end{equation}
If the Wannier coefficients $w_{j-p,\tau}$ are localized around $p=j$, the sum is dominated by $p\simeq j$, leading to the asymptotic behavior
\begin{equation}
\bigl|\langle 0|\hat{\mathfrak T}_{j\tau}\hat T_{j'\tau}^\dagger|0\rangle\bigr|
\;\propto\;
e^{-2m|j-j'|}.
\end{equation}
In other words, the overlap between $\hat{\mathfrak T}^{\dagger}_{j\tau}\ket{0}$ and $\hat T_{j'\tau}^\dagger\ket{0}$ decays exponentially in the separation $|j-j'|$, with a correlation length $\xi = {1}/{2m}$.

\begin{widetext}
\section{\label{Appe:Matrix_elems} Matrix elements for hoppings amplitudes}

This appendix presents the calculations for the matrix elements $h_{\tau}(s)$ defined in the main text in Sec.~\ref{Sec:tb_ham}. 

The vacuum state corresponds to the fully polarized state along the $x$ direction, given by

\begin{equation}
    \ket{0} =  \bigotimes_{l=1}^L  \frac{1}{\sqrt{2}}\Big[\ket{\uparrow}_l + \ket{\downarrow}_l\Big]
\end{equation}
The action of the soliton (antisoliton) operator centered at site $j$ over the fullypolarized state is given by
\begin{equation}
    \ket{\varphi_{\tau}^{(j)}} \equiv \hat{T}_{j\tau}^{\dagger} \ket{0} =\bigotimes_{l=1}^L  \frac{1}{\sqrt{2}}\Big[e^{-i\varphi_{\tau}(l-j)/2}\ket{\uparrow}_l + e^{i\varphi_{\tau}(l-j)/2}\ket{\downarrow}_l\Big] \equiv \bigotimes_{l=1}^L \ket{\mathbf{n}_{l}[\varphi_{\tau}(l-j)]},
\label{Eq:soliton_coh_state}
\end{equation}
where the last term defines the local coherent state $\ket{\mathbf{n}_{l}[\varphi_{\tau}(l-j)]}$ at site $l$, associated with the soliton (antisoliton) centered at site $j$. 
Note that the $\mathbf{n}_{l}[\varphi_{\tau}(l-j)]$ unit vectors are perpendicular to the rotation $z$ axis and $\varphi_{\tau}(l-j)$ is the corresponding azimuthal angle.
The Hamiltonian of \eqref{Eq:Ham} can be decomposed into a sum of bilinear terms acting on pairs of neighboring sites $n$ and $n+1$, along with the Zeeman term that acts only on site $n$, 

\begin{equation}
    \hat{H}_S = \sum_{l} \hat{h}^{({\rm Bi})}_{l,l+1} + \sum_l \hat{h}^{(Z)}_l
\end{equation}

Now, the matrix element of the Hamiltonian in the basis of the single soliton can be decomposed as

\begin{equation}
\begin{split}
    \mel{\varphi_{\tau}^{(j')}}{H}{\varphi_{\tau}^{(j)}} 
    &=\sum_l \mel{\varphi_{\tau}^{({j'})}}{\hat{h}^{({\rm Bi})}_{l,l+1}}{\varphi_{{\tau}}^{({j})}}  
    + \mel{\varphi_{\tau}^{({j'})}}{\hat{h}^{(Z)}_{l}}{\varphi_{\tau}^{({j})}} \\ 
    &= \sum_l \bra{\mathbf{n}_{l+1}[{\varphi_\tau( l+1-j')}]} \bra{\mathbf{n}_{l}[{\varphi_\tau( l-j')}]} \hat{h}^{(\rm{Bi})}_{l,l+1} \ket{\mathbf{n}_{l}[{\varphi_\tau( l-j)}]} \ket{\mathbf{n}_{l+1}[{\varphi_\tau( l+1-j)}]} \prod_{p\neq l, l+1}{\cal O}^{j'j}_p   \\
    &+ \sum_l \mel{\mathbf{n}_{l}[{\varphi_\tau( l-j')}]}{\hat{h}^{(Z)}_{l}} {\mathbf{n}_{l}[{\varphi_\tau( l-j)}]} \prod_{p\neq l} {\cal O}^{j'j}_p  
\end{split}
\label{Eq:sanguche_ham}
\end{equation}
with 
\begin{equation} {\cal O}^{j'j}_p \equiv \langle \mathbf{n}_{p}[{\varphi_\tau( p-j')}]|\mathbf{n}_{p}[{\varphi_\tau( p-j)}]\rangle = \cos{\left[\frac{\varphi_{\tau}(p-j')-\varphi_{\tau}(p-j)}{2}\right]}.
\end{equation}
We define,
\begin{equation}
    F_{jj'}^{\tau}(x) \equiv \prod_{p\neq x} {\cal O}^{j'j}_p  
     = \prod_{p\neq x} \cos{\left[\frac{\varphi_{\tau}(p-j')-\varphi_{\tau}(p-j)}{2}\right]}.
    \label{Eq:overlaps}
\end{equation}

Now we compute the matrix elements of the spin operators between two coherent states. Although only the expression for $\hat S^{x}$ is needed, we include the remaining components for completeness:

\begin{equation}
    \begin{split}
        \mel{\mathbf{n}_l[\varphi_{\tau}(l-j')]}{\hat{S}^x_n}{\mathbf{n}_l[\varphi_{\tau}(l-j)]} &= \frac{1}{2} \cos{\left[\frac{\varphi_{\tau}(l-j')+\varphi_{\tau}(l-j)}{2
        }\right]}, \\ 
        \mel{\mathbf{n}_l[\varphi_{\tau}(l-j')]}{\hat{S}^y_n}{\mathbf{n}_l[\varphi_{\tau}(l-j)]} &= \frac{1}{2} \sin{\left[\frac{\varphi_{\tau}(l-j')+\varphi_{\tau}(l-j)}{2
        }\right]}, \\ 
        \mel{\mathbf{n}_l[\varphi_{\tau}(l-j')]}{\hat{S}^z_n}{\mathbf{n}_l[\varphi_{\tau}(l-j)]} &= \frac{i}{2} \sin{\left[\frac{\varphi_{\tau}(l-j')-\varphi_{\tau}(l-j)}{2
        }\right]}.
    \end{split}
\end{equation}

Using this result, the single-site matrix element for the Zeeman part of the Hamiltonian is given by,

\begin{equation}
            \mel{\mathbf{n}_l[\varphi_{\tau}(l-j')]}{\hat{S}^x_n}{\mathbf{n}_l[\varphi_{\tau}(l-j)]} = -\frac{H}{2} \cos{\left[\frac{\varphi_{\tau}(l-j')+\varphi_{\tau}(l-j)}{2}\right]}
\end{equation}

The next step is to compute the matrix elements of the bilinear part of the Hamiltonian, 
\begin{equation}
\bra{\mathbf{n}_{l+1}[{\varphi_\tau( l+1-j')}]} \bra{\mathbf{n}_{l}[{\varphi_\tau( l-j')}]} \hat{h}^{(\rm{Bi})}_{l,l+1} \ket{\mathbf{n}_{l}[{\varphi_\tau( l-j)}]} \ket{\mathbf{n}_{l+1}[{\varphi_\tau( l+1-j)}]}
\end{equation}
We note that the product state of two coherent states can be expressed as 
\begin{equation}
        \ket{\mathbf{n}_{l}[{\varphi_\tau( l-j)}]} \ket{\mathbf{n}_{l+1}[{\varphi_\tau( l+1-j)}]} = \frac{1}{2}
    \begin{pmatrix}
e^{-i\varphi_{\tau}(l-j)/2} e^{-i\varphi_{\tau}(l+1-j)/2}\\
e^{-i\varphi_{\tau}(l-j)/2} e^{i\varphi_{\tau}(l+1-j)/2} \\
e^{i\varphi_{\tau}(l-j)/2} e^{-i\varphi_{\tau}(l+1-j)/2} \\
e^{i\varphi_{\tau}(l-j)/2} e^{i\varphi_{\tau}(l+1-j)/2}
\end{pmatrix} ,
\end{equation}
and the matrix representation of the two-site Heisenberg and DM Hamiltonian is given by
\begin{equation}
    [\hat{\mathbf{S}}_l \cdot \hat{\mathbf{S}}_{l+1}] = 
    \begin{pmatrix}
1/4 & 0 & 0 & 0 \\
0 & -1/4 & 1/2 & 0 \\
0 & 1/2 & -1/4 & 0 \\
0 & 0 & 0 & 1/4
\end{pmatrix},  
\quad 
[\hat{S}^x_l\hat{S}^y_{l+1} - \hat{S}^y_l\hat{S}^x_{l+1}] = 
    \begin{pmatrix}
0 & 0 & 0 & 0 \\
0 & 0 & i/2 & 0 \\
0 & -i/2 & 0 & 0 \\
0 & 0 & 0 & 0
\end{pmatrix}.  
\end{equation}

By evaluating the matrix elements and substituting them into \eqref{Eq:sanguche_ham}, we obtain
\begin{equation}
    \begin{split}
         \mel{\varphi_{\tau}^{({j'})}}{\hat{H}_S}{\varphi_{\tau}^{({j})}} = &-\frac{J}{8}\sum_n   F^{\tau}_{j,j'}(l,l+1) \cos{\left[\frac{\varphi_{\tau}(l-j')+\varphi_{\tau}(l+1-j')-\varphi_{\tau}(l-j)-\varphi_{\tau}(l+1-j)}{2}\right]} \\ 
         & -\frac{J}{4}\sum_l   F^{\tau}_{j,j'}(l,l+1)\cos{\left[\frac{\varphi_{\tau}(l-j')+\varphi_{\tau}(l-j)-\varphi_{\tau}(l+1-j')-\varphi_{\tau}(l+1-j)}{2}\right]} \\ 
         & +\frac{J}{8}\sum_l  F^{\tau}_{j,j'}(l,l+1)\cos{\left[\frac{\varphi_{\tau}(l-j')-\varphi_{\tau}(l+1-j')-\varphi_{\tau}(l-j)+\varphi_{\tau}(l+1-j)}{2}\right]} \\ 
         & -\frac{D}{4}\sum_l   F^{\tau}_{j,j'}(l,l+1)\sin{\left[\frac{\varphi_{\tau}(l+1-j')-\varphi_{\tau}(l-j')-\varphi_{\tau}(l-j)+\varphi_{\tau}(l+1-j)}{2}\right]} \\ 
          & -\frac{H}{2} \sum_l F^{\tau}_{j,j'}(l)\cos{\left[\frac{\varphi_{\tau}(l-j')+\varphi_{\tau}(l-j)}{2}\right]}.
    \end{split}
    \label{Eq:matrix_elem}
\end{equation}
The hopping amplitudes can now be evaluated using \eqref{Eq:calc_hop} of the main text.
\end{widetext}

\section{\label{App:sign_hop} Berry-phase origin of the integer/half-integer hopping sign alternation}

This appendix shows that, within our lattice formulation, the minimum of the soliton band lies at $k=\pi$ for half-integer spin chains and at $k=0$ for integer ones, in agreement with the continuum formulation discussed in Sec.~\ref{Sec:conti}. 

As shown in Appendix \ref{Appe:Matrix_elems}, the soliton centered at site $j$ is represented as a product of ($S=1/2$) coherent spin states [see \eqref{Eq:soliton_coh_state}], a construction that extends straightforwardly to spin-$S$ coherent states. 

In general, the phase of the overlap between two spin-$S$ coherent states at a
given lattice site $l$ is given exactly by the oriented solid angle on the
Bloch sphere spanned by their corresponding spin directions and a chosen
quantization axis. Keeping the fixed unit vector $\hat{z}$, parallel to the rotation axis of the skyrmion, as the quantization
axis, let $\ket{\mathbf n_l}$ and $\ket{\mathbf n'_l}$ denote two coherent
states satisfying
\[
\langle \mathbf n_l | \hat{\boldsymbol S}_l | \mathbf n_l \rangle 
= S\,\mathbf n_l,
\qquad
\langle \mathbf n'_l | \hat{\boldsymbol S}_l | \mathbf n'_l \rangle 
= S\,\mathbf n'_l,
\]
where 
$\hat{\boldsymbol S}_l=(\hat S_l^x,\hat S_l^y,\hat S_l^z)$ 
is the vector of spin-$S$ operators at site $l$.  Note that this is just a generalization for arbitrary spin $S$ of the 
$S=1/2$ coherent states introduced in the previous sections.

The argument of the overlap between a pair of coherent states is 
\begin{equation}
\arg\!\big(\braket{\mathbf n_l}{\mathbf n'_l}\big)
~=~ S\,\Omega(\hat{z},\mathbf n_l,\mathbf n'_l)
\qquad (\mathrm{mod}\;2\pi),
\end{equation}
where $\Omega(\hat{z},\mathbf n_l,\mathbf n'_l)$ is the oriented solid angle of
the spherical triangle with vertices $\hat{z}$, $\mathbf n_l$, and 
$\mathbf n'_l$.  
For the classical solitons with arbitrary spin $S$, the vectors $\mathbf n_l$ and $\mathbf n'_l$ are still \emph{perpendicular} to the rotation $\hat{z}$ axis, implying that we can still parametrize them with the single azimuthal angle $\varphi_l$ because $\theta_l=\pi/2$.
In this case, the overlap between two coherent states with $\mathbf n_l$ and $\mathbf n'_l$  differing by a small azimuthal shift $\delta\varphi$ satisfies
\begin{equation}
\arg\!\big(\braket{\mathbf{n}_l[\varphi]}{\mathbf{n}_l[\varphi+\delta\varphi]}\big)
~=~ S\,\delta\varphi \;\; (\text{mod } 2\pi),
\label{eq:coh_overlap_small}
\end{equation}

The soliton ($\tau=+$) and antisoliton $(\tau=-)$  states centered around the site $j$ can then be expressed as
\begin{equation}
\ket{\varphi_{\tau}^{(j)}}    = \bigotimes_l |\mathbf{n}_l[\varphi_{\tau}(l-j)] \rangle,
\end{equation}
where 
$\varphi_{\tau}(l-j)$ is the azimuthal angle of the classical soliton solution.
Then, the total overlap between the soliton ($\tau=+$) or antisoliton $(\tau=-)$ product states,
$\ket{\varphi_{\tau}^{(j)}}$ and $\ket{\varphi_{\tau}^{(j+1)}}$ centered at sites $j$ and $j+1$,
 is
\begin{eqnarray}
\zeta &\equiv& \langle \varphi_{\tau}^{(j)} | \varphi_{\tau}^{(j+1)}\rangle
= \prod_l \langle \mathbf n_{l} [\varphi_{\tau}(l-j)]| \mathbf n_{l} [\varphi_{\tau}(l-j-1)] \rangle
\nonumber \\
&=& |\zeta|\,e^{\,i S\,\Delta\Phi},
\end{eqnarray}
with
\begin{equation}
\Delta\Phi \equiv \sum_l \big[\,\varphi_{\tau}(l-(j+1)) - \varphi_{\tau}(l-j)\,\big].
\end{equation}

For a chiral soliton or antisoliton with unit winding, $\Delta\Phi=\pm 2\pi$ for $\tau=\pm 1$, we have
\begin{equation}
\zeta = |\zeta|\,e^{\pm i2\pi S} = |\zeta|(-1)^{2S}.
\label{eq:s_final}
\end{equation}
Thus the sign of the soliton overlap alternates between integer and half-integer spin.
Note that the real character of $\zeta$ makes it independent of the $\tau$ index: $\zeta_+=\zeta^*_{-}=\zeta_-=\zeta$.

Consider now the DM part of $\hat{H}_S$,
\begin{equation}
\hat H_{\rm DM} = -D\sum_l \hat h_l,
\qquad
\hat h_l = (\hat{\mathbf S}_l\times  \hat{\mathbf S}_{l+1}) \cdot \hat{e}_z,
\end{equation}
and define,
\begin{equation}
h_0=\bra{{\varphi_{\tau}}^{(j)}}\hat H_{\rm DM}\ket{\varphi_{\tau}^{(j)}},\quad
h_1=\bra{{\varphi}^{(j)}_{\tau}}\hat H_{\rm DM}\ket{{\varphi}_{\tau}^{(j+1)}}.
\end{equation}
Using the exact identity for coherent states~\cite{Radcliffe_1971},
\begin{equation}
\frac{\bra{\mathbf n_l } \hat{S}^\alpha_l\ket{{\mathbf n}_l'}}{\braket{{\mathbf n}_l} {{\mathbf n_l'}}}
= S\,\mathcal A^\alpha(\mathbf n_l,\mathbf n_l'),
\end{equation}
where,
\begin{equation}
\boldsymbol{\mathcal A}(\mathbf n_l,\mathbf n_l')
=
\frac{\mathbf n_l+\mathbf n_l'-i\,\mathbf n_l\times\mathbf n_l'}{1+\mathbf n_l\cdot\mathbf n_l'},
\end{equation}
one finds that,
\begin{eqnarray}
\frac{h^{\tau}_1}{\zeta} &=& -DS^2
\sum_l \big(\boldsymbol{\mathcal A}^{\tau}_l\times\boldsymbol{\mathcal A}^{\tau}_{l+1}\big)_z,
\nonumber \\ 
h^{\tau}_0 &=& -DS^2
\sum_l (\mathbf n_l[\varphi_{\tau}(l-j)]\times \mathbf n_{l+1}[\varphi_{\tau}(l+1-j)])_z.  
\nonumber \\
\end{eqnarray}
with
\begin{equation}
\boldsymbol{\mathcal A}^{\tau}_l \equiv
\boldsymbol{\mathcal A}(\mathbf n_l[\varphi_{\tau}(l-j)],\mathbf n_l[\varphi_{\tau}(l-j-1)]), 
\end{equation}
To find the sign of the nearest-neighbor soliton hopping we introduce  the difference
\begin{equation}
\delta E_{\tau} \equiv 
\frac{\bra{{\varphi}_{\tau}^{(j)}}\hat H_{\rm DM}\ket{{\varphi}_{\tau}^{(j+1)}}}{\braket{{\varphi}_{\tau}^{(j)}}{{\varphi}_{\tau}^{(j+1)}}}
-
\bra{{\varphi}_{\tau}^{(j)}}\hat H_{\rm DM}\ket{{\varphi}_{\tau}^{(j)}}.
\label{eq:deltaE_def}
\end{equation}
This quantity is real because there is a symmetry transformation that exchanges the solitons centered at sites $j$ and $j+1$. 
According to ~\eqref{eq:Fmn_asymptotic}, the overlaps $\braket{{{\varphi}_{\tau}^{(j)}}}{{{\varphi}_{\tau}^{(j+n)}}} =\cal{O}(\zeta^n)$, which implies that, to leading order in $\zeta$, the ``Wannierization'' of the soliton state gives
\begin{equation}
\ket{{\Phi}_{\tau}^{(j)}}=\ket{{\varphi}_{\tau}^{(j)}}
-\frac{\zeta}{2}\left(\ket{{\varphi}_{\tau}^{(j+1)}}+\ket{{\varphi}_{\tau}^{(j-1)}}\right) 
+ \mathcal O(\zeta^2).
\end{equation}
where the coefficients of the expansion result from imposing the orthogonality condition $\braket{{{\Phi}_{\tau}^{(j)}}}{{{\Phi}_{\tau}^{(j')}}} =\delta_{jj'}$.
Then, the DM-induced hopping matrix element is, 
\begin{equation}
t^{\tau}_{\rm DM}
= \bra{{\Phi}_{\tau}^{(j)}}\hat H_{\rm DM}\ket{{\Phi}_{\tau}^{(j+1)}}
= \zeta\,\delta E_{\tau} + \mathcal O(\zeta^2).
\label{eq:tDM_general}
\end{equation}
Hence, the sign of $t_{\rm DM}$ is determined by the Berry-phase factor in $\zeta$ and the real prefactor $\delta E_{\tau}$.

Using \eqref{eq:s_final} in \eqref{eq:tDM_general}, we obtain
\begin{equation}
t^{\tau}_{\rm DM} = |\zeta|\,(-1)^{2S}\,\delta E_{\tau},
\label{eq:f16}
\end{equation}
which implies
\begin{equation}
\operatorname{sgn}(t^{\tau}_{\rm DM})=
\begin{cases}
+\operatorname{sgn}(\delta E_{\tau}) & S\ \text{integer}\\[4pt]
-\operatorname{sgn}(\delta E_{\tau}) & S\ \text{half-integer}.
\end{cases}
\label{eq:f17}
\end{equation}

The remaining task is to determine the sign of $\delta E_{\tau}$.  To this end, we
expand to second order in
\begin{equation}
d_l \equiv \varphi_{+}(l-j-1)-\varphi_{+}(l-j).
\label{eq:F18}
\end{equation} 
The coherent-state connection expands as
\begin{equation}
\boldsymbol{\mathcal A}^{\tau}_l
=
\mathbf n_l^{(j\tau)}
+ \tau \tan{\frac{d_l}{2}} (\hat{\mathbf z} \times \mathbf n_l^{(j)}
-i\hat{\mathbf z}),
\end{equation}
where we have used $\varphi_{+} = - \varphi_-$ and $n_l^{(j\tau)} \equiv \mathbf n_{l+1}[\varphi_{\tau}(l+1-j)]$.
This equation leads to
\begin{eqnarray}
\big(\boldsymbol{\mathcal A}^{\tau}_l\times\boldsymbol{\mathcal A}^{\tau}_{l+1}\big)_z
&=&
\big(\mathbf n_l^{(j\tau)}\times\mathbf n_{l+1}^{(j\tau)}\big)_z
- \tau \tan{\frac{d_l}{2}} \tan{\frac{d_{l+1}}{2}}\,\sin d_{l+1}
\nonumber \\
&+&\tau \left( \tan{\frac{d_{l+1}}{2}} - \tan{\frac{d_l}{2}} \right) \cos{d_{l+1}} .
\label{eq:Al_cross_final_Text_app}
\end{eqnarray}
Substituting this expansion into \eqref{eq:deltaE_def} gives
\begin{eqnarray}
\delta E_{\tau} &=& \tau DS^2\sum_l
 \tan{\frac{d_l}{2}} \tan{\frac{d_{l+1}}{2}}\,\sin d_{l+1} 
 \nonumber \\
 &-& \tau DS^2\sum_l \left( \tan{\frac{d_{l+1}}{2}} - \tan{\frac{d_l}{2}} \right) \cos{d_{l+1}}.    
\end{eqnarray}
For smooth soliton profiles, \eqref{eq:F18} can be Taylor expanded. In this limit, the second term is parametrically subleading Its lowest nonvanishing contribution in the gradient expansion arises only at fifth order, whereas the leading term contributes already at third order:
\begin{equation*}
\tau \sum_l \!\left( \tan\!\frac{d_{l+1}}{2} - \tan\!\frac{d_l}{2} \right) \!\cos d_{l+1} 
~\rightarrow~
\frac{a^{4}}{4}\int\!dx\, \partial_x\varphi_{\tau}\,(\partial^2_x\varphi_{\tau})^{2}.
\end{equation*}
Therefore, to leading  order in $d_l$, we have
\begin{equation}
\delta E_{\tau} \simeq \tau DS^2\sum_l
\tan{\frac{d_l}{2}} \tan{\frac{d_{l+1}}{2}}\,\sin d_{l+1},
\end{equation}
which, in the long wavelength (continuum) limit, reduces to 
\begin{equation}
\delta E_{\tau} \simeq \frac{ a^2 D S^2}{4} \int\!dx\, (\partial_x \varphi_{\tau}(x))^3.
\end{equation}
By using the sine-Gordon solution identity $(\partial_x \varphi_{\tau})^2 = 2 m^2 (1-\cos\varphi_{\tau})$, this can be further simplified to
\begin{equation}
\delta E_{\tau} \simeq 
- \tau a^2 \pi D^2 S^2m^2. 
\end{equation}

We then conclude that $\delta E<0$ for the soliton solution ($d_l D <0$), while  
$\delta E>0$ ($d_l D > 0$) for the antisoliton solution. 
Because $\delta E_{\tau}$ is odd under the transformation $\tau \to -\tau$, the DM contribution to the hopping amplitude acquires opposite signs for soliton and antisoliton configurations. This result follows from the fact that $\hat{H}_{\rm DM}$ is odd under spatial inversion about site $j$,
$\hat{\mathcal I}_j \hat{H}_{\rm DM} \hat{\mathcal I}_j = -\hat{H}_{\rm DM}$,
and that this inversion transforms a soliton into an antisoliton,
$\hat{\mathcal I}_j \ket{\varphi^{(j)}_{+}} = \ket{\varphi^{(j)}_{-}}$.  
Thus,
\begin{eqnarray}
\mel{\varphi^{(j)}_{+}}
     {\hat{H}_{\rm DM}}
     {\varphi^{(j+1)}_{+}}
     &=& \mel{\varphi^{(j)}_{+}}
     {\hat{\mathcal I}_j^{\,2}\hat{H}_{\rm DM}\hat{\mathcal I}_j^{\,2}}
     {\varphi^{(j+1)}_{+}}
     \nonumber \\
&=&
-\,\mel{\varphi^{(j)}_{-}}
        {\hat{H}_{\rm DM}}
        {\varphi^{(j+1)}_{-}},
\end{eqnarray}
confirming that the DM-induced hopping amplitudes for solitons and antisolitons differ by a minus sign.
\color{black}

Combined with Eqs. (\ref{eq:f16}) and (\ref{eq:f17}) this result yields the integer and half-integer
alternation of the DM-induced soliton hopping amplitude. Solitons of the half-integer spin chain have $t_{\rm DM} > 0$, while for those of the integer one $t_{\rm DM} < 0$. Consequently, the
Berry-phase structure enforces soliton band minima at $k=0$ for integer $S$ and at
$k=\pi$ for half-integer $S$. The situation is reversed for antisolitons.

Note that, in this appendix, we have evaluated only the DM
contribution to the nearest-neighbor hopping amplitude. This restriction is justified 
because the DM term yields a contribution whose magnitude exceeds the combined 
contributions from the Heisenberg and Zeeman terms; consequently, the sign of $t_1$ 
is fixed by the DM part computed here. That said, by following an analogous 
derivation, one can show that the Heisenberg and Zeeman contributions to $t_1$ 
also exhibit an alternation between half-integer and integer spin, since both are 
linear in $\zeta$ at leading order.

\section{\label{Appe:dmrg} Details about TEBD calculations}

To compute the dynamical correlation functions, we employe the TEBD method as implemented in the \texttt{ITensor.jl} (v0.9) library~\cite{itensor,itensor-r0.3}. The algorithmic steps are as follows:

\begin{enumerate}
    \item Use DMRG to obtain the ground state $\ket{0}$.
    \item At $t=0$, apply the spin operator at site $j'$ to the ground state: $\ket{\psi(0)} = \hat{S}_{j'}^\beta \ket{0}$.
    \item Evolve the state in time with a small time step $\delta t \ll 1$ using TEBD to obtain $\ket{\psi(t=\delta t)} = e^{-i\hat{H}\delta t}\ket{\psi(t=0)}$.  
    \item Compute ${\cal S}^{\alpha\beta}_{jj'}(t=\delta t) = \mel{0}{\hat{S}^\alpha_{j}}{\psi(t=\delta t)}$.
    \item Continue the time evolution of the initial state by repeating step 3 a total of $n_{\textrm{steps}} = T_f / \delta t$ times, with $T_f$ the total evolution time.
\end{enumerate}

After completing these steps, a Fourier transformation in both space and time is applied to obtain the dynamical function. However, since the total evolution time $T_f$ is finite ($T_f < \infty$), the Fourier transform of ${\cal S}^{\alpha\beta}_{jj'}(t)$ will exhibit nonphysical oscillations due to the truncation in time. To avoid this effect, we multiply the time-dependent correlator by a Parzen function \cite{Li2022}, 

\begin{equation}
W(x,a) = \left\lbrace
\begin{array}{ll}
 1-6\left|\frac{x}{a}\right|^2 + 6\left|\frac{x}{a}\right|^3 & \textup{if}~  |x|\leq \frac{a}{2}\\
 2\left( 1 - \left|\frac{x}{a}\right|^3 \right)^3 & \textup{if}~  \frac{a}{2} <|x|\leq a\\ 
  0 & \textup{if}~  |x|>a\\
\end{array}
\right.
\end{equation}

Hence, we compute,
\begin{equation}
    \begin{split}
        {\cal S}^{\alpha\beta}_{jj'}(\omega) &= \int_{-T_f}^{{T_f}} dt~ e^{i\omega t} {\cal S}^{\alpha\beta}_{jj'}(t)W(t,T_f) \\ 
        &=\int_{-\infty}^{{\infty}} dt~ e^{i\omega t} {\cal S}^{\alpha\beta}_{jj'}(t)W(t,T_f) \\ 
        &=\frac{1}{2\pi}\int_{-\infty}^{{\infty}} d\omega'~ {\cal F}[ {\cal S}^{\alpha\beta}_{jj'}(t)](\omega') {\cal F}[W(t,T_f)](\omega - \omega') \\ 
        &=\frac{1}{2\pi}  {\cal F}[ {\cal S}^{\alpha\beta}_{jj'}(t)](\omega) *{\cal F}[W(t,T_f)](\omega )
    \end{split}
\end{equation}
with the convolution kernel
\begin{equation}
    \frac{1}{2\pi} {\cal F}[W(t,T_f)](\omega ) = \frac{96\sin^4{(T_f\omega/4)}}{\pi T_f^4\omega^4}.
\end{equation}
The frequency resolution is determined by the standard error of the distribution, which is equal to $2\sqrt{3}/T_f$. 

In addition, several symmetry properties of the correlators can be used to reduce the computational cost. Summarizing them for the case $\alpha=\beta$ (the one relevant for us), we have

\begin{enumerate}
    \item ${\cal S}_{jj'}(t) =  {\cal S}^{*}_{j'j}(-t)$ \quad \textrm{(hermiticity of $\hat{H}$)}
    \item ${\cal S}_{jj'}(-t) =  {\cal S}^{*}_{jj'}(t)$ \quad \textrm{[symmetry ${\cal T}_R R^z(\pi)$]}
    \item ${\cal S}_{jj'}(t) =  {\cal S}_{j'j}(t)$ \quad \textrm{(1 and 2)}
    \item ${\cal S}_{jj'}(t) = {\cal S}(j-j',t)$ \quad \textrm{(translational invariance)}
\end{enumerate}

\section{\label{Appe:bound} Hybridization between the single-magnon and the two-magnon bound state in the high field limit}

In the high-field limit $H \geq J$, the soliton size becomes of the order of a few lattice spacings. In this case, we can approximately think of it as a few-magnon bound state. Specifically, here we consider the two-magnon bound state and its hybridization with the single-magnon state through the DM interaction, which, as always in our paper, is directed perpendicular to the magnetic field axis. Our consideration of the magnon bound state closely follows Sec.III of the Supplemental Material of Ref.\cite{Keselman2020}, which analyzed the bound magnon pair in the polarized state of the antiferromagnetic chain. This approach is easily adapted to the current case by changing the sign of the exchange $J_1 \to -J$. Finally, here we assume $D \ll J$ and treat the DM interaction as a perturbation that mixes one- and two-magnon subspaces. 

Under these assumptions, the one-magnon state is described by the standard
\begin{equation}
    |k\rangle = \frac{1}{\sqrt{L}} \sum_{l=1 }^L e^{i k l} \hat{S}^{-}_l |\uparrow \uparrow...\rangle = \frac{1}{\sqrt{L}} \sum_{l=1}^L e^{i k l} |l\rangle
    \label{eq:b1}
\end{equation}
where $\ket{l}$ describes the state with down spin at location $l$ and all other spins pointing up. The magnon dispersion is given by the standard $\epsilon_1(k) = H + J(1-\cos{(k)}) = H + 2J \sin^2(k/2)$. 

The two-magnon bound state with the center-of-mass momentum $K$ is given by~\cite{Keselman2020}
\begin{eqnarray}
|2 K\rangle &=& \sum_{l,l'=1}^L \psi_{ll'} \hat{S}^{-}_l \hat{S}^{-}_{l'} |\uparrow \uparrow...\rangle, \nonumber\\
\psi_{ll'} &=& \frac{1}{2\sqrt{L}} \sin(K/2) e^{i K (l+l')/2} e^{-\gamma(|l-l'|-1)}, \nonumber\\
\epsilon_2(K) &=& 2H + J \sin^2(K/2), 
\label{eq:b2}
\end{eqnarray}
where the bound-state radius $1/\gamma$ is determined by $K$ via $e^{-\gamma} = \cos(K/2)$.

Our goal is to find the overlap $\langle 2K| \hat{H}_{\rm DM} |k\rangle$. Consider
\begin{eqnarray}
    \hat{H}_{\rm DM} |k\rangle &=& \frac{D}{2 i \sqrt{L}}\sum_{l=1}^L e^{i k l} \sum_{l'=1}^L (\hat{S}^z_{l'} - \hat{S}^z_{l'+2})(\hat{S}^+_{l'+1} - \hat{S}^{-}_{l'+1}) |l\rangle \nonumber\\
    &=& \frac{D}{2 i \sqrt{L}}\sum_{l=1}^L e^{i k l} (|l,l+1\rangle - |l-1,l\rangle),
    \label{eq:b3}
\end{eqnarray}
where $|l,l'\rangle$ denotes the state with down spins at $l$ and $l'$. Using \eqref{eq:b2}, it is easy to find that
\begin{equation}
    \langle 2K| \hat{H}_{\rm DM} |k\rangle = D \sin^2(k/2) \delta_{k,K}
    \label{eq:b4}
\end{equation}
Clearly, the hybridization is strongest at the zone boundary, $k = \pi$ and it is absent in the zone center, $k=0$.

Therefore, the effective Hamiltonian describing the mixing of the one- and two-magnon sectors is given by the $2\times 2$ matrix with diagonal elements $\epsilon_1(k)$ and $\epsilon_2(k)$, and the off-diagonal ones $ D\sin^2 (k/2)$. The new eigenvalues follow immediately,
\begin{widetext}
\begin{eqnarray}
    E(k) &=& \tfrac{1}{2}\Big[\epsilon_1(k) + \epsilon_2(k) 
    \pm \sqrt{(\epsilon_1(k) - \epsilon_2(k))^2 + 4D^2 \sin^4(k/2)} ~\Big]\nonumber
    \approx
    \begin{cases} 
        \epsilon_2(k) + \dfrac{D^2}{H}\sin^4\!\big(k/2\big), \\[6pt]
        \epsilon_1(k) - \dfrac{D^2}{H}\sin^4\!\big(k/2\big).
    \end{cases}
\end{eqnarray}
\end{widetext}
where the last line is obtained in the $H \gg J, D$ limit. It is also clear that the DM interaction couples the two-magnon sector to the three-magnon one, and so on. Thus, one should be able to reconstruct the soliton and antisoliton excitations as multimagnon bound states. At this stage, we refrain from a more general analysis of this interesting problem.

\section{\label{Appe:Lowdin}L\"owdin symmetric orthogonalization}

In Sec.~\ref{Sec:hybridization}, the hybridized band structure at intermediate fields is obtained by projecting the Hamiltonian onto a low-energy subspace $\mathcal{S}_0$ spanned by nonorthogonal variational states. Here, we provide the details of the orthogonalization procedure used in that construction.

Let $\{\ket{\phi_j}\}_{j=1}^{N}$ denote the set of (in general, nonorthogonal) variational states spanning $\mathcal{S}_0$---for instance, the soliton, antisoliton, single-magnon, and composite magnon-soliton states introduced in the main text. Their mutual overlaps define the $N \times N$ overlap matrix
\begin{equation}
    S_{jj'} = \braket{\phi_j}{\phi_{j'}}.
    \label{Eq:overlap_matrix}
\end{equation}
Since the states $\{\ket{\phi_j}\}$ are assumed to be linearly independent, $S$ is a positive-definite Hermitian matrix and its inverse square root $S^{-1/2}$ is well defined.

The L\"owdin symmetric orthogonalization~\cite{Lowdin1950,Mayer2002} constructs an orthonormal basis $\{\ket{\psi_j}\}$ via the transformation
\begin{equation}
    \ket{\psi_j} = \sum_{l=1}^{N} (S^{-1/2})_{jl}\,\ket{\phi_l}.
    \label{Eq:Lowdin_transform}
\end{equation}
By construction, these states satisfy

\begin{equation}
\begin{split}
    \braket{\psi_j}{\psi_{j'}} 
    &= \sum_{l,l'} (S^{-1/2})^{*}_{jl}\,(S^{-1/2})_{j'l'}\, \braket{\phi_l}{\phi_{l'}} \\
    &= \big(S^{-1/2}\, S\, S^{-1/2}\big)_{jj'} 
    = \delta_{jj'}.
\end{split}
\end{equation}

A key property of the L\"owdin procedure, which distinguishes it from other orthogonalization methods such as Gram-Schmidt, is that it is symmetric: It treats all basis vectors on equal footing and does not depend on an arbitrary ordering of the states. Moreover, it can be shown~\cite{Lowdin1950} that the L\"owdin-orthogonalized basis minimizes the total deviation from the original set of vectors in the least-squares sense,
\begin{equation}
    \sum_{j=1}^{N} \big\| \ket{\psi_j} - \ket{\phi_j} \big\|^2 \quad \text{is minimized},
\end{equation}
among all orthonormal bases of the form $\ket{\psi_j} = \sum_l U_{jl} \ket{\phi_l}$. This property ensures that each orthogonalized state $\ket{\psi_j}$ retains, as closely as possible, the physical character of the corresponding original state $\ket{\phi_j}$.

In practice, we compute $S^{-1/2}$ by diagonalizing the overlap matrix, $S = V \Lambda V^{\dagger}$, where $\Lambda = \mathrm{diag}(\lambda_1, \ldots, \lambda_N)$ contains the eigenvalues and $V$ the corresponding eigenvectors, so that
\begin{equation}
    S^{-1/2} = V\, \Lambda^{-1/2}\, V^{\dagger}, 
    \ \  
    \Lambda^{-1/2} = \mathrm{diag}\!\big(\lambda_1^{-1/2}, \ldots, \lambda_N^{-1/2}\big).
    \label{Eq:Sinvhalf}
\end{equation}

Once the orthonormal basis $\{\ket{\psi_j}\}$ is obtained, the projected Hamiltonian matrix is
\begin{equation}
    h^{\mathrm{multi}}_{jj'}(k) 
    = \mel{\psi_j}{\hat{H}_{S}}{\psi_{j'}} - E_{0}\,\delta_{jj'},
    \label{Eq:projected_H}
\end{equation}
where $E_0$ is the ground-state energy. The hybridized excitation energies are then given by the eigenvalues of $h^{\mathrm{multi}}(k)$ at each momentum $k$, and the eigenvectors encode the admixture of solitonic, antisolitonic, and magnonic character in each hybridized mode.

\bibliography{soliton}

\end{document}